\documentclass[longauth]{aa}   

\usepackage{graphicx}
\usepackage{txfonts}
\usepackage[
hidelinks=true,
colorlinks=true,
colorlinks=blue,
citecolor=blue]{hyperref}

\usepackage{gensymb}
\usepackage{multirow}

\usepackage{amsmath}
\usepackage{float}
\usepackage{makecell}
\usepackage{afterpage}
\usepackage{needspace}
\usepackage{subcaption}
\usepackage{placeins}
\usepackage[table]{xcolor}

\newcommand{\clean}{\texttt{CLEAN}\xspace} 
\newcommand{\frank}{\texttt{frank}\xspace}
\newcommand{\rave}{\texttt{rave}\xspace}
\newcommand{\fave}{\texttt{fave}\xspace}
\newcommand{\arksia}{\texttt{arksia}\xspace}
\defcitealias{Krivov_2021}{K21}

\graphicspath{{./}{Figures/}}
\begin{document}

\title{The ALMA survey to Resolve exoKuiper belt Substructures (ARKS)}
\subtitle{III: The vertical structure of debris disks}

\author{B.~Zawadzki\textsuperscript{1}\fnmsep\thanks{E-mail: bzawadzki@wesleyan.edu} \and A.~Fehr\textsuperscript{2} \and A.~M.~Hughes\textsuperscript{1} \and E.~Mansell\textsuperscript{1} \and J.~Kittling\textsuperscript{3} \and Y.~Han\textsuperscript{4} \and C.~Hou\textsuperscript{1} \and M.~Pan\textsuperscript{5} \and J.~Milli\textsuperscript{6} \and J.~Olofsson\textsuperscript{7} \and T.D.~Pearce\textsuperscript{8} \and A.~A.~Sefilian\textsuperscript{9} \and A.~Nurmohamed\textsuperscript{1} \and J.~Lee\textsuperscript{1} \and Y.~Mpofu\textsuperscript{1} \and M.~Bonduelle\textsuperscript{6} \and M.~Booth\textsuperscript{10} \and A.~Brennan\textsuperscript{11} \and C.~del~Burgo\textsuperscript{12,13} \and J.~M.~Carpenter\textsuperscript{14} \and G.~Cataldi\textsuperscript{15,16} \and E.~Chiang\textsuperscript{17} \and S.~Ertel\textsuperscript{9,18} \and Th.~Henning\textsuperscript{19} \and M.~R.~Jankovic\textsuperscript{20} \and G.~M.~Kennedy\textsuperscript{21,8} \and \'A.~K\'osp\'al\textsuperscript{22,23,19} \and A.~V.~Krivov\textsuperscript{24} \and J.~B.~Lovell\textsuperscript{2} \and P.~Luppe\textsuperscript{11} \and M.~A.~MacGregor\textsuperscript{25} \and S.~Mac~Manamon\textsuperscript{11} \and S.~Marino\textsuperscript{26} \and J.~P.~Marshall\textsuperscript{27} \and L.~Matr\`a\textsuperscript{11} \and A.~Mo\'or\textsuperscript{22} \and S.~P\'erez\textsuperscript{28,29,30} \and P.~Weber\textsuperscript{28,29,30} \and D.~J.~Wilner\textsuperscript{2} \and M.~C.~Wyatt\textsuperscript{31}} 

\institute{
Department of Astronomy, Van Vleck Observatory, Wesleyan University, 96 Foss Hill Dr., Middletown, CT, 06459, USA \and
Center for Astrophysics | Harvard \& Smithsonian, 60 Garden St, Cambridge, MA 02138, USA \and
Kavli Institute for Particle Astrophysics and Cosmology, 382 Via Pueblo Mall Stanford, CA 94305-4060, USA \and
Division of Geological and Planetary Sciences, California Institute of Technology, 1200 E. California Blvd., Pasadena, CA 91125, USA \and
Smithsonian Astrophysical Observatory, 60 Garden St, Cambridge, MA 02138, USA \and
Univ. Grenoble Alpes, CNRS, IPAG, F-38000 Grenoble, France \and
European Southern Observatory, Karl-Schwarzschild-Strasse 2, 85748 Garching bei M\"unchen, Germany \and
Department of Physics, University of Warwick, Gibbet Hill Road, Coventry CV4 7AL, UK \and
Department of Astronomy and Steward Observatory, The University of Arizona, 933 North Cherry Ave, Tucson, AZ, 85721, USA \and
UK Astronomy Technology Centre, Royal Observatory Edinburgh, Blackford Hill, Edinburgh EH9 3HJ, UK \and
School of Physics, Trinity College Dublin, the University of Dublin, College Green, Dublin 2, Ireland \and
Instituto de Astrof\'isica de Canarias, Vía L\'actea S/N, La Laguna, E-38200, Tenerife, Spain \and
Departamento de Astrof\'isica, Universidad de La Laguna, La Laguna, E-38200, Tenerife, Spain \and
Joint ALMA Observatory, Avenida Alonso de C\'ordova 3107, Vitacura 7630355, Santiago, Chile \and
National Astronomical Observatory of Japan, Osawa 2-21-1, Mitaka, Tokyo 181-8588, Japan \and
Department of Astronomy, Graduate School of Science, The University of Tokyo, Tokyo 113-0033, Japan \and
Department of Astronomy, University of California, Berkeley, Berkeley, CA 94720-3411, USA \and
Large Binocular Telescope Observatory, The University of Arizona, 933 North Cherry Ave, Tucson, AZ, 85721, USA \and
Max-Planck-Insitut f\"ur Astronomie, K\"onigstuhl 17, 69117 Heidelberg, Germany \and
Institute of Physics Belgrade, University of Belgrade, Pregrevica 118, 11080 Belgrade, Serbia \and
Malaghan Institute of Medical Research, Gate 7, Victoria University, Kelburn Parade, Wellington, New Zealand \and
Konkoly Observatory, HUN-REN Research Centre for Astronomy and Earth Sciences, MTA Centre of Excellence, Konkoly-Thege Mikl\'os \'ut 15-17, 1121 Budapest, Hungary \and
Institute of Physics and Astronomy, ELTE E\"otv\"os Lor\'and University, P\'azm\'any P\'eter s\'et\'any 1/A, 1117 Budapest, Hungary \and
Astrophysikalisches Institut und Universit\"atssternwarte, Friedrich-Schiller-Universit\"at Jena, Schillerg\"a{\ss}chen 2-3, 07745 Jena, Germany \and
Department of Physics and Astronomy, Johns Hopkins University, 3400 N Charles Street, Baltimore, MD 21218, USA \and
Department of Physics and Astronomy, University of Exeter, Stocker Road, Exeter EX4 4QL, UK \and
Academia Sinica Institute of Astronomy and Astrophysics, 11F of AS/NTU Astronomy-Mathematics Building, No.1, Sect. 4, Roosevelt Rd, Taipei 106319, Taiwan. \and
Departamento de Física, Universidad de Santiago de Chile, Av. V\'ictor Jara 3493, Santiago, Chile \and
Millennium Nucleus on Young Exoplanets and their Moons (YEMS), Chile \and
Center for Interdisciplinary Research in Astrophysics Space Exploration (CIRAS), Universidad de Santiago, Chile \and
Institute of Astronomy, University of Cambridge, Madingley Road, Cambridge CB3 0HA, UK
}

\date{Received YYY; accepted XXX}

\abstract
    {
    Debris disks --- collisionally sustained belts of dust and sometimes gas around main sequence stars --- are remnants of planet formation processes and are found in systems ${\gtrsim}10$ Myr old.
    Millimeter-wavelength observations are particularly important, as the grains probed by these observations are not strongly affected by radiation pressure and stellar winds, allowing them to probe the dynamics of large bodies producing dust.
    The ALMA survey to Resolve exoKuiper belt Substructures (ARKS) is analyzing high-resolution observations of 24 debris disks to enable the characterization of debris disk substructures across a large sample for the first time.
    }
    {
    For the most highly inclined disks, it is possible to recover the vertical structure of the disk. We aim to model and analyze the most highly inclined systems in the ARKS sample in order to uniformly extract the vertical dust distributions for a sample of well-resolved debris disks.
    }
    {
    We employed both parametric and nonparametric methods to constrain the vertical dust distributions for the most highly inclined ARKS targets.
    }
    {We find a broad range of aspect ratios, revealing a wide diversity in vertical structure, with a range of best-fit parametric values of $0.0026 \leq h_{\rm HWHM} \leq 0.193$ and a median best-fit value of $h_{\rm HWHM}=0.021$. The results obtained by nonparametric modeling are generally consistent with the parametric modeling results. We find that five of the 13 disks are consistent with having total disk masses less than that of Neptune (17~$M_{\oplus}$), assuming stirring by internal processes (self-stirring and collisional and frictional damping). Furthermore, most systems show a significant preference for a Lorentzian vertical profile rather than a Gaussian.
    }
    {}


\maketitle

\section{Introduction} \label{sec:intro}

Debris disks offer a window into the final stages of planet formation, revealing details about the dynamical evolution of systems on timescales from tens of millions to billions of years \citep[for reviews, see: ][]{Wyatt_2008, Matthews_2014, Hughes_2018}. In particular, the vertical distribution of dust in debris disks contains information about the orbital distribution of planetesimals and collision rates within the disk \citep[e.g.,][]{Thebault_2007, Quillen_2007, Moore_2023}. A measurement of the disk scale height at (sub)millimeter wavelengths is effectively a measurement of the inclination dispersion of the dust grains, which in turn enables constraints to be placed on the system dynamics \citep[e.g.,][]{Quillen_2007}. At (sub)millimeter wavelengths, dust grains are largely unaffected by radiation pressure and allow direct probing of the mass distribution and level of dynamical excitation of large planetesimals in the system. 

With high angular resolution (sub)millimeter observations from the Submillimeter Array (SMA) and the Atacama Large Millimeter/submillimeter Array (ALMA), dozens of debris disks have now been resolved, including some with marginally resolved or partially constrained vertical structures \citep[e.g.,][]{Kennedy_2018, Daley_2019, Matra_2019, Vizgan_2022, Hales_2022, Marshall_2023, Terrill_2023, Matra_2025}. \citet{Matra_2019} resolved the vertical structure in the disk around $\beta$~Pictoris and found that a two-component vertical structure including both a dynamically cold and hot population was required to match the data --- a structure that is reminiscent of the dual populations of planetesimals in the classical Kuiper belt caused by Neptune's migration early in the history of the Solar System \citep[e.g.,][]{Brown_2001, Morbidelli_2008, Petit_2011}. \citet{Daley_2019} resolved the vertical structure of the edge-on debris disk around AU Mic using 1.3 mm data and used grain-size dependent velocity dispersions in the collisional cascade models of \citet{Pan_2012} to infer a total mass of stirring bodies in the debris belt no greater than 1.8\,M$_{\oplus}$. Incorporating additional observations, \citet{Vizgan_2022} examined the relationship between grain size and scale height by obtaining scale height values using both 1.3 mm and 450 ${\mu}$m data. \citet{Kennedy_2018} modeled the HR~4796A (HD~109573) debris disk and determined that their modeling consistently returned a marginally resolved value for the scale height; however, this value was somewhat inconsistent with the indications from scattered light of a dynamically cold population of planetesimals \citep[e.g.,][]{Kenyon_1999}. These observations revealed the promise of using vertical structure to interpret the dynamics of debris disk systems, although larger samples are needed to draw more robust conclusions.

While (sub)millimeter observations closely trace the masses and dynamical histories of debris disks, scattered light observations probe smaller grains that are sensitive to both radiation forces and collisions. At these wavelengths, debris disks are expected to have a minimum vertical aspect ratio full width at half maximum (FWHM) of $h_{\rm{min}} = 0.04\pm0.02$, which is driven by radiation pressure and collisions inflating inclinations and eccentricities of the small grains \citep{Thebault_2009}. The vertical structures of debris disks have already been characterized for several disks in scattered light, with most scale height measurements at or above the theoretically predicted minimum \citep[e.g.,][]{Milli_2014, MillarBlanchaer_2015, MillarBlanchaer_2016, Engler_2019, Boccaletti_2019, Olofsson_2020, Olofsson_2022}.

However, a larger sample of scale height measurements at (sub)millimeter wavelengths is needed to complement this sample of scattered light scale heights and illuminate connections between grain size and system dynamics. For example, small grains may be stirred to higher inclinations by a giant planet in the disk, inflating the observed scale height \citep{Quillen_2006, Pan_2012}. Furthermore, comparing the vertical structures from millimeter and scattered light observations may enable constraints on the gas mass of debris disks as well as on the origins of the gas \citep{Olofsson_2022}. However, if grains are fully coupled to the gas or turbulent diffusion is significant, inclination damping could be reduced, preventing micron-sized grains from fully settling into the midplane and resulting in a more vertically extended grain distribution \citep{Marino_2022}. More observational measurements of vertical disk structures at multiple wavelengths will reveal how these processes sculpt debris disks.

The ALMA survey to Resolve exoKuiper belt Substructures (ARKS) sample contains ALMA observations of 24 debris disks at a high enough resolution and sensitivity to carry out the detailed characterization of radial and vertical dust structures \citep{Marino_2026}. In order to resolve the vertical distribution of dust, observations of highly inclined (i.e., close to edge-on) debris disks at high angular resolution are necessary. The most highly inclined debris disks in the ARKS sample provide the opportunity to begin building a larger sample of vertically well-resolved debris disks at millimeter wavelengths.

In this paper, we analyze these highly inclined ARKS debris disks in order to place new constraints on their vertical dust distributions. The observations and sample of debris disks are described in Sect. \ref{sec:observations}. Our modeling framework is detailed in Sect. \ref{sec:modeling}, and our analyses are presented in Sect. \ref{sec:analysis}. In Sect. \ref{sec:discuss} we relate our parametric modeling results to theoretical models of debris disk dynamics to infer the masses of stirring bodies and the dynamical history of the system, compare the parametric and nonparametric modeling results, and investigate the relationship between vertical structure at millimeter and scattered light wavelengths. Our main conclusions are summarized in Sect. \ref{sec:conclude}. Some technical details are provided in the appendices.

\section{Observations} \label{sec:observations}

ARKS (2022.1.00338.L, PI: S.\ Marino) aimed to resolve the FWHM of the vertical distribution of the most highly inclined ($i>75$\degree) disks in the sample. The new Band 7 observations were configured to resolve the vertical FWHM of the disks with at least two resolution elements, assuming a vertical aspect ratio\footnote{Several common formulations of the aspect ratio, including $h_\sigma$, are defined in Sect. \ref{sec:modeling:hdef}.} of $h_\sigma = 0.05$. The observing strategy for ARKS targets with $i<75$\degree was not optimized for vertical structure recovery, and observations of these targets often lack the resolution or S/N to firmly constrain vertical dust distributions (assuming the same vertical aspect ratio of $h_\sigma = 0.05$). However, parametric and/or nonparametric methods can recover vertical structure constraints for some disks at $i<75$\degree. In addition to the methods described in this work, for example, azimuthal brightness variations in moderately inclined, optically thin dust rings can be exploited to estimate the disk scale height, provided that the belt is relatively vertically thick and radially narrow \citep{Doi_2021, Villenave_2025}. In this paper, we include all disks with vertical structure constraints, though the focus is on the most highly inclined disks, which are amenable to detailed vertical characterization via parametric modeling.

The ARKS sample also includes six archival observations of well-resolved debris disks. The archival observations for HD~39060 ($\beta$~Pic), HD~107146, HD~197481 (AU~Mic), and HD~206893 were taken in Band 6. The vertical aspect ratio of a disk is known to vary with observing wavelength; this variation has been measured in AU~Mic between 1.3 mm and 450 \textmu m, with the scale height decreasing by a factor of $\sim 1.3$ with a factor of about three decrease in wavelength \citep{Vizgan_2022}. The difference between scale height measurements at Band 7 (0.89 mm) and Band 6 (1.3 mm) is likely more subtle given that the wavelength differs by less than a factor of two, but there are no Band 7 observations of these targets at sufficient resolution to confirm this. The full target selection and observing strategy is justified in \citet{Marino_2026}, as well as further details about the observations and data calibration. In this study, we use the corrected data produced and presented in \citet{Marino_2026}, which account for weights and phase center and flux offsets between execution blocks.

\section{Modeling} \label{sec:modeling}

\subsection{Definitions of scale height} \label{sec:modeling:hdef}
In general, a vertical scale height characterizes the thickness of the disk by quantifying the size of the region around the disk midplane from which the majority of disk emission originates. This can be defined in a number of ways, and can depend on the exact vertical distribution of disk material. Here, we parameterize the disk scale height, $H$, as
\begin{equation} \label{eqn:const_h}
    H(r) = hr    
\end{equation}
such that the vertical aspect ratio, $h$, is constant at all radii, $r$.

It is common to express a disk scale height as the standard deviation of a Gaussian distribution, $H_{\sigma}$. In this work, we consider multiple vertical distributions, including Lorentzian and multi-component vertical dust distributions which do not have a clearly defined standard deviation. Therefore, we report scale heights in terms of the half width at half maximum (HWHM) of the vertical profile, $H_{\rm HWHM}$. We use the HWHM rather than the full width at half maximum (FWHM) to be more directly comparable with legacy $H_{\sigma}$ values. The different scale height formats are related by
\begin{align} \label{eqn:Hformats}
    H_{\rm HWHM} &= \frac{1}{2} H_{{\rm FWHM}} = \sqrt{2 \ln 2}\, H_{\sigma} \\
    h_{\rm HWHM} &= \frac{1}{2} h_{{\rm FWHM}}\; = \sqrt{2 \ln 2}\, h_{\sigma}. \nonumber
\end{align}

We use the above notation to express which scale height or aspect ratio definition is used. For a more direct comparison across different methodologies and archival studies, we preferentially use $H_{\rm HWHM}$ and $h_{\rm HWHM}$, converting to this format when possible. A complete list of aspect ratio values in their original and converted formats is available in Table \ref{tab:all_h}.

\subsection{Parametric modeling}

We use a parametric debris disk modeling code adapted from the publicly available code described in \citet{Flaherty_2015},\footnote{\href{https://github.com/kevin-flaherty/disk model3}{\url{https://github.com/kevin-flaherty/disk_model3}}} which was developed from earlier disk modeling works \citep{Dartois_2003, Rosenfeld_2012, Rosenfeld_2013}. The modeling framework, optimized for modeling the continuum (dust) emission of axisymmetric debris disks,\footnote{While some ARKS debris disks are known to contain gas, we do not calculate hydrostatic equilibrium or the velocity field as we expect the dust mass to dominate over the gas mass.} is described in detail in \citet{Fehr_2023} and summarized below.

As debris disks are optically thin, their temperatures and densities are fully degenerate. Since we are modeling observations at a single wavelength, the geometric results are insensitive to the temperature structure and we assume a blackbody equilibrium temperature structure, allowing us to fit for the density structure. We computed the dust temperature by setting the flux received by the disk from its host star at some distance $d$ from the star ($d=\sqrt{r^2+z^2}$, where $r$ and $z$ are the radial and vertical distances in cylindrical coordinates, respectively) equal to the flux emitted by the dust grains at that distance. The bolometric flux emitted at the surface of a dust grain is 
\begin{equation}
    F_{\text{dust}}(r)=\frac{L_*}{16 \pi d^2},
\end{equation}
where $L_*$ is the bolometric luminosity of the host star. Assuming grains emit and absorb in a manner similar to black bodies, the dust temperature is then
\begin{equation}
    T_{\text {dust }}(r)=\left(\frac{L_*}{16 \pi \sigma d^2}\right)^{\frac{1}{4}},
\end{equation}
where $\sigma$ is the Stefan-Boltzmann constant.

\begin{table*}[t]
\caption{Summary of the density functional forms used in the parametric modeling.}
\label{tab:formsfree}
\centering
\renewcommand{\arraystretch}{1.5}
\begin{tabular}{|c|l|c|}
\hline
Radial Form       & Surface Density Profile & Free Parameters \\
\hline 
\rule{0pt}{5ex} \parbox[c]{2.5cm}{\centering Double Power Law (DPL)} & $\Sigma(r)=10^{\Sigma_{\mathrm{c}}}\left[\left(\dfrac{r}{R_{\mathrm{c}}}\right)^{-\alpha_{\mathrm{in}} \gamma}+\left(\dfrac{r}{R_{\mathrm{c}}}\right)^{-\alpha_{\mathrm{out}} \gamma}\right]^{-1 / \gamma}$ & $R_{\mathrm{c}}$, $\alpha_{\mathrm{in}}$, $\alpha_{\mathrm{out}}$, $\gamma$, $\Sigma_{\mathrm{c}}$                 \\ [3ex]
\rowcolor{gray!15}
\parbox[c]{2.5cm}{\centering Double Gaussian (DG)} & $\Sigma(r)=10^{\Sigma_{\mathrm{c}}}\left[C_1 \exp\left[ \dfrac{-(r-R_1)^2}{2\sigma_1^2} \right] + (1-C_1) \exp\left[ \dfrac{-(r-R_2)^2}{2\sigma_2^2} \right]  \right]$ & \parbox[c]{2.0cm}{\centering $R_1$, $R_2$, $\sigma_1$, $\sigma_2$, $C_1$, $\Sigma_{\mathrm{c}}$} \\ [3ex]
\parbox[c]{2.5cm}{\centering Error Function + Power Law (Erf)} & $\Sigma(r) = 10^{\Sigma_{\mathrm{c}}}\left[1-\text{erf}\left(\dfrac{R_{\mathrm{c}}-r}{\sqrt{2} \sigma_{\text{in}} R_{\mathrm{c}}}\right)\right] \left(\dfrac{r}{R_{\mathrm{c}}}\right)^{-\alpha_{\text{out}}}$ & $R_{\mathrm{c}}, \sigma_{\text{in}}, \alpha_{\text{out}}, \Sigma_{\mathrm{c}} $ \\ [3ex]
\rowcolor{gray!15}
\parbox[c]{2.5cm}{\centering Asymmetric Gaussian (AG)} &
    \(
    \begin{array}{l} 
        \Sigma(r) = 
        \begin{cases}
            10^{\Sigma_{\mathrm{c}}}\exp\left[-\dfrac{(r-R_{\mathrm{c}})^2}{2 \sigma_{\text{in}}^2}\right] & \text{if } r<R_{\mathrm{c}}\\
    10^{\Sigma_{\mathrm{c}}}\exp\left[-\dfrac{(r-R_{\mathrm{c}})^2}{2 \sigma_{\text{out}}^2}\right] & \text{if } r \geq R_{\mathrm{c}}
        \end{cases}
    \end{array}
    \)
    & $R_{\mathrm{c}}, \sigma_{\text{in}}, \sigma_{\text{out}}, \Sigma_{\mathrm{c}} $ \\ [5ex]
\parbox[c]{2.3cm}{\centering Triple Power Law (TPL)} & \parbox[c]{11.0cm}{$\Sigma(r) = 10^{\Sigma_{\mathrm{c}}}\left(\dfrac{R_{\text{in}}}{R_{\text{out}}}\right)^{-\alpha_{\text{mid}}} \left[\left(\dfrac{r}{R_{\text{in}}}\right)^{-\alpha_{\text{in}} \gamma_{\text{in}}} + \left(\dfrac{r}{R_{\text{in}}}\right)^{-\alpha_{\text{mid}} \gamma_{\text{in}}}\right]^{-1/\gamma_{\text{in}}}$ \\ $\textcolor{white}{\Sigma(r) =} \hspace{4cm} \times \left[\left(\dfrac{r}{R_{\text{out}}}\right)^{-\alpha_{\text{mid}} \gamma_{\text{out}}} + \left(\dfrac{r}{R_{\text{out}}}\right)^{-\alpha_{\text{out}} \gamma_{\text{out}}}\right]^{-1/\gamma_{\text{out}}}$} & \parbox[c]{2.0cm}{\centering $R_{\text{in}}$, $R_{\text{out}}$, $\alpha_{\text{in}}$, $\alpha_{\text{mid}}$, $\alpha_{\text{out}}$, $\gamma_{\text{in}}$, $\gamma_{\text{out}}$, $\Sigma_{\mathrm{c}}$ } \\ [5ex]
\rowcolor{gray!15}
Gaussian (G) & $\Sigma(r) = 10^{\Sigma_{\mathrm{c}}}\exp\left[-\dfrac{(r-R)^2}{2 \sigma^2}\right]$ & $R, \sigma, \Sigma_{\mathrm{c}}$ \\ [3ex]
\hline
Vertical Form       & Density Profile & Free Parameters \\ 
\hline 
\rule{0pt}{5ex} Gaussian (vG) & $\rho(r,z) = \Sigma(r) \times \left[ \exp\left[-\dfrac{1}{2} \left( \dfrac{z}{H_{\sigma}(r)} \right)^2 \right] \, \middle/ \sqrt{2\pi}H_{\sigma}(r) \right]$   & $h_{\rm HWHM}$                \\ [3ex]
\rowcolor{gray!15}
Exponential (vE) & $\rho(r,z) = \Sigma(r) \times \left[ \exp\left[-\dfrac{1}{2} \left( \dfrac{|z|}{H_{\sigma}(r)} \right)^{\zeta} \, \right] \, \middle/ \sqrt{2\pi}H_{\sigma}(r) \right]$ & $h_{\rm HWHM}$, $\zeta$  \\ [3ex]
\parbox[c]{2.5cm}{\centering Double Gaussian (vDG)} & $\rho(r,z) = \Sigma(r) \times  \left[\dfrac{C_{\rm{vert}} \exp\left[-\frac{1}{2}(z \middle/ H_{\sigma1}(r))^2\right]}{\sqrt{2\pi}H_{\sigma1}(r)} +  \dfrac{(1-C_{\rm{vert}}) \exp\left[-\frac{1}{2}(z \middle/ H_{\sigma2}(r))^2\right]}{\sqrt{2\pi}H_{\sigma 2}(r)}  \right]$ & $h_{{\rm HWHM} 1}$, $h_{{\rm HWHM} 2}$, $C_{\rm{vert}}$ \\ [3ex]
\rowcolor{gray!15}
Lorentzian (vL) & $\rho(r, z)=\Sigma(r) \times \dfrac{1}{\pi} \dfrac{H_{\rm HWHM}(r)}{z^2+H_{\rm HWHM}(r)^2}$ & $h_{\rm HWHM}$   \\ [3ex]
\hline
\end{tabular}
\tablefoot{Top: Radial forms and free parameters. Bottom: Vertical forms and free parameters. For all radial DPL models, we fix $\gamma=2$. We note that while the vertical aspect ratio free parameters are all listed as $h_{\rm HWHM}$, some vertical forms are more conveniently expressed in terms of $H_{\sigma}$. The conversion to $h_{\rm HWHM}$ is described by Equations \ref{eqn:Hformats} and \ref{eqn:const_h}. In addition to the parameters shown in the table, all models fit for the disk inclination, PA, and stellar flux, as well as RA and declination offsets.}
\end{table*}

The density structure $\rho(r,z)$ of the disk is defined by two components: a radial profile form and a vertical profile form. Careful treatment of the radial structures is critical, as they can be degenerate with the vertical structure of the disk. Thus, the set of radial functional forms applied for the vertical parametric modeling presented here was informed by the ARKS radial structure analysis \citep{Han_2026}. Most of our models use either a radial double power law or a radial double Gaussian; the functional forms and associated free parameters are shown in the top section of Table \ref{tab:formsfree}. For a more complete examination of the disk radial structures and functional forms, refer to \cite{Han_2026}.

For the disk vertical structure, we parameterize the disk scale height according to Eqn. \ref{eqn:const_h}, such that the vertical aspect ratio $h$ is constant throughout the disk in order to minimize additional free parameters. We discuss this assumption and its implications in Sect. \ref{sec:discuss:limits}. The vertical functional forms used in this study are shown in the bottom section of Table \ref{tab:formsfree}, and are motivated by the array of vertical functional forms used in other debris disk studies \citep[e.g.,][]{Golimowski_2006, Lagrange_2012, Apai_2015, Matra_2019}.

\subsubsection{Radiative transfer}
Once the temperature and density structures of the disk have been defined, we compute the flux throughout the disk in order to obtain a model image of the disk with projected major axis $x$, projected minor axis $y$, and distance from disk center along the line of sight $s$. The intensity of the dust emission is
\begin{equation}
    I_\nu(x, y)=\int_0^{\infty} S_\nu(x, y, s) \exp \left[-\tau_\nu(x, y, s)\right] K_\nu(x, y, s) d s    ,
\end{equation}
where $S_\nu(x, y, s)$ is the source function, $\tau_\nu(x, y, s)$ is the optical depth, and $K_\nu(x, y, s)$ is the absorption coefficient. We approximated the source function with the Planck function:
\begin{equation}
    S_\nu(x, y, s)=\frac{2 h_p \nu^3}{c^2} \frac{1}{\exp \left[\frac{h_p \nu}{k_B T(x, y, s)}\right]-1},
\end{equation}
where $h_p$ is Planck’s constant, $c$ is the speed of light, and $k_B$ is the Boltzmann constant.

The optical depth is
\begin{equation}
    \tau_\nu(x, y, s)=\int_0^s K_\nu\left(x, y, s^{\prime}\right) d s^{\prime}.
\end{equation}
We assumed
\begin{equation}
    K_\nu\left(x, y, s\right) = \kappa_\nu \rho\left(x, y, r\right),
\end{equation}
where $\kappa_\nu$ is the dust mass opacity. We adopted $\kappa_\nu=$1.9~cm$^{2}$~g$^{-1}$ at 0.89~mm and $\kappa_\nu=$1.3~cm$^{2}$~g$^{-1}$ at 1.3~mm as in \citet{Marino_2026}, who assumed a power law grain size distribution with a slope of -3.5, and a composition of astrosilicates (70\%), crystalline water ice (15\%), and amorphous carbon (15\%) by mass \citep{Draine_2003, Li_1998}.

After generating the model disk image and multiplying it by a Gaussian approximation of the primary beam,\footnote{The archival Band 6 observations of HD~39060 are mosaicked. For this source, we apply a primary beam correction at each location in the mosaic field.} we used \texttt{galario} to obtain synthetic visibilities sampled at the same spatial frequencies as the data \citep{Galario_2018}. The stellar flux is added to the image in the visibility domain at the disk center, which is set by free parameters that enable offsets in right ascension (dRA) and declination (dDec) from the phase center.

\subsubsection{Model parameters}

We fit the synthetic model visibilities to the observed visibilities using Markov Chain Monte Carlo (MCMC) methods. We used the affine-invariant MCMC implementation in \texttt{emcee} to obtain best-fit model parameters and their posterior distributions \citep{emcee_2013}. For each disk model, the free parameters included the disk inclination ($i$), position angle (PA), surface density normalization ($\Sigma_{\mathrm{c}}$), stellar flux ($F_{\star}$) at the observing wavelength\footnote{AU Mic exhibited significant variations in the stellar flux across the three observations used here. Thus, we fit three separate stellar flux values as in \citet{Daley_2019}.} ($F_{*}$), and offsets from the phase center (dRA and dDec). The additional free parameters for each model depend on the choice of radial and vertical density forms, presented in Table \ref{tab:formsfree}. 

For each model, we initialized the MCMC with a 64 walker ensemble (more than double the number of free parameters for each model). To ensure that the parameter space was sufficiently sampled, we required at least $10^{5}$ samples after convergence, which corresponds to approximately $1500$ steps for each of the 64 walkers. Convergence was determined by visually inspecting the evolution of all walkers in both parameter space and log-likelihood space. We identified convergence once the log-likelihood stabilized and the walkers fluctuated around consistent parameter ranges, discarding earlier steps as burn-in. We further discarded walkers with log-likelihood values at least $5 \sigma$ lower than the maximum log-likelihood value in order to prevent individual unconverged walkers from affecting the results.

The log-likelihood is proportional to $-0.5 \chi^2$, where the model and data are compared using a $\chi^2$ metric\footnote{
\href{https://mtazzari.github.io/galario/}{Using \texttt{galario.double.chi2Image()}}}:
\begin{equation}
    \chi^2=\sum_{j=1}^N w_j \left[\left(\operatorname{ReV}_{{\rm obs}\, j}-\operatorname{ReV}_{\bmod j}\right)^2+\left(\operatorname{ImV}_{{\rm obs}\, j}-\operatorname{ImV}_{\bmod j}\right)^2\right],
\end{equation}
where $w_j$ are the visibility weights and $V_{{\rm obs}\, j}$ and $V_{\bmod j}$ are the observed and model visibilities, respectively.

\subsubsection{Model comparison}
We tested several model parameterizations for each source, typically beginning with two simple cases: a vertical Gaussian with a radial double power law, and a vertical Gaussian with a radial double Gaussian (Tables \ref{tab:formsfree} and \ref{tab:formsall}). If radial parametric modeling results favored a different radial density distribution, we included additional models \citep[see][]{Han_2026}. After running the initial suite of models with a vertical Gaussian density distribution, we expanded the range of models to include additional vertical complexities. 

We compared different model parameterizations using the Akaike and Bayesian Information Criteria (AIC and BIC, respectively), which quantify the trade-off between how well the different models fit the data and how many free parameters are needed \citep[e.g.,][]{Akaike_1974, Schwarz_1978, Burnham_2002, Liddle_2007}. These criteria are defined as
\begin{align} \label{eqn:aicbic}
    \mathrm{AIC} &\equiv-2 \ln \mathcal{L}_{\max }+2 k , \\
    \mathrm{BIC} &\equiv-2 \ln \mathcal{L}_{\max }+ k\ln N , \nonumber
\end{align}
where $\mathcal{L}_{\max}$ is the maximum likelihood achievable by the model, $k$ is the number of parameters of the model, and $N$ is the number of data points (in this application, the number of visibilities).

The set of model parameterizations tested for each source are shown in Table \ref{tab:formsall}, including the AIC and BIC comparisons. These criteria were computed such that
\begin{align}
    \Delta \rm{AIC}_{i} &= \rm{AIC}_{i} - \rm{AIC}_{min} , \\
    \Delta \rm{BIC}_{i} &= \rm{BIC}_{i} - \rm{BIC}_{min}, \nonumber
\end{align}
where $\rm{AIC}_{min}$ and $\rm{BIC}_{min}$ are the minimum values of the criteria across the full suite of models, indicating the best fit to the data. Thus, for both $\Delta \rm{AIC}$ and $\Delta \rm{BIC}$, a value of 0 indicates a best-fit model. AIC comparisons can be converted to a relative likelihood, which is defined as $p_i = \exp\left[{-\Delta{\rm{AIC_i}}/2}\right]$. Assuming a normal distribution with a mean of 0, we compute a symmetric interval which contains probability $1-p_i$. We present AIC comparisons using the upper bound of this interval, which gives the relative likelihood in terms of a confidence level $\sigma$. BIC results are directly presented as $\Delta{\rm{BIC}}$. Generally, $\Delta{\rm{BIC}}>10$ is regarded as strong evidence against a particular model \citep[>99\% confidence, see][]{Kass_1995, Raftery_1995}.

We use the AIC and BIC to identify a fiducial model for each source. When the $\Delta{\rm{AIC}}$ and $\Delta{\rm{BIC}}$ are minimized for the same model, this model is selected as the fiducial model. When the $\Delta{\rm{AIC}}$ and $\Delta{\rm{BIC}}$ are minimized for different models, the model which is least strongly ruled out by the AIC and BIC (AIC confidence $<3 \sigma$ and $\Delta{\rm{BIC}}<10$) is selected as the fiducial model. 

\subsection{Nonparametric modeling}

Vertical information can also be extracted using nonparametric methods. We used both \frank, which operates in the visibility domain, and \rave, which operates in the image domain. Each is briefly summarized below. (For more detailed descriptions in the context of radial profile recovery, see \citet{Han_2026}.)

\subsubsection{\frank}

\frank is an open-source code which reconstructs the 1D (azimuthally averaged) radial intensity profile of a disk by fitting the visibilities with a Fourier-Bessel series \citep{Jennings_2020}. While \frank is primarily a tool for extracting radial profiles, we use an extension which enables constraints to be placed on the vertical aspect ratio of the disk \citep[originally implemented in][]{Terrill_2023}. The specific implementation of \frank used for analyzing the ARKS sample \citep[\arksia,][]{Jennings_arksia} is publicly available\footnote{\href{https://github.com/jeffjennings/arksia}{https://github.com/jeffjennings/arksia}}.

For all fits, we adopt the position angle, inclination, and RA and declination offsets presented in \citet{Marino_2026}. We assumed the disk to be vertically Gaussian, with the scale height $H_{\sigma}$ proportional to the radius, thereby defining an aspect ratio of $h_{\sigma} = H_{\sigma}(r) / r$. Following the approach in \citet{Terrill_2023}, we repeated the radial profile fit over a range of scale height assumptions, computing the $\chi^2$ value for each height and estimating its likelihood as $\exp(-\chi^2/2)$. The radial profile fits are presented in \citet{Han_2026}; here, we focus only on the vertical constraints obtained with \frank.

Starting from a wide range in $h_{\sigma}$, we narrowed the height grid to more densely sample the region of $h_{\sigma}$ in which the bulk of the probability mass is centered. The sample points were then interpolated quadratically to obtain a finer probability density distribution (PDF). For approximately half of the disks, the PDF exhibits a clear peak and approaches 0 or a value close to 0 on either side of the peak, and the best-fit height was estimated with the median and the uncertainties with the 16th and 84th percentiles. For the rest of the disks, the PDF does not approach 0 as $h_{\sigma}$ approaches 0. In such cases, an upper limit was estimated with the 99.7th percentile if the PDF approaches 0 toward large values of $h_{\sigma}$. The fitted scale heights are listed in Table~\ref{table:frankraveheight}, and the PDFs are shown in Fig \ref{fig:frank_h}. 

\subsubsection{\rave}

\rave is an open-source package for recovering the radial brightness profiles of edge-on disks by fitting models of concentric annuli in the image domain \citep{Han_2022}. Like \frank, \rave also enables vertical height aspect ratio fitting. Assuming the disk to be vertically Gaussian, the radial profile was fit at a range of $h_{\sigma}$ assumptions including 0, and nine additional points spaced uniformly in logarithmic space between 0.01 and 0.5. The $\chi^2$ value for the model fit at each height assumption was calculated using the root mean square (rms) noise of the \clean image and assuming the number of independent elements to be equal to the number of beams in the region of the image containing disk flux \citep[defined as a rectangle within $r_\text{max}$ from the star along the major axis and $y_\text{max}$ along the minor axis, as listed in the appendix of][]{Han_2026}.
The likelihood of each height was estimated to be proportional to $\exp(-\chi^2/2)$. 

As with \frank, the sample points were interpolated to obtain a finer PDF. For disks with PDFs that show a clear peak, the best-fit height was estimated with the median and the lower and upper uncertainties were estimated with the 16th and 84th percentiles. For the remainder of disks, the PDF does not approach 0 even when $h_{\sigma}$ is 0. In such cases, if the PDF only drops to 0 at $h_{\sigma} \gtrapprox 0.3$, we take the 84th percentile to be an upper limit on $h_{\sigma}$; otherwise, the height was unconstrained. The scale heights fitted with \rave are listed in Table~\ref{table:frankraveheight}, and the PDFs are shown in Fig \ref{fig:rave_h}. 

\section{Analysis} \label{sec:analysis}

\begin{table*}
\caption{Summary of the fiducial parametric models for each source along with the best-fit and median values of $h_{\rm HWHM}$ and inclination (errors showing the 16th and 84th percentiles).}
\label{tab:formsshort}
\centering
\renewcommand{\arraystretch}{1.5}
\begin{tabular}{
    |>{\centering\arraybackslash}m{2.0cm}
    |>{\centering\arraybackslash}m{1.3cm}
    |>{\centering\arraybackslash}m{1.3cm}
    ||>{\centering\arraybackslash}m{2.5cm}
    >{\centering\arraybackslash}m{3.1cm}
    ||>{\centering\arraybackslash}m{1.8cm}
    >{\centering\arraybackslash}m{1.8cm}|
}
\hline
\multirow{2}{*}{Source} & \multirow{2}{*}{\shortstack{Radial \\ Form}} & \multirow{2}{*}{\shortstack{Vertical \\ Form}} & \multicolumn{2}{c||}{$h_{\rm HWHM}$} & \multicolumn{2}{c|}{$i$} \\ \cline{4-7}
& & & Best-fit & 50th Percentile & Best-fit & 50th Percentile \\
\hline
\rowcolor{gray!15}
HD~9672    & DPL & vL  & 0.041 & $0.046^{+0.008}_{-0.008}$     & 80.03\degree & $80.18$\degree$^{+0.12}_{-0.14}$ \\
\rowcolor{white}
\multirow{2}{*}{HD~10647} & \multirow{2}{*}{DG} & \multirow{2}{*}{vDG} 
                       & 0.052 ($91$\%) & $0.053^{+0.006}_{-0.007}$ ($88\pm6$\%) & \multirow{2}{*}{78.74\degree} & \multirow{2}{*}{$8.83$\degree$^{+0.21}_{-0.18}$} \\ 
  &  &                 & 0.226 ($9$\%)  & $0.225^{+0.002}_{-0.002}$ ($12\pm6$\%) &  &  \\ 
\rowcolor{gray!15}
HD~14055   & DPL & vG  & 0.059  & $0.053^{+0.007}_{-0.008}$    & 80.0\degree  & $80.0$\degree$^{+0.2}_{-0.3}$      \\
HD~15115   & DG  & vL  & 0.021  & $0.022^{+0.002}_{-0.002}$    & 86.74\degree & $86.79$\degree$^{+0.05}_{-0.04}$   \\
\rowcolor{gray!15}
HD~32297   & DPL & vL  & 0.0098 & $0.0092^{+0.0012}_{-0.0012}$ & 88.48\degree & $88.48$\degree$^{+0.04}_{-0.03}$   \\ 
HD~39060   & DPL & vL  & 0.112  & $0.112^{+0.002}_{-0.001}$    & 87.65\degree & $87.66$\degree$^{+0.17}_{-0.13}$   \\ 
\rowcolor{gray!15}
HD~61005   & DG  & vL  & 0.0129 & $0.0143^{+0.0020}_{-0.0021}$ & 86.23\degree & $86.24$\degree$^{+0.03}_{-0.04}$   \\ 
HD~76582   & AG  & vL  & 0.193  & $0.192^{+0.009}_{-0.009}$    & 72.1\degree  & $72.6$\degree$^{+0.7}_{-0.6}$      \\ 
\rowcolor{gray!15}
HD~109573  & DG  & vG  & 0.0065 & $0.0082^{+0.0033}_{-0.0038}$ & 76.57\degree & $76.62$\degree$^{+0.09}_{-0.10}$   \\ 
HD~131488  & DG  & vG  & 0.0048 & $0.0059^{+0.0011}_{-0.0010}$ & 84.91\degree & $84.87$\degree$^{+0.05}_{-0.05}$   \\ 
\rowcolor{gray!15}
HD~131835  & DG  & vL  & 0.017  & $0.022^{+0.007}_{-0.006}$    & 74.5\degree  & $74.6$\degree$^{+0.2}_{-0.2}$      \\ 
HD~161868  & DPL & vL  & 0.192  & $0.192^{+0.012}_{-0.016}$    & 66.7\degree  & $66.7$\degree$^{+0.9}_{-0.9}$      \\ 
\rowcolor{gray!15}
HD~197481  & DPL & vL  & 0.0026 & $0.0028^{+0.0009}_{-0.0002}$ & 88.16\degree & $88.13$\degree$^{+0.08}_{-0.09}$   \\ 
\hline
\end{tabular}
\tablefoot{The full set of parametric modeling results, including all free parameters for the fiducial models and the AIC/BIC for each tested model, is presented in Table \ref{tab:formsall}.}
\end{table*}

\begin{figure*}
\centering
\includegraphics[width=\linewidth]{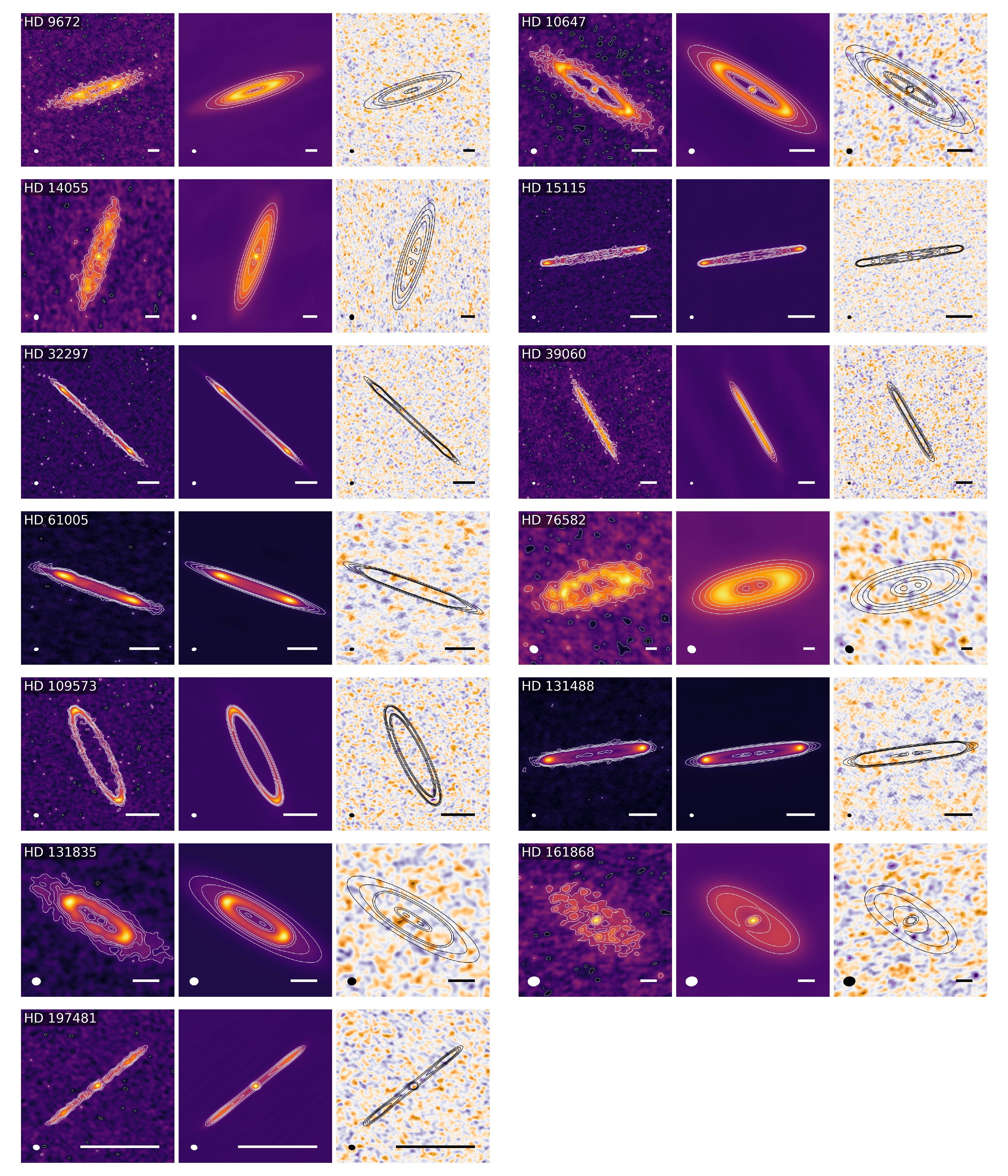}
\caption{Gallery of the data (left), models (center), and residuals (right, with model contours overplotted for reference) for the fiducial parametric vertical structure models for each source (models shown in Table \ref{tab:formsshort}). The effective beam is denoted by the shaded ellipse in the bottom-left corner of each panel, and the scale bars in the bottom right of each panel are 50 au in length. In the data and model panels, contours show three, five, seven, and nine times the rms (positive in light gray, negative in dark gray). In the residual panels, the purple and orange contours show $+3$ and $-3$ times the rms, respectively, though none of the sources show significant structure in the residuals. For each source, the data and model images are plotted on the same color scale, which spans the full dynamic range of each data image.}
\label{fig:dmr_gallery}
\end{figure*}

\begin{figure}[htbp!]
\centering
\includegraphics[width=\linewidth]{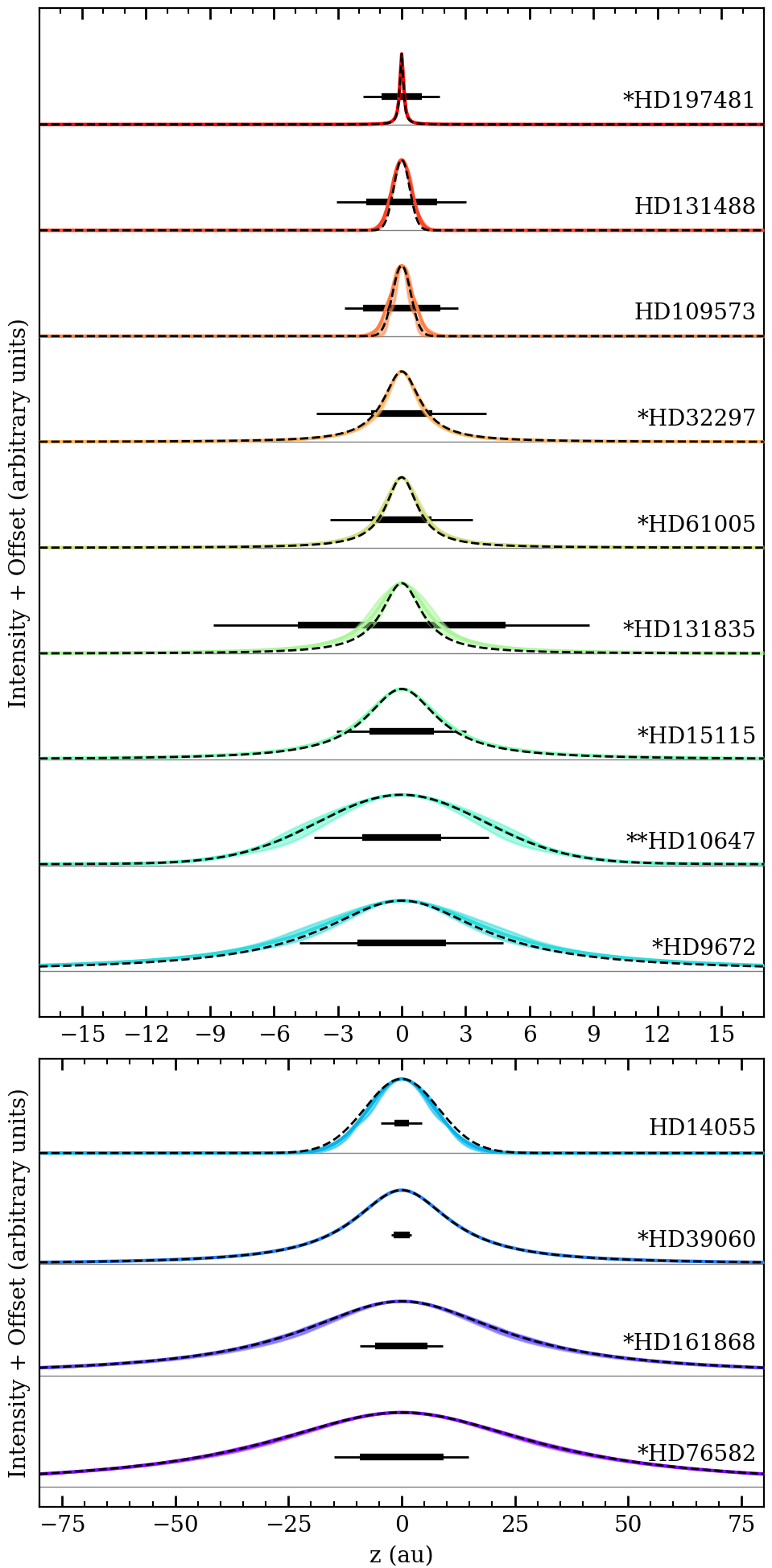}
\caption{Comparison of the fiducial vertical profiles for each disk at the reference radius $R_{\rm{ref}}$, defined as the location of peak intensity in the fiducial model (Table \ref{tab:masses}). Sources are ordered by $H_{\rm HWHM}(R_{\rm{ref}})$ and divided into two panels for visual clarity, with disks with $H_{\rm HWHM}(R_{\rm{ref}})<15$ au in the top panel and disks with $H_{\rm HWHM}(R_{\rm{ref}})>15$ au in the bottom panel. We denote extended or multi-component vertical profiles with asterisks (one asterisk for a Lorentzian profile, and two asterisks for a double Gaussian vertical profile). The remaining sources have Gaussian vertical profiles. Colored lines and the shaded regions denote the median profile $\pm 1 \sigma$, while the dashed black lines show the best-fit model. The thick horizontal black bars show the nominal resolution of the interferometer, $\lambda_{\rm obs}/b_{\rm max}$, where $\lambda_{\rm obs}$ is the observing wavelength and $b_{\rm max}$ is the longest baseline in the array. The thin horizontal black bars show the width of one beam along the vertical axis of the disk using the images shown in Figure \ref{fig:dmr_gallery}.}
\label{fig:vertical_profiles}
\end{figure}

The fiducial parametric models are summarized briefly in Table~{\ref{tab:formsshort}}. The fiducial images and residuals for each disk are shown in Figure \ref{fig:dmr_gallery} using the same imaging parameters as in \citet{Marino_2026}. For most sources the fiducial models closely match the observations, with no positive or negative features in the residuals at the 5$\sigma$ level. Figure \ref{fig:vertical_profiles} shows the best-fit vertical profile for each disk at its reference radius (see Table \ref{tab:masses}), defined as the radial location of peak intensity in our models, which are axisymmetric. We find aspect ratios in the range between roughly $0.003 \leq h_{\rm HWHM} \leq0.19$. We note that while we present $h$ in terms of the HWHM of the vertical distribution, many previous studies (particularly those which assume a Gaussian vertical distribution) present $h_{\sigma}$, the standard deviation of the vertical profile. For comparison with other studies, we have converted the published $h_{\sigma}$ values mentioned in this work to $h_{\rm HWHM}$ values.

\begin{figure*}
\centering
\includegraphics[width=\linewidth]{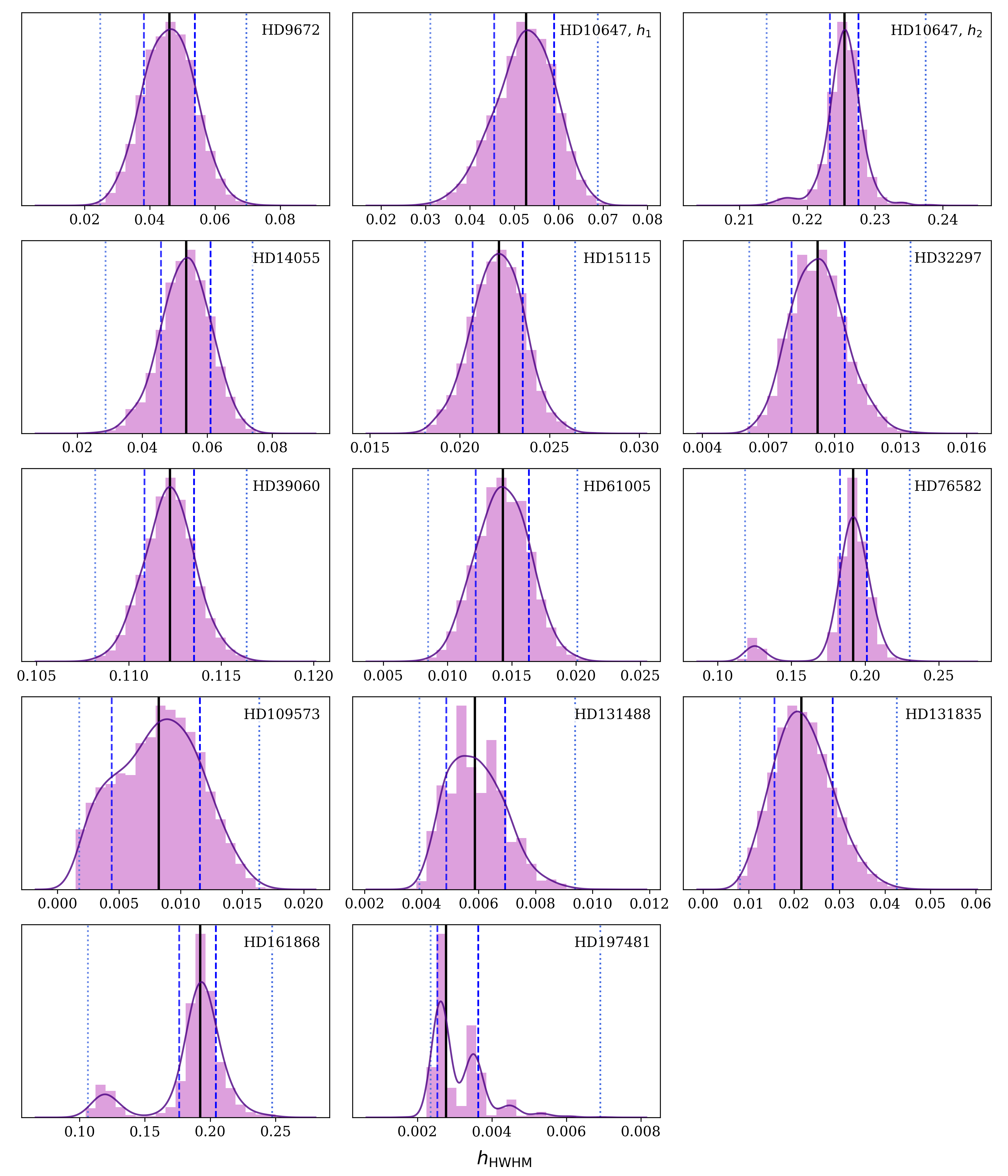}
\caption{Posterior distributions of the vertical aspect ratios for the fiducial parametric models. The histograms and KDEs are shown in light and dark purple, respectively. The median $h_{\rm HWHM}$ is shown by the vertical black line. The dark blue dashed line shows the 16th and 84th percentiles (equivalent to $1 \sigma$ for Gaussian distributions), while the light blue dotted line shows the 0.15th and 99.85th percentiles (equivalent to $3 \sigma$ for Gaussian distributions). Since the fiducial model for HD~10647 has two distinct vertical components, we show the posterior distributions for both $h_{\rm HWHM1}$ (top center panel) and $h_{\rm HWHM1}$ (top right panel). }
\label{fig:h_post}
\end{figure*}

For each source, the aspect ratio is meaningfully constrained, i.e., the posterior probability distribution of $h_{\rm HWHM}$ values has a nonzero peak (Figure \ref{fig:h_post}). To determine whether the scale heights are fully or marginally resolved, we take the difference between the median and 16th percentile $h$ values in order to approximate a $1\sigma$ confidence interval (this difference is equivalent to a $1\sigma$ confidence interval for a perfectly Gaussian posterior probability distribution, though not all of the posteriors are Gaussian). We use this interval to extrapolate to a $3\sigma$ confidence interval; for sources with a $h_{\rm HWHM}\leq0$ at the $3\sigma$ lower bound, we consider $h_{\rm HWHM}$ to be marginally resolved, placing only an upper limit on $h_{\rm HWHM}$. We apply this extrapolation because $h_{\rm HWHM}\leq0$ is not allowed by our model. Therefore, the 0.15th percentile (corresponding to $\sim3\sigma$ for a Gaussian distribution) will always be positive and non-zero.

We additionally examine how well-resolved $h_{\rm HWHM}$ is with respect to the resolution of the observations. Importantly, the beams shown in Figure \ref{fig:dmr_gallery} do not reflect the smallest scales resolvable in our analyses. Because our parametric modeling occurs in the visibility domain, we retain spatial information up to the nominal resolution of the interferometer, $\theta_{\rm{min}} = \lambda_{\rm obs}/b_{\rm max}$, where $\lambda_{\rm obs}$ is the observing wavelength and $b_{\rm max}$ is the longest baseline in the array during the observations. As a result, $h_{\rm HWHM}$ is more easily resolved in Fourier space than in image space, and some values of $h_{\rm HWHM}$ may be resolvable even if the vertical extent of the disk is on roughly the same scale as the \clean beam (Figure \ref{fig:vertical_profiles}). Furthermore, it is not necessary that $\theta_{\rm min}<h_{\rm HWHM}$ in order to resolve the aspect ratio, as $h_{\rm HWHM}$ does not represent the full vertical extent of the disk. Rather, $h_{\rm HWHM}$ only describes the distance from the disk mid-plane to the height at which the disk has half of its peak density. Depending on the vertical parametrization, the disk could thus host a substantial amount of material at several times $h_{\rm HWHM}$.

For example, a vertical Gaussian distribution contains about $95\%$ of the material within $\pm 2 \, h_{\sigma}$ from the midplane, equivalent to $\sim 3.4 \, h_{\rm HWHM}$. In this case, a scale height may not be resolved by the observations if $\theta_{\rm min}>3.4 \, h_{\rm HWHM}$. A vertical Lorentzian distribution has thicker tails, with more disk material further from the midplane: $\sim95\%$ within $\sim 25 \, h_{\rm HWHM}$ centered on the midplane. In this case, a scale height may not be resolved by the observations if $\theta_{\rm min}>25 \, h_{\rm HWHM}$. Measurements of $h_{\rm HWHM}$ that are limited by the data resolution are diskussed in the source-specific sections, and values of $\theta_{\rm min}$ are provided in Table \ref{tab:masses} for reference.

\subsection{Marginally resolved scale heights} \label{sec:marginal}

\subsubsection{HD~109573 (HR~4796A)}
Parametric models in \citet{Han_2026} suggest that the radial density distribution of this source is well-fit by a double Gaussian, assuming vertical Gaussian profile with a fixed vertical aspect ratio of $h_{\rm HWHM}=0.015$. Incorporating scale height as a free parameter, we examined double power law, double Gaussian, Gaussian, and asymmetric Gaussian radial profiles. We find that the double Gaussian continues to outperform the other radial density distributions even when incorporating vertical flexibility into the model. We then modeled HD~109573 with a radial double Gaussian density distribution and the four different vertical parameterizations presented in Table \ref{tab:formsfree}. 

The AIC and BIC strongly favor the vertical Gaussian and vertical Lorentzian models over the other vertical forms, with the vertical Gaussian model preferred over the vertical Lorentzian only at low confidence (AIC $\sigma=0.09$). However, the posterior probability distributions are broad, with a vertical aspect ratio of $h_{\rm HWHM}=0.0082^{+0.0033}_{-0.0038}$ (vertical Gaussian) or $h_{\rm HWHM}=0.0049^{+0.0016}_{-0.0014}$ (vertical Lorentzian). Both of these models exhibit a non-Gaussian posterior probability distribution for $h_{\rm HWHM}$, truncating near zero. As a result, we only marginally resolve the scale height for HD~109573. We take the upper limit to be the 84th percentile of the posteriors, $h_{\rm HWHM}<0.0116$, using the fiducial vertical Gaussian model. Nonparametric fitting also constrains only upper limits, with \frank finding $h_{\rm HWHM}<0.012$ and \rave finding $h_{\rm HWHM}<0.022$.

The vertical structure of HD~109573 has previously been studied using Band 7 ALMA observations \citep[$\sim 0$\farcs17 resolution,][]{Kennedy_2018}. \citet{Kennedy_2018} modeled the disk with a radial Gaussian and a vertical Gaussian, finding $H_{\rm FWHM}=7 \pm 1$ au at the radial location of the ring (78.6 au), corresponding to an aspect ratio of $h_{\rm HWHM}=0.045^{+0.006}_{-0.006}$. \citet{Matra_2025} reanalyzed these observations, also modeling HD~109573 with radial and vertical Gaussians. At the radial ring location of 77.8 au, they obtain $H_{\rm HWHM}=5.3$ au, or $h_{\rm HWHM}=0.068^{+0.002}_{-0.002}$. \citet{Han_2025a} used \fave, an extension of \rave optimized for less inclined disks, to fit the aspect ratio of HD~109573, finding $h_{\rm HWHM}=0.057^{+0.016}_{-0.022}$. With the new Band 7 ARKS data (0\farcs08), we find a similar ring location (77.6 au), but a much smaller vertical extent ($H_{\rm HWHM}<0.9$ au). This discrepancy is likely due to the lower resolution of previous observations combined with the different best-fit functional form found here (double rather than single Gaussian); the degeneracy between radial and vertical structure makes it difficult to place strong constraints on the scale height, and this issue is more pronounced with lower-resolution data. Despite the improved resolution of the ARKS observations, we may still be limited by the data resolution of the data ($\theta_{\rm min}=3.63$ au). At a reference radius of 77.6 au (the location of peak brightness in the fiducial model), we obtain $H_{\rm HWHM}=0.636$ au. For a vertical Gaussian profile, 95\% of the disk material would then lie within 2.16 au centered on the disk midplane, about 60\% of the distance that could be resolved by the observations.

\begin{table}[htbp]
    \centering
    \renewcommand{\arraystretch}{1.4}
    \caption{Scale height aspect ratios obtained from parametric modeling, \frank, and \rave.}
    \label{table:frankraveheight}
\rowcolors{2}{gray!15}{white}
\begin{tabular}{l|lll}
\hline \
       &                & $h_{\rm HWHM}$ values &           \\
Target & Parametric & \frank & \rave \\
\hline
HD~9672   & $0.046^{+0.008}_{-0.008}$   & $0.061^{+0.004}_{-0.004}$ & $0.058^{+0.017}_{-0.024}$ \\ 
HD~10647  & $0.053^{+0.007}_{-0.007}$   & $<0.008$                  & $0.040^{+0.016}_{-0.021}$ \\ 
HD~14055  & $0.054^{+0.008}_{-0.008}$   & $0.037^{+0.010}_{-0.016}$ & $0.047^{+0.038}_{-0.031}$ \\ 
HD~15115  & $0.022^{+0.002}_{-0.002}$   & $0.019^{+0.001}_{-0.001}$ & $0.022^{+0.002}_{-0.002}$ \\ 
HD~15257  & N/A                         & Unconst.                  & Unconst.                  \\ 
HD~32297  & $0.009^{+0.002}_{-0.001}$   & $0.010^{+0.001}_{-0.001}$ & $0.015^{+0.003}_{-0.004}$ \\ 
HD~39060  & $0.112^{+0.002}_{-0.001}$   & $0.073^{+0.002}_{-0.002}$ & $0.086^{+0.005}_{-0.005}$ \\ 
HD~61005  & $0.015^{+0.002}_{-0.003}$   & $0.019^{+0.001}_{-0.001}$ & $0.020^{+0.005}_{-0.007}$ \\ 
HD~76582  & $0.192^{+0.010}_{-0.009}$   & $0.065^{+0.012}_{-0.012}$ & $<0.097$                  \\ 
HD~84870  & N/A                         & Unconst.                  & Unconst.                  \\ 
HD~92945  & N/A                         & $0.039^{+0.011}_{-0.013}$ & $<0.079$                  \\ 
HD~95086  & N/A                         & Unconst.                  & Unconst.                  \\ 
HD~107146 & N/A                         & $0.134^{+0.021}_{-0.022}$ & $<0.167$                  \\ 
HD~109573 & $<0.0116$                   & $<0.012$                  & $<0.022$                  \\
HD~121617 & N/A                         & $0.106^{+0.011}_{-0.011}$ & $0.074^{+0.039}_{-0.044}$ \\ 
HD~131488 & $<0.0070$                   & $<0.008$                  & $0.016^{+0.005}_{-0.006}$ \\ 
HD~131835 & $0.022^{+0.007}_{-0.006}$   & $0.039^{+0.006}_{-0.006}$ & $0.057^{+0.020}_{-0.028}$ \\ 
HD~145560 & N/A                         & $0.070^{+0.015}_{-0.016}$ & $<0.103$                  \\ 
HD~161868 & $0.19^{+0.02}_{-0.02}$      & $<0.097$                  & $<0.135$                  \\ 
HD~170773 & N/A                         & $<0.116$                  & $<0.143$                  \\ 
HD~197481 & $0.0028^{+0.0009}_{-0.0002}$& $0.023^{+0.002}_{-0.002}$ & $0.026^{+0.005}_{-0.006}$ \\ 
HD~206893 & N/A                         & $0.113^{+0.034}_{-0.025}$ & $<0.212$                  \\ 
HD~218396 & N/A                         & Unconst.                  & Unconst.                  \\ 
TYC~9340  & N/A                         & Unconst.                  & Unconst.                  \\ 
\hline
\end{tabular}
\tablefoot{For all resolved scale heights, we show the median $h_{\rm HWHM}$, with uncertainties showing the 16th and 84th percentiles. The parametric scale heights are from the fiducial models shown in Table \ref{tab:formsshort}; sources marked with N/A do not have any vertical parametric models. For scale heights fitted with \frank and \rave, we show upper limits if the PDF only approaches zero at large $h_{\sigma}$, and we note the aspect ratio as unconstrained if the PDF does not approach zero at large $h_{\sigma}$.}
\end{table}

\subsubsection{HD~131488}
Parametric models in \citet{Han_2026} suggest that the radial density distribution of this source is best fit by a double Gaussian, assuming a vertical Gaussian profile with a fixed vertical aspect ratio of $h_{\rm HWHM}=0.005$. We find that a double Gaussian radial density distribution is strongly preferred over a double power law or single Gaussian, assuming a vertical Gaussian profile with the vertical aspect ratio as a free parameter. We additionally modeled HD~131488 using a radial double Gaussian density distribution and both a vertical exponential and a vertical Lorentzian.

We find that the vertical Gaussian model performs best when considering both the AIC and BIC (AIC $\sigma=0.90$ over the exponential, $\Delta \rm{BIC}=0$), so we select this as the fiducial model for HD~131488. The radial double Gaussian with a vertical exponential is most strongly favored by the AIC, but disfavored by the BIC ($\Delta \rm{BIC}=11.63$). The radial double Gaussian with a vertical Lorentzian form is not strongly ruled out by either the AIC or BIC tests (AIC $\sigma=1.26$, $\Delta \rm{BIC}=1.14$), but because the fiducial Gaussian model provides a better fit with the same number of parameters, we do not increase model complexity any further (i.e., we do not run a vertical double Gaussian model, which could have a similar vertical profile to the Lorentzian model at the cost of two additional free parameters). The aspect ratio is well-constrained, with $h_{\rm HWHM}=0.0059^{+0.0011}_{-0.0010}$. The aspect ratio constraints from \frank and \rave are in agreement, finding $h_{\rm HWHM}<0.008$ and $h_{\rm HWHM}=0.016^{+0.005}_{-0.006}$, respectively. These values are consistent with the uncertain constraint reported in \citet{Worthen_2024}, $h_{\rm HWHM}=0.09^{+0.05}_{-0.05}$, which was derived from Band 6 ALMA observations of HD~131488 ($\sim$0\farcs5 resolution).

We note that while our measurements of $h_{\rm HWHM}$ are constrained according to the posterior probability distributions, the interpretation of our measured values of $h_{\rm HWHM}$ may be limited by the resolution of the data. The observations of HD~131488 have a nominal resolution of $\theta_{\rm min}=3.32$ au. At a reference radius of 90.8 au, we obtain $H_{\rm HWHM}=0.536$ au. For a vertical Gaussian profile, 95\% of the disk material would then lie within 1.82 au centered on the disk midplane, nearly half the distance that could be resolved by the observations. We take the upper limit of $h_{\rm HWHM}$ to be the 84th percentile of the posteriors for the fiducial model, $h_{\rm HWHM}<0.0070$, though follow-up observations at higher resolution are necessary to confirm this measurement.

\subsection{Resolved Gaussian scale heights: HD~14055} \label{sec:resolved}

The only source in our sample which is best fit with a vertical Gaussian distribution and has a fully resolved scale height is HD~14055. Parametric models in \citet{Han_2026} suggest that the radial density distribution of this source is well-fit by a double power law, though a double Gaussian is not ruled out at high confidence. In the radial models, a vertical Gaussian profile was used with a fixed vertical aspect ratio of $h_{\rm HWHM}=0.0475$. Incorporating scale height as a free parameter, we examined double power law and double Gaussian radial density distributions with a vertical Gaussian. The AIC favored the double Gaussian model, while the BIC favored the double power law model, so we modeled both radial double power law and radial double Gaussian density distributions with the four different vertical parameterizations presented in Table \ref{tab:formsfree}. Of these eight models, we select the radial double power law with a vertical Gaussian as the fiducial model, as it is favored at high confidence with the BIC and not strongly ruled out with the AIC ($\sigma=0.56$ over a radial double Gaussian with a vertical Gaussian).

We measure $h_{\rm HWHM}=0.053^{+0.007}_{-0.008}$ with the fiducial parametric model, using the 0\farcs23 ARKS observations in Band 7. The vertical aspect ratio is also resolved with \frank and \rave: \frank finds $h_{\rm HWHM}=0.037^{+0.010}_{-0.016}$, while \rave finds $h_{\rm HWHM}=0.047^{+0.038}_{-0.031}$. These values are in agreement with \citet{Terrill_2023}, \citet{Matra_2025}, and \citet{Han_2025a}, who used Band 6 observations ($\sim$2'' resolution) of HD~14055 to find limits of $h_{\rm HWHM}<0.12$, $h_{\rm HWHM}=0.04^{+0.02}_{-0.02}$, and $h_{\rm HWHM}=0.131^{+0.073}_{-0.075}$, respectively.

\subsection{Resolved complex structures} \label{sec:complex}
In this section we present sources that are consistent with having an extended or multi-component millimeter vertical emission profile (i.e., statistically preferred models use either a Lorentzian or a double Gaussian). We note that while we report a single $h_{\rm HWHM}$ for each vertical Lorentzian model, the broad tails of a Lorentzian can approximate a combination of broad and narrow components with different widths (see Sect. \ref{sec:discuss:lor}, where we also discuss the possible mechanisms that could generate these distributions).

\subsubsection{HD~9672 (49~Ceti)}
Parametric models in \citet{Han_2026} suggest that the radial density distribution of this source is best fit by a double power law, assuming a vertical Gaussian profile with a fixed vertical aspect ratio of $h_{\rm HWHM} = 0.07$. We modeled HD~9672 using a radial double power law density distribution and the four different vertical parametrizations presented in Table \ref{tab:formsfree}, as well as with a radial double Gaussian and vertical Gaussian. We find that the vertical Lorentzian model is most favored by the BIC, and not strongly ruled out by the AIC ($\sigma=0.27$). The vertical double Gaussian model is most strongly favored by the AIC, and disfavored by the BIC ($\Delta \rm{BIC}=25.81$). We identify the vertical Lorentzian model as the fiducial model for HD~9672.

We obtained a vertical aspect ratio of $h_{\rm HWHM}=0.046^{+0.008}_{-0.008}$ for the fiducial model. As the double Gaussian model is only disfavored by the BIC, which penalizes additional parameters more heavily than the AIC, we may tentatively resolve two distinct dynamical populations within the disk. The first has a measured aspect ratio of $h_{\rm HWHM}=0.031^{+0.022}_{-0.013}$ containing $68^{+11}_{-15} \;\%$ of the dust mass, and the other has an aspect ratio of $h_{\rm HWHM}=0.105^{+0.021}_{-0.040}$ containing the rest of the mass. The nonparametric aspect ratios fall in between, with $h_{\rm HWHM}=0.061^{+0.004}_{-0.004}$ from the \frank fit and $h_{\rm HWHM}=0.058^{+0.017}_{-0.024}$ from the \rave fit.

The aspect ratio measured for the dynamically cooler population of the parametric vertical double Gaussian model is in agreement with the values reported in \citet{Delamer_2021} from an analysis of Band 8 (614 $\mu$m, $\sim$0\farcs25 resolution) and Band 7 (868 $\mu$m, $\sim$0\farcs13 resolution) continuum emission of HD~9672 ($h_{\rm HWHM}=0.088^{+0.005}_{-0.005}$ and $h_{\rm HWHM}=0.081^{+0.006}_{-0.008}$, respectively), assuming a Gaussian vertical profile. 

With the same Band 8 observations, \citet{Terrill_2023} found $h_{\rm HWHM}=0.059^{+0.008}_{-0.008}$ and \citet{Han_2025a} found $h_{\rm HWHM}=0.039^{+0.016}_{-0.021}$, with both studies using a nonparametric model with an assumed Gaussian vertical density distribution. \citet{Matra_2025} reanalyzed the Band 7 observations with radial and vertical Gaussians, finding an aspect ratio of $h_{\rm HWHM}=0.089^{+0.011}_{-0.011}$. Finally, \citet{Delamer_2021} also analyzed Band 6 (1.33 mm, $\sim$0\farcs85 resolution) observations, finding a slightly larger aspect ratio of $h_{\rm HWHM}=0.18^{+0.02}_{-0.02}$.

These differences could be partially driven by the range of observations represented, as the dust scale height may vary with observing wavelength. However, the variation among aspect ratios derived from Band 7 observations is likely driven by the fact that a Lorentzian or two Gaussians have broader tails, while a single Gaussian drops off more quickly away from the midplane. When fitting the disk with only one vertical Gaussian, the model may settle on a larger value of $h_{\rm HWHM}$ as it attempts to capture the more dynamically active population in the disk. HD~9672 also hosts a likely warp asymmetry, discussed in detail in \citet{Lovell_2026}. While we do not see significant features in the residual map (Fig. \ref{fig:dmr_gallery}), we note that a subtle warp feature could bias our axisymmetric models to larger scale heights.

\subsubsection{HD~10647}

Parametric models in \citet{Han_2026} suggest that the radial density distribution of HD~10647 is well-fit by a double Gaussian, assuming a vertical Gaussian profile with a fixed vertical aspect ratio of $h_{\rm HWHM}=0.055$. We modeled the disk using radial double power law and double Gaussian density distributions with the four different vertical parametrizations presented in Table \ref{tab:formsfree}. The AIC strongly favors the model with both radial and vertical double Gaussian density distributions, while $\Delta \rm{BIC}=9.042$. The BIC favors the radial double power law with a vertical Lorentzian, though the AIC strongly disfavors this model ($\rm{AIC}=6.151 \sigma$). We thus identify the radial and vertical double Gaussian model as the fiducial model. 

We resolve two distinct dynamical populations with the fiducial model of HD~10647, one with a measured scale height of $h_{\rm HWHM}=0.053^{+0.006}_{-0.007}$ containing $88^{+6}_{-6} \;\%$ of the dust mass, and the other with a scale height of $h_{\rm HWHM}=0.225^{+0.002}_{-0.002}$ containing the rest of the mass. This is in line with the constraint from \rave, $h_{\rm HWHM}=0.040^{+0.016}_{-0.021}$. However, the upper limit derived from the \frank fit is much smaller, $h_{\rm HWHM}<0.008$. While we do not see significant features in the residual map of HD~10647 (Fig. \ref{fig:dmr_gallery}, this source has a major axis asymmetry that could bias the results of our axisymmetric modeling \citep{Lovell_2026}.

Several studies have already made measurements which are in agreement with our measurement of the dynamically cooler population of the fiducial model (which contains most of the disk mass): \citet{Lovell_2021} found the scale height to be consistent with $h_{\rm HWHM}=0.05^{+0.01}_{-0.01}$, while \citet{Han_2025a} found $h_{\rm HWHM}=0.054^{+0.011}_{-0.012}$. \citet{Terrill_2023} and \citet{Matra_2025} found marginal measurements of $h_{\rm HWHM}=0.044^{+0.009}_{-0.008}$ and $h_{\rm HWHM}=0.06^{+0.01}_{-0.01}$, respectively.

\subsubsection{HD~15115}
The parametric modeling in \citet{Han_2026} shows that HD~15115 is well-represented by a double Gaussian radial density distribution, assuming a vertical Gaussian profile with a fixed vertical aspect ratio of $h_{\rm HWHM}=0.02$. We modeled HD~15115 using a radial double Gaussian density distribution and the four different vertical parametrizations presented in Table \ref{tab:formsfree}. We identify the vertical Lorentzian model as the fiducial model for HD~15115, as it is strongly favored by both the AIC and BIC. The scale height is well-constrained, with a vertical aspect ratio of $h_{\rm HWHM}=0.0222^{+0.0013}_{-0.0015}$ for the vertical Lorentzian model. The \frank and \rave values are comparable, $h_{\rm HWHM}=0.019^{+0.001}_{-0.001}$ and $h_{\rm HWHM}=0.022^{+0.002}_{-0.002}$, respectively.

\citet{MacGregor_2019} found that HD~15115 is not vertically resolved in 0\farcs6 (29 au) Band 6 observations. Using the same Band 6 observations, \citet{Matra_2025} marginally detected the scale height ($h_{\rm HWHM}=0.06^{+0.01}_{-0.02}$) by modeling the disk with a radial and vertical Gaussian. This is in good agreement with results from \citet{Terrill_2023} and \citet{Han_2025a}, who used nonparametric methods with a vertical Gaussian assumption to measure $h_{\rm HWHM}=0.056^{+0.008}_{-0.008}$ and $h_{\rm HWHM}=0.079^{+0.035}_{-0.040}$, respectively. The differences between these values from previous observations and our measurement are likely due to the improved resolution of the ARKS data (0\farcs12), which is 5 times higher than the previous observations of HD~15115, combined with the different assumed radial and vertical distributions. In addition, there is evidence that HD~15115 is eccentric \citep{Lovell_2026}, though this may be compensated for by the offset parameters dRA and dDec in our modeling (see Table~\ref{tab:formsall}).

\subsubsection{HD~32297}
Parametric models in \citet{Han_2026} suggest that the radial density distribution of HD~32297 is well-fit by a double power law, assuming a vertical Gaussian profile with a fixed vertical aspect ratio of $h_{\rm HWHM}=0.012$. We modeled the disk using a radial double power law density distribution and the four different vertical parametrizations presented in Table \ref{tab:formsfree}. The vertical Lorentzian model is favored by both the AIC and BIC, so we select it as the fiducial model. The aspect ratio of the fiducial model is well-constrained ($h_{\rm HWHM}=0.0092^{+0.0012}_{-0.0012}$) and in line with fits from \frank ($h_{\rm HWHM}=0.010^{+0.001}_{-0.001}$) and \rave ($h_{\rm HWHM}=0.015^{+0.003}_{-0.004}$)

The AIC score of the vertical double Gaussian model indicates that it also reproduces the data well ($\sigma=0.97$); however, due to the additional parameters in the vertical double Gaussian model, the BIC score is significantly worse compared to the Lorentzian model ($\Delta \rm{BIC}=27.95$). Nevertheless, we marginally resolve two independent dynamical populations within HD~32297: one with a scale height of $h_{\rm HWHM}=0.0083^{+0.0018}_{-0.0034}$ containing $92^{+2}_{-3} \;\%$ of the dust mass, and another with a scale height of $h_{\rm HWHM}=0.146^{+0.027}_{-0.047}$ containing the remainder.   

There exist several previous scale height measurements for HD~32297 at $\sim$millimeter wavelengths. \citet{Terrill_2023} and \citet{Han_2025a} use nonparametric methods (assuming a Gaussian vertical distribution) to find $h_{\rm HWHM}=0.09^{+0.01}_{-0.01}$  and $h_{\rm HWHM}=0.139^{+0.041}_{-0.052}$, respectively. \citet{Worthen_2024} find $h_{\rm HWHM}=0.018^{+0.002}_{-0.002}$ and \citet{Matra_2025} find $h_{\rm HWHM}=0.09^{+0.01}_{-0.02}$, both modeling the disk with Gaussian radial and vertical density distributions. Our measured $h_{\rm HWHM}$ is smaller than all of the above, likely due to several factors. First, the ARKS data have a higher angular resolution (0\farcs06). Second, the vertical and radial distributions of disk material can be degenerate with disk inclination; uncertainties in the disk inclination make it more difficult to constrain the scale height and belt width, and all of the above studies find/use smaller inclinations than the value we obtain from our fiducial model ($i=88.48$\degree$^{+0.04}_{-0.03}$, which does not appear to be affected by this degeneracy). We find that HD~32297 is best modeled with a vertical Lorentzian profile and a radial double power law profile; the common assumption of vertical and/or radial Gaussian profiles used for previous scale height measurements may be biasing the measurement of $h_{\rm HWHM}$, resulting in artificially high values. Lastly, there is evidence that HD~32297 is eccentric \citep{Lovell_2026}, though this may be compensated for by the offset parameters dRA and dDec in our modeling (see Table~\ref{tab:formsall}).

\subsubsection{HD~39060 (Beta Pic)}
Parametric models in \citet{Han_2026} suggest that the radial density distribution of this source is well-represented by a double power law, assuming a vertical Gaussian profile with a fixed vertical aspect ratio of $h_{\rm HWHM}=0.0445$. We modeled HD~39060 using a radial double power law density distribution and the four different vertical parametrizations presented in Table \ref{tab:formsfree}. The vertical Lorentzian (AIC $\sigma=1.55$, $\Delta \rm{BIC}=0$) and vertical double Gaussian (AIC $\sigma=0$, $\Delta \rm{BIC}=22.85$) models are both good fits to the data. For the fiducial vertical Lorentzian model, the aspect ratio is well-constrained with, $h_{\rm HWHM}=0.1122^{+0.0013}_{-0.0013}$. The \frank fit yields $h_{\rm HWHM}=0.073^{+0.002}_{-0.002}$, while \rave finds $h_{\rm HWHM}=0.086^{+0.005}_{-0.005}$.

We identified two distinct dynamical populations within the disk with the vertical double Gaussian model; the first has a measured aspect ratio of $h_{\rm HWHM}=0.046^{+0.005}_{-0.003}$ containing $58^{+4}_{-3} \;\%$ of the dust mass, and the other has a scale height of $h_{\rm HWHM}=0.212^{+0.002}_{-0.003}$ containing the rest of the mass. \citet{Matra_2019} also measured two distinct aspect ratios for this source, finding $h_{\rm HWHM}=0.130^{+0.009}_{-0.007}$ containing $80^{+4}_{-5} \;\%$ of the dust mass, and $h_{\rm HWHM}=0.016^{+0.007}_{-0.007}$ containing the rest of the mass.

The differences are likely partially driven by the choice in radial density distribution. Here we use a double power law, finding a relatively shallow slope in the inner disk ($\alpha_{\rm{in}}=1.78^{+0.14}_{-0.10}$) with a sharper decrease in density in the outer disk ($\alpha_{\rm{out}}=-5.2^{+0.6}_{-0.5}$). \citet{Matra_2019} use a radially Gaussian density distribution with $\sigma=36.7^{+1.0}_{-1.3}$ au. Interestingly, \citet{Matra_2019} find that the broad component contains most of the disk mass, while we find that the narrow component contains slightly more mass. \citet{Matra_2019} also find marginal evidence of a radially varying aspect ratio in HD~39060, with $h(r) = h_0(r/r_0)^{\beta-1}$ and $\beta = 0.7^{+0.2}_{-0.2}$, which our model specification does not account for.

Asymmetries in the disk could also contribute to differences in the measured scale heights. Though the models presented both here and in \citet{Matra_2019} are axisymmetric, there is evidence of asymmetric structures in the millimeter emission of HD~39060 \citep{Lovell_2026}. An asymmetric feature at the $\sim 3 \sigma$ level can be seen in the northeast of the residual map in Figure~\ref{fig:dmr_gallery}.

\subsubsection{HD~61005}
The parametric modeling in \citet{Han_2026} shows that HD~61005 is well-represented by a double Gaussian radial density distribution, assuming a vertical Gaussian profile with a fixed vertical aspect ratio of $h_{\rm HWHM}=0.02$. We modeled HD~61005 using a radial double Gaussian and the four different vertical parametrizations presented in Table \ref{tab:formsfree}. We find that the vertical Lorentzian model is favored by both the AIC and the BIC at high significance, corresponding to an aspect ratio of $h_{\rm HWHM}=0.0143^{+0.0020}_{-0.0021}$. Compared to the vertical double Gaussian model, the AIC only favors the Lorentzian model at $1.31 \sigma$. We thus find marginal evidence of two resolved populations in the disk; the first has a measured scale height of $h_{\rm HWHM}=0.014^{+0.005}_{-0.006}$ containing $90^{+6}_{-12} \;\%$ of the dust mass, and the other has a scale height of $h=0.117^{+0.043}_{-0.087}$ containing the rest of the mass. Aspect ratio fits from \frank and \rave find $h_{\rm HWHM}=0.019^{+0.001}_{-0.001}$ and $h_{\rm HWHM}=0.020^{+0.005}_{-0.007}$, respectively.

Assuming Gaussian radial and vertical density distributions, \citet{Matra_2025} find $h_{\rm HWHM}=0.052^{+0.004}_{-0.005}$. Similarly, \citet{Terrill_2023} and \citet{Han_2025a} find $h_{\rm HWHM}=0.046^{+0.004}_{-0.005}$ and $h_{\rm HWHM}=0.034^{+0.008}_{-0.011}$, respectively, using nonparametric methods and assuming a Gaussian vertical distribution. These values may be inflated due to the models fitting a more dynamically active disk to a simple vertical Gaussian, driving up the scale height measurement to account for the material further from the disk midplane. 

It is also possible that asymmetries in the disk could bias our axisymmetric model toward a vertical structure indicative of multiple populations. While we do not identify structure in our model residuals (see Figure \ref{fig:dmr_gallery}), \citet{Lovell_2026} find evidence of asymmetry in HD~61005, which could affect the vertical structure measurements presented here.

\subsubsection{HD~76582}
Parametric models in \citet{Han_2026} suggest that the radial density distribution of HD~76582 is well-fit by an asymmetric Gaussian (which is defined by some central radius and two different $\sigma$ values inside and outside of the central radius), assuming a vertical Gaussian profile with a fixed vertical aspect ratio of $h_{\rm HWHM}=0.015$. We examined the vertical structure of HD~76582 by first modeling the vertical structure of the disk with a Gaussian, and the radial structure with a variety of different radial forms. The best-fitting radial forms were a double power law, asymmetric Gaussian, and error function. We modeled HD~76582 using each of these radial density structures and the four different vertical parametrizations presented in Table \ref{tab:formsfree}. The model with a radial asymmetric Gaussian and vertical Lorentzian is favored by both the AIC and BIC, yielding $h_{\rm HWHM}=0.192^{+0.009}_{-0.009}$; we thus selected this as the fiducial model for HD~76582. The \frank aspect ratio fit ($h_{\rm HWHM}=0.065^{+0.012}_{-0.012}$) and the \rave upper limit ($h_{\rm HWHM}\leq0.097$) are smaller.

The \rave and \frank results are in agreement with the upper limits identified by \citet{Matra_2025} ($h_{\rm HWHM}<0.1$) and \citet{Han_2025a} ($h_{\rm HWHM}<0.180$). All of these values were obtained by assuming Gaussian vertical density distributions. Furthermore, it is likely that the radial Gaussian density distribution assumed by \citet{Matra_2025} resulted in a biased scale height value. We found that an asymmetric Gaussian yields a good fit to the data, with substantially different widths on each side of the Gaussian peak ($\sigma_{\rm{in}}=45 \pm 8$ au and $\sigma_{\rm{out}}=114^{+23}_{-15}$ au). This corresponds to a FWHM of $\Delta R = 187^{+37}_{-27}$ au, while the Gaussian model of \citet{Matra_2025} has $\Delta R = 210 \pm 20$ au. Our asymmetric Gaussian model has its peak at $R=181 \pm 11$ au, in contrast to $R=219^{+9}_{-8}$ au in \citet{Matra_2025}. As the Gaussian model of \citet{Matra_2025} is both marginally wider and significantly farther out compared to the best-fit asymmetric Gaussian model presented here, it is likely that the scale height measurement was underestimated to compensate for the extra radial width in the model.

\subsubsection{HD~131835}
Parametric models in \citet{Han_2026} suggest that the radial density distribution of this source is well-fit by a double Gaussian, assuming vertical Gaussian profile with a fixed vertical aspect ratio of $h_{\rm HWHM}=0.015$. We find that a radial double Gaussian continues to perform significantly better than a radial double power law or a radial triple Gaussian when the vertical aspect ratio is included as a free model parameter. We then modeled HD~131835 using a radial double Gaussian density distribution and the four different vertical parameterizations presented in Table \ref{tab:formsfree}.

The AIC and BIC strongly favor the vertical Gaussian and vertical Lorentzian models, with the Lorentzian model favored at low confidence over the Gaussian model (AIC $\sigma=0.548$, $\Delta \rm{BIC}=1.077$). Therefore, we select the Lorentzian model as the fiducial model for HD~131835. The posterior probability distributions are broad for both the fiducial and vertical Gaussian model, with an aspect ratio of $h_{\rm HWHM}=0.022^{+0.007}_{-0.006}$ for the fiducial Lorentzian model and $h_{\rm HWHM}=0.030^{+0.010}_{-0.010}$ for the vertical Gaussian model. \frank and \rave find $h_{\rm HWHM}=0.039^{+0.006}_{-0.006}$ and $h_{\rm HWHM}=0.057^{+0.020}_{-0.028}$, respectively.

\subsubsection{HD~161868}
Parametric models in \citet{Han_2026} suggest that the radial density distribution of this source is well-fit by a double power law, assuming a vertical Gaussian profile with a fixed vertical aspect ratio of $h_{\rm HWHM}=0.015$. We modeled HD~161868 with a radial double power law and the four different vertical parametrizations presented in Table \ref{tab:formsfree}. We find that the vertical Lorentzian model is favored by the BIC (AIC $\sigma=1.255$), while the double Gaussian model is favored by the AIC ($\Delta \rm{BIC}=20.786$). We select the vertical Lorentzian model as the fiducial model for HD~161868.

For the fiducial model, we measure a scale height of $h_{\rm HWHM}=0.192^{+0.012}_{-0.016}$. The double Gaussian model is likely disfavored by the BIC more because of the additional parameters in the model rather than due to a poor fit to the data, as evidenced by the AIC. We thus find evidence that we resolve two distinct dynamical populations within the disk. The first has a measured aspect ratio of $h_{\rm HWHM}=0.099^{+0.034}_{-0.045}$ containing $89^{+3}_{-2} \;\%$ of the dust mass, and the other has an aspect ratio of $h_{\rm HWHM}=0.380^{+0.037}_{-0.023}$ containing the rest of the mass. Fits with \frank and \rave yield only upper limits, finding $h_{\rm HWHM}<0.097$ and $h_{\rm HWHM}<0.135$, respectively.

\citet{Matra_2025} marginally resolve the scale height of HD~161868, finding $h_{\rm HWHM}=0.15^{+0.05}_{-0.06}$ with a parametric fit assuming Gaussian radial and vertical density distributions. By nonparametrically fitting the visibilities and assuming a Gaussian vertical density distribution, \citet{Terrill_2023} find a marginal measurement of $h_{\rm HWHM}=0.18^{+0.04}_{-0.05}$ and \citet{Han_2025a} find $h_{\rm HWHM}=0.16^{+0.04}_{-0.05}$. All three of these measurements are consistent with the aspect ratio of our fiducial model.

\subsubsection{HD~197481 (AU Mic)}
The parametric modeling in \citet{Han_2026} shows that models of HD~197481 perform comparably well with a variety of radial density structures, assuming a vertical Gaussian profile with a fixed vertical aspect ratio of $h_{\rm HWHM}=0.015$. We modeled HD~197481 using both radial double power law and double Gaussian density distributions with the four different vertical parametrizations presented in Table \ref{tab:formsfree}. There is no single model which stands out as a true best fit, but we identify the radial double power law and vertical Lorentzian model as the fiducial model due to its moderate AIC confidence ($\sigma=2.491$) and strong BIC ($\Delta \rm{BIC}=0$). For this model, we resolve an aspect ratio of $h_{\rm HWHM}=0.0028^{+0.0009}_{-0.0002}$, the smallest in our sample.

The radial double Gaussian and vertical exponential model was favored by the AIC ($\Delta \rm{BIC}=23.328$), yielding a scale height measurement of $h_{\rm HWHM}=0.0338^{+0.0014}_{-0.0015}$ with a profile that drops off sharply ($\gamma=16^{+3}_{-3}$). \frank and \rave find similar aspect ratios: $h_{\rm HWHM}=0.023^{+0.002}_{-0.002}$ and $h_{\rm HWHM}=0.026^{+0.005}_{-0.006}$, respectively. Lastly, our model with a radial double power law and vertical double Gaussian density distribution also cannot be ruled out (AIC $\sigma=1.196$, $\Delta \rm{BIC}=15.567$), corresponding to two measured scale heights: one with $h_{\rm HWHM}=0.0009^{+0.0005}_{-0.0003}$ containing $78^{+9}_{-14} \;\%$ of the dust mass, and the other with $h_{\rm HWHM}=0.058^{+0.022}_{-0.017}$ containing the rest of the mass. 

Many of these measurements are comparable to other aspect ratio measurements of HD~197481; for example, \citet{Daley_2019} model the 1.35 mm observations with a radial power law and vertical Gaussian, finding $h_{\rm HWHM}=0.037^{+0.006}_{-0.005}$. \citet{Vizgan_2022} repeated this analysis, with a resulting measurement of $h_{\rm HWHM}=0.029^{+0.005}_{-0.006}$. They also analyze 450 $\mu$m observations of HD~197481, yielding a ratio of $h_{1.35 \rm{mm}}/h_{450 \mathrm{\mu m}} = 1.35$ ($h_{\rm HWHM}\approx 0.022$ at 450 $\mu$m). Recently, \citet{Matra_2025} reanalyzed the 1.35 mm data assuming a radially Gaussian belt with a vertical Gaussian density distribution, finding $h_{\rm HWHM}=0.020^{+0.004}_{-0.004}$. \citet{Terrill_2023} used a nonparametric approach to fit the radial structure of the 1.35 mm observations; assuming a vertical Gaussian profile, they find $h_{\rm HWHM}=0.024^{+0.002}_{-0.002}$. Similarly, \citet{Han_2025a} find $h_{\rm HWHM}=0.026^{+0.004}_{-0.004}$.

The variety of measured $h_{\rm HWHM}$ for HD~197481 suggest that the characterization of its vertical structure is highly dependent on the assumed density structure. Furthermore, while the fiducial measurement of $h_{\rm HWHM}$ --- the smallest aspect ratio measured for this source --- is resolved according to the criteria outlined in Sect. \ref{sec:analysis}, it remains unclear whether the data resolution is sufficient to fully resolve such a small aspect ratio. The Band 6 observations of HD~197481 have a nominal resolution of $\theta_{\rm min}=1.86$ au. At a reference radius of 36.3 au, an aspect ratio of $h_{\rm HWHM}=0.0028$ yields a scale height of $H_{\rm HWHM} \approx 0.1$ au. A vertical Lorentzian profile with a HWHM of 0.1 au would contain 95\% of the disk material within 2.5 au centered on the disk midplane. This is less than a factor of two larger than $\theta_{\rm min}$; while such small spatial scales could be supported by the longest baseline observations, they cannot be resolved with two resolution elements. Given the spread of measured $h_{\rm HWHM}$ values under different density distribution assumptions, higher resolution follow-up observations may be helpful in breaking this degeneracy.

\begin{figure}[t]
\centering
\includegraphics[width=\linewidth]{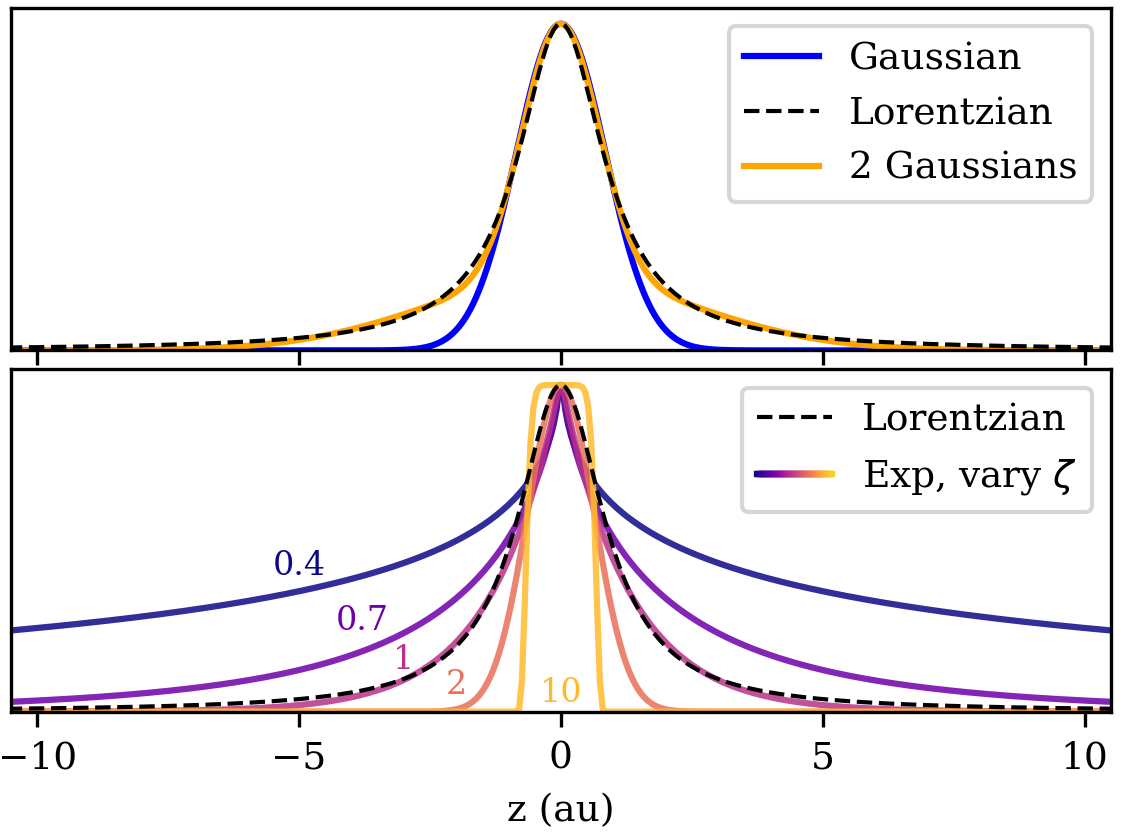}
\caption{Top: Arbitrary Gaussian, Lorentzian, and double Gaussian profiles that demonstrate how a Lorentzian can mimic a double Gaussian without the extra model parameters. The Gaussian (blue line) and Lorentzian (black line) profiles have a FWHM of 2 au, while the double Gaussian (orange dotted line) is composed of one Gaussian with a FWHM of 1.7 au and another with a FWHM of 6 au. Bottom: Various exponential forms with different $\zeta$ (colored lines, $\zeta$ values annotated on the panel). The wings of the Lorentzian (black line) are closely matched with an exponential with $\zeta=1$.
All profiles are normalized to the same maximum value.}
\label{fig:profile_compare}
\end{figure}

\section{Discussion} \label{sec:discuss}
\subsection{Non-Gaussian vertical distributions as the new norm} \label{sec:discuss:lor}

Previous debris disk studies commonly assume that the vertical density structure follows a Gaussian distribution \citep[e.g.,][]{Kennedy_2018, Daley_2019, Terrill_2023, Matra_2025}. This common assumption arises because self-stirring processes can generate a Rayleigh inclination distribution \citep{Ida_1992, Lissauer_1993}, which corresponds to a Gaussian vertical density distribution \citep{Matra_2019}. 
Furthermore, distinguishing between different vertical parameterizations may not have been possible in studies analyzing lower-resolution data, particularly since the radial dust distribution is also less certain in these conditions.

However, high resolution observations of debris disks have enabled a more detailed exploration of the precise dust distributions. We find that most sources analyzed here (10/13) are significantly better fit with a thick-tailed vertical density distribution (i.e., either a Lorentzian or a double Gaussian) compared to a single Gaussian distribution. For each of the three remaining disks, vertical Lorentzian models are not ruled out at high confidence (HD~14055 with a radial double power law and vertical Lorentzian, AIC $\sigma=1.616$ and $\Delta \rm{BIC}=3.376$; HD~109573 with a radial double Gaussian and vertical Lorentzian, AIC $\sigma=0.090$ and $\Delta \rm{BIC}=0.149$; and HD~131488 with a radial double Gaussian and vertical Lorentzian, AIC $\sigma=1.255$ and $\Delta \rm{BIC}=1.141$; see Table \ref{tab:formsall}).

\begin{table*}
\caption{Stellar, disk, and parametric model properties with derived solid mass and stirring estimates.}
\label{tab:masses}
\centering
\renewcommand{\arraystretch}{1.4}
\renewcommand\cellgape{\Gape[2pt]}
\rowcolors{2}{gray!15}{white}
\begin{tabular}{l|ccccl|llll}
\hline
Source & \makecell{$M_{\star}$ \\ {[$M_{\odot}$]}} & \makecell{$\theta_{\rm min}$ \\ {[au]}} & \makecell{$R_{\rm{ref}}$ \\ {[au]}} & \makecell{$H_{\rm{HWHM}}(R_{\rm{ref}})$ \\ {[au]}} & \makecell{$M_{\rm{mm}}$ \\ {[$M_{\oplus}$]}} & \makecell{$M_{\rm{disk}}$ \\ {[$M_{\oplus}$]}} & \makecell{$M_{\rm{stir}}$ \\ {[$M_{\oplus}$]}} & \makecell{$R_{\rm{stir}}$ \\ {[km]}} & \makecell{$T_{\rm{stir}}$ \\ {[Myr]}} \\
\hline
HD~9672   &  2.00  & 4.16 & 110.1    & 5.06  & $4.95\times 10^{-1}$ & $86.6 \pm 15.5$ & $3.11^{+1.85}_{-1.34} \times 10^{-4}$ & $606^{+102}_{-104}$  & 1.23 \\ 
HD~10647  &  1.12  & 3.67 & 88.2     & 4.67  & $4.07\times 10^{-2}$ & $19.8 \pm 2.6$  & $2.72^{+1.10}_{-0.97} \times 10^{-4}$ & $579^{+69}_{-79}$    & 0.43 \\ 
HD~14055  &  2.19  & 3.09 & 160.6    & 8.51  & $5.56\times 10^{-2}$ & $25.3 \pm 4.2$  & $3.18^{+1.53}_{-1.18} \times 10^{-4}$ & $609^{+85}_{-88}$    & 40.0 \\ 
HD~15115  &  1.43  & 3.05 & 97.9     & 2.15  & $6.17\times 10^{-3}$ & $3.72 \pm 0.26$ & $2.51^{+0.48}_{-0.46} \times 10^{-5}$ & $261^{+16}_{-17}$    & 13.5 \\ 
HD~32297  &  1.57  & 2.85 & 107.3    & 0.987 & $1.22\times 10^{\,0}$& $21.9 \pm 3.0$  & $1.81^{+0.83}_{-0.61} \times 10^{-6}$ & $109^{+15}_{-14}$    & 76.6 \\ 
HD~39060  &  1.72  & 3.46 & 116.1    & 13.0  & $1.83\times 10^{-1}$ & $104 \pm 5$     & $3.32^{+0.12}_{-0.12} \times 10^{-3}$ & $1333^{+16}_{-16}$   & 0.10 \\ 
HD~61005  &  0.95  & 2.81 & 70.0     & 1.00  & $1.26\times 10^{-1}$ & $10.2 \pm 1.7$  & $6.02^{+2.94}_{-2.28} \times 10^{-6}$ & $163^{+23}_{-24}$    & 6.38 \\ 
HD~76582  &  1.61  & 18.51& 188.5    & 36.2  & $2.25\times 10^{-1}$ & $114 \pm 12$    & $7.26^{+1.12}_{-0.96} \times 10^{-3}$ & $1730^{+85}_{-80}$   & 1.20 \\ 
HD~109573 &  2.14  & 3.63 & 77.6     & 0.636 & $4.68\times 10^{-2}$ & $7.21 \pm 3.1$  & $3.31^{+5.92}_{-2.80} \times 10^{-6}$ & $133^{+54}_{-62}$    & 6.86 \\ 
HD~131488 &  1.81  & 3.32 & 90.8     & 0.536 & $2.75\times 10^{-1}$ & $9.01 \pm 1.64$ & $7.41^{+4.80}_{-3.18} \times 10^{-7}$ & $81^{+15}_{-14}$     & 70.5 \\ 
HD~131835 &  1.70  & 9.75 & 69.2     & 1.52  & $4.08\times 10^{-1}$ & $49.7 \pm 15.3$ & $5.30^{+6.53}_{-3.12} \times 10^{-5}$ & $330^{+105}_{-90}$   & 0.07 \\ 
HD~161868 &  2.11  & 11.57& 141.9    & 27.2  & $2.29\times 10^{-2}$ & $87.1 \pm 27.0$ & $1.68^{+0.33}_{-0.39} \times 10^{-2}$ & $2288^{+143}_{-190}$ & 0.10 \\ 
HD~197481 &  0.61  & 1.86 & 36.3     & 0.102 & $1.49\times 10^{-2}$ & $0.830\pm 0.179$& $6.01^{+7.62}_{-1.38} \times 10^{-8}$ & $35^{+11}_{-3}$      & 117  \\ 
\hline
\end{tabular}
\tablefoot{Left: The stellar mass $M_{\star}$, nominal data resolution $\theta_{\rm min}$, reference radius $R_{\rm{ref}}$, scale height $H_{\rm{HWHM}}$, and millimeter dust mass $M_{\rm{mm}}$ for each source. $R_{\rm{ref}}$ is the location of peak brightness in the fiducial parametric model, and the location at which we report the fiducial scale height. $M_{\rm{mm}}$ is obtained from our fiducial parametric models. Right: Estimated total mass of disk solids $M_{\rm{disk}}$ assuming only internal processes (self-stirring, damping, etc.) are derived from the \citet{Pan_2012} collisional cascade model. Estimates on the lower limits of the largest stirring bodies in the disks ($M_{\rm{stir}}$ and $R_{\rm{stir}}$) are obtained from Eqns. \ref{eqn:vrel} and \ref{eqn:vesc}. The stirring timescale $T_{\rm{stir}}$ is computed as described in Sect. \ref{sec:discuss:solids}.}
\end{table*}

Some studies have already used non-Gaussian vertical structures for debris disks, in particular for HD~39060 ($\beta$ Pic). For example, scattered light observations of HD~39060 have been fit using two vertical Lorentzians \citep[e.g.,][]{Golimowski_2006, Lagrange_2012} and two vertical Gaussians \citep[e.g.,][]{Apai_2015}, though the authors emphasize that these parameterizations were not physically motivated, but rather chosen by qualitatively assessing the suitability of different profiles. Similarly, \citet{Matra_2019} find that modeling HD~39060 using two vertical Gaussians provides a significantly better fit to the 1.33 mm dust observations than a single Gaussian, and use this parameterization to find marginal evidence of a radially varying vertical aspect ratio.

Now equipped with a larger sample of highly resolved debris disk observations, we have begun teasing out trends in the vertical dust distributions of disks. In our detailed modeling of 13 disks, we find evidence that the millimeter-sized dust typically exhibits a more gradual decline from the midplane density compared to the commonly used, steeper Gaussian profile. Rather, the best models tend to have vertical dust distributions with the heavy-tailed behavior seen in a Lorentzian distribution, or in the combination of a narrow Gaussian with a broader Gaussian. While we do not resolve two distinct scale height values in many disks at high confidence (only HD~10647 is identified as having a fiducial model with a vertical double Gaussian), we note that a Lorentzian distribution can be qualitatively similar to 1) the sum of two Gaussians with different widths and 2) an exponential distribution with $\zeta<2$, both of which are shown in Figure \ref{fig:profile_compare}. Thus, some disks that are best fit by a Lorentzian may have multiple dynamical populations even if we are unable to measure multiple distinct scale heights at high confidence.

When comparing a model with fewer free parameters (e.g., one with a Gaussian vertical density distribution) to a more complex model like the vertical double Gaussian models presented here, the choice of model selection criteria may have a large impact on which model is identified as the best fit. Here, four of the 13 sources (HD~9672, HD~10647, HD~39060, and HD~161868) have a vertical double Gaussian model which is most strongly favored by the AIC, while no sources have a vertical double Gaussian model which is most strongly favored by the BIC. As the BIC penalizes model complexity more aggressively than the AIC, particularly for large datasets like ALMA observations (Eqn. \ref{eqn:aicbic}), it may be useful to employ multiple selection criteria (e.g., both the AIC and BIC) to robustly evaluate model fit.

Multi-component fits have long been used for the classical Kuiper belt objects (KBOs), including double Gaussians, a Gaussian and a Lorentzian, and two von Mises–Fisher functions \citep{Brown_2001, Kavelaars_2009, Gulbis_2010, VanLaerhoven_2019, Malhotra_2023}. Single Lorentzians have also been used to fit the classical KBOs \citep{Adams_2014}. The inclinations and eccentricities of KBOs were likely inflated by Neptune's outward migration during the early history of the Solar System \citep[e.g.,][]{Malhotra_1995, Hahn_2005, Levison_2008, Nesvorny_2015}. Therefore, one possible interpretation of the apparent preference for Lorentzian or multi-component distributions in debris disks is that Neptune-like migration histories are common in these systems.

However, such structures can also arise from secular (i.e., long-term) planet-disk interactions without invoking planetary migration. For instance, \citet{Farhat_2023} find a two-component inclination distribution emerges for the case of HD~106906, accounting for the central binary star as well as the known external planetary companion, while assuming a massless disk. \citet{Sefilian_2025} consider a scenario involving a massive, back-reacting (but not self-gravitating) debris disk with an internal planetary perturber, finding that non-Rayleigh inclinations and non-Gaussian vertical profiles emerge naturally regardless of the disk mass. The range of disk processes and conditions which could generate multiple vertical populations is still an active area of study. For example, most previous works on planet-debris disk interactions neglect at least one or more of: the disk's gravitational back-reaction and/or self-gravity, self-stirring processes, the possibility of multiple perturbers, and collisions within the disk. Gas-bearing debris disks should also be treated with care, as gas effects can also have a large impact on grain dynamics depending on the uncertain gas densities \citep[though this may not have an observable effect on the vertical profile at millimeter wavelengths,][]{Olofsson_2022}. Regardless of the mechanisms driving the vertical grain distribution, non-Gaussianity appears to be nearly ubiquitous.

Furthermore, if the vertical distribution of solids changes with observing wavelength, then even a Gaussian vertical distribution of solids could produce a non-Gaussian emission profile \citep[e.g.,][]{Terrill_2023}. This degeneracy could be broken by modeling well-resolved observations at multiple wavelengths. Currently, most debris disks which have been observed at multiple wavelengths lack the resolution for detailed vertical characterization at all wavelengths (Appendix~\ref{appendix:h}).

\subsection{Constraints on the disk mass} \label{sec:discuss:solids}

\begin{figure}
\centering
\includegraphics[width=\linewidth]{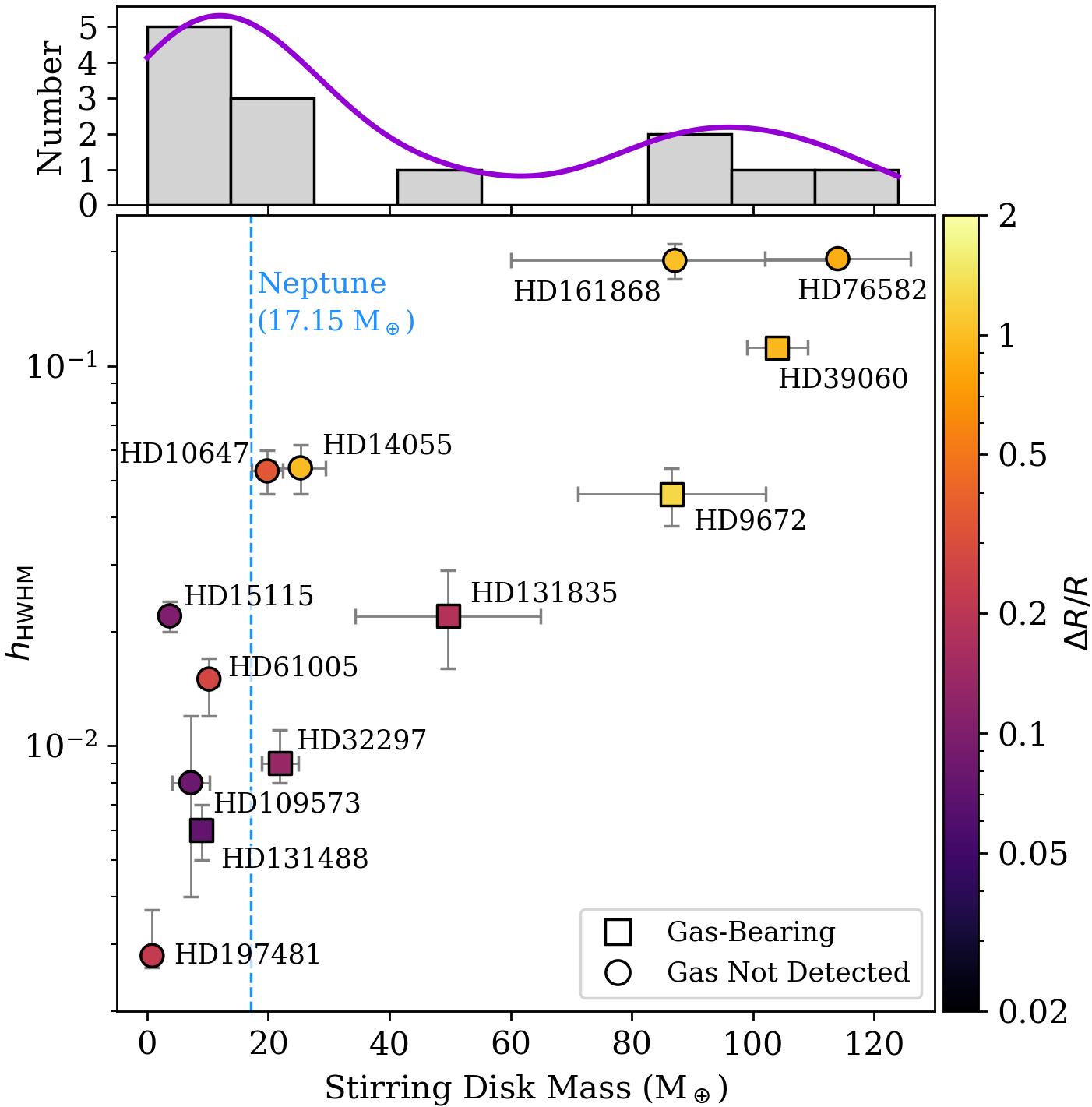}
\caption{Stirring disk masses obtained from the \citep{Pan_2012} stirring model with the fiducial $h_{\rm HWHM}$ measurements for both the gas-bearing (square) and gas-poor (circles) debris disks. The color bar corresponds to the parametrically derived fractional widths ($\Delta R/R$) presented in \citet{Han_2026}. We took the average $\Delta R/R$ for disks with multiple reported rings (HD~15115 and HD~197481). The upper panel shows a histogram of disk masses in our sample overlaid with a KDE (purple line).}
\label{fig:discmasses}
\end{figure}

The vertical thickness of debris disks can be influenced by a number of mechanisms, including early stirring in the protoplanetary disk phase \citep[e.g.,][]{Walmswell_2013, Booth_2016}, stirring by one or more planetary perturbers \citep[e.g.,][also see Sect. \ref{sec:discuss:planets}]{Mustill_2009, Pearce_2014, Jilkova_2015, Dong_2020, Chai_2024}, or stellar flybys \citep[e.g.,][]{Kenyon_2002b, Lestrade_2011, Moore_2020, Batygin_2020}. Here, we primarily consider self-stirring of the disk, which is commonly invoked as a key mechanism driving disk scale heights \citep[e.g.,][]{Ida_1990, Ida_1993, Kenyon_2008, Kennedy_2010}. 

Under the assumption that the vertical thickness of disks is set by only self-stirring, measured scale height values can be used to place an upper limit on the mass of bodies stirring the disk. Following \citet{Daley_2019}, we defined the rms eccentricities and inclinations as $\bar{e}=\sqrt{\left\langle e^2\right\rangle}$ and $\bar{i}=\sqrt{\left\langle i^2\right\rangle}$, respectively, where $\langle e^2\rangle$ and $\langle i^2\rangle$ are the eccentricity and inclination dispersions of the grain population. We assume that grains are gravitationally excited isotropically in space such that $\bar{i}=\bar{e}/2$ \citep{Inaba_2001}. At millimeter wavelengths, the inclination of a self-stirred disk is related to the observed aspect ratio by $\bar{i} \sim \sqrt{2}h_{\rm HWHM}$ \citep{Quillen_2007}.
The relative velocities of particles are then given by
\begin{equation} \label{eqn:vrel}
    \left\langle v_{\rm rel}\right\rangle \approx v_{\rm Kep}(r) \sqrt{\bar{i}^2+1.25 \bar{e}^2} \approx 2\sqrt{3} v_{\rm Kep}(r) h_{\rm HWHM}, 
\end{equation}
where $v_{\rm Kep}(r)$ is the Keplerian velocity at radius $r$ \citep{Lissauer_1993, Wetherill_1993, Wyatt_2002}.

In the absence of significant damping, $\langle v_{\rm rel}\rangle$ will be excited to roughly the escape velocity, $v_{\rm{esc}}$, of the largest bodies in the system \citep[which dominate the disk dynamics and drive most of the stirring, e.g.,][]{Pan_2012, Schlichting_2014}, at which point collisions begin to dominate, limiting the growth of $\langle v_{\rm rel}\rangle$. The escape velocity is given by
\begin{equation} \label{eqn:vesc}
    v_{\rm{esc}} = \sqrt{\dfrac{2\pi G D^{2}\rho_d}{3}}
\end{equation}
for a spherical object of diameter $D$ and bulk density $\rho_d$. By equating $\langle v_{\rm rel}\rangle$ and $v_{\rm{esc}}$, we can obtain lower limits on the size and mass of the largest bodies in the disk. We assumed a bulk density of $\rho_d=2.0$ g/cm$^3$, corresponding to the bulk densities of trans-Neptunian objects \citep{Carry_2012}. The minimum radii $R_{\rm stir}$ and masses $M_{\rm stir}$ of the largest bodies stirring each disk are presented in Table \ref{tab:masses}.

We also estimated the total mass of each disk using the collisional cascade model of \citet{Pan_2012}. The model assumes a steady-state disk in which grains are stirred by close encounters with large bodies and damped by dynamical friction and collisions. Mass is conserved in all catastrophic collisions, and we assumed efficient collisional damping such that collisions with small bodies contribute significantly to the velocity damping \citep[see Sect. 2 of][]{Pan_2012}. We adopted an eccentricity of 0.03 for all perturbing bodies. For each disk, we used Eqn. \ref{eqn:vrel} to estimate the typical random velocity of millimeter-sized grains within the disk, using the fiducial scale height values presented in Table \ref{tab:formsshort}. We used the stellar masses shown in Table \ref{tab:masses}, which were derived in \citet{Marino_2026}. Solving for the steady-state sizes and velocity distributions enabled the dust characteristics obtained from our parametric modeling (e.g., the millimeter dust mass and scale height) to be extrapolated to estimate the total amount of mass in the disk. 

The derived masses, $M_{\rm{disk}}$, for each disk are presented in Table \ref{tab:masses} and plotted in Figure \ref{fig:discmasses}. The main panel shows the connection between $h_{\rm HWHM}$ and $M_{\rm{disk}}$ (we note that the positive correlation occurs due to the model specification, as $h_{\rm HWHM}$ is used to obtain the typical grain velocity, which directly affects the derived disk mass), while the top panel shows a histogram of the disk masses. Under half of the disks (5/13) contain less than the mass of Neptune, while the more massive disks reach up to several times the mass of Neptune. The color bar corresponds to the parametrically derived fractional widths ($\Delta R/R$) presented in \citet{Han_2026}. We find that belts which are vertically thicker and radially wider tend to be more massive. Furthermore, we see that the gas-bearing disks in the sample tend to be more massive than gas-poor disks with similar scale heights; however, gas is not considered in the collisional cascade model used to obtain our estimates on $M_{\rm{disk}}$.

To verify whether the disk masses are large enough to stir the disks to their observed morphologies within the age of the system, we compute the stirring timescale $T_{\rm stir}$ using Eqn. 28 in \citet{Krivov_2018}. This calculation requires assumptions about the maximum planetesimal radius, belt location and fractional width, disk mass, stellar mass, bulk density, and fragmentation velocity $v_{\rm frag}$. We use the values of $R_{\rm stir}$, $R_{\rm ref}$, $M_{\rm disk}$, and $M_{\star}$ in Table~\ref{tab:masses}, and the $\Delta R/R$ values from \citet{Han_2026}. As with our stirring mass estimates, we assume a bulk density of $\rho_d=2.0$ g/cm$^3$. Because $v_{\rm frag}$ is not well-constrained, we adopt the nominal value of $v_{\rm frag}=30$ m/s used in \citet{Krivov_2018}. Lastly, the equation includes the parameter $\gamma$, which ranges from $1 \leq \gamma \leq 2$ and bounds the upper and lower limits on $T_{\rm stir}$; we choose an intermediate $\gamma=1.5$. The stirring timescales are shown in Table~\ref{tab:masses}. We find that all but three disks (HD~32297, HD~131488, and HD~197481) could have been self-stirred within the age of the system. The disks that cannot be explained solely by self-stirring require an alternative mechanism, such as the planet-disk stirring scenarios discussed in Sect. \ref{sec:discuss:planets}.

There exist other methods of estimating $M_{\rm{disk}}$. For example, \citet[hereafter \citetalias{Krivov_2021}]{Krivov_2021} compute a minimum disk mass $M_{\rm disk}^{\rm min}$ contained within the collisional cascade by extrapolating the observed mm-sized dust, assuming a power law size distribution with $q=3.7$ up to some maximum size at the top of the cascade ($s_{\rm max}$, estimated using observables). The \citet{Pan_2012} model employed here, on the other hand, estimates the mass of the largest bodies driving the stirring within the disk, which may not necessarily contribute directly to the collisional cascade (as the collision timescale becomes larger than the age of the system at these sizes). In this case, we assume that the dust mass is negligible, such that $M_{\rm disk} \approx M_{\rm large}$.

These different modeling frameworks result in different mass estimates; for many of the disks, our estimates of $M_{\rm{disk}}$ are incompatible with the minimum masses reported in \citetalias{Krivov_2021}. However, much of this difference may stem from the choice of assumed millimeter dust mass, as the $M_{\rm disk}^{\rm min}$ estimates from \citetalias{Krivov_2021} use different millimeter dust masses than found here. To mitigate this effect, we rescaled the \citetalias{Krivov_2021} estimates of $M_{\rm disk}^{\rm min}$ to use the millimeter dust masses derived from our fiducial models (Table \ref{tab:masses}), finding that the estimates of $M_{\rm disk}^{\rm min}$ are compatible with the estimates of $M_{\rm{disk}}$ found using the \citet{Pan_2012} model for four disks, and incompatible for the remaining three disks. These results are presented and discussed in more detail in Appendix~\ref{appendix:krivovcompare}.

\subsection{Comparison to other vertical measurements}
\begin{figure}[t]
\centering
\includegraphics[width=\linewidth]{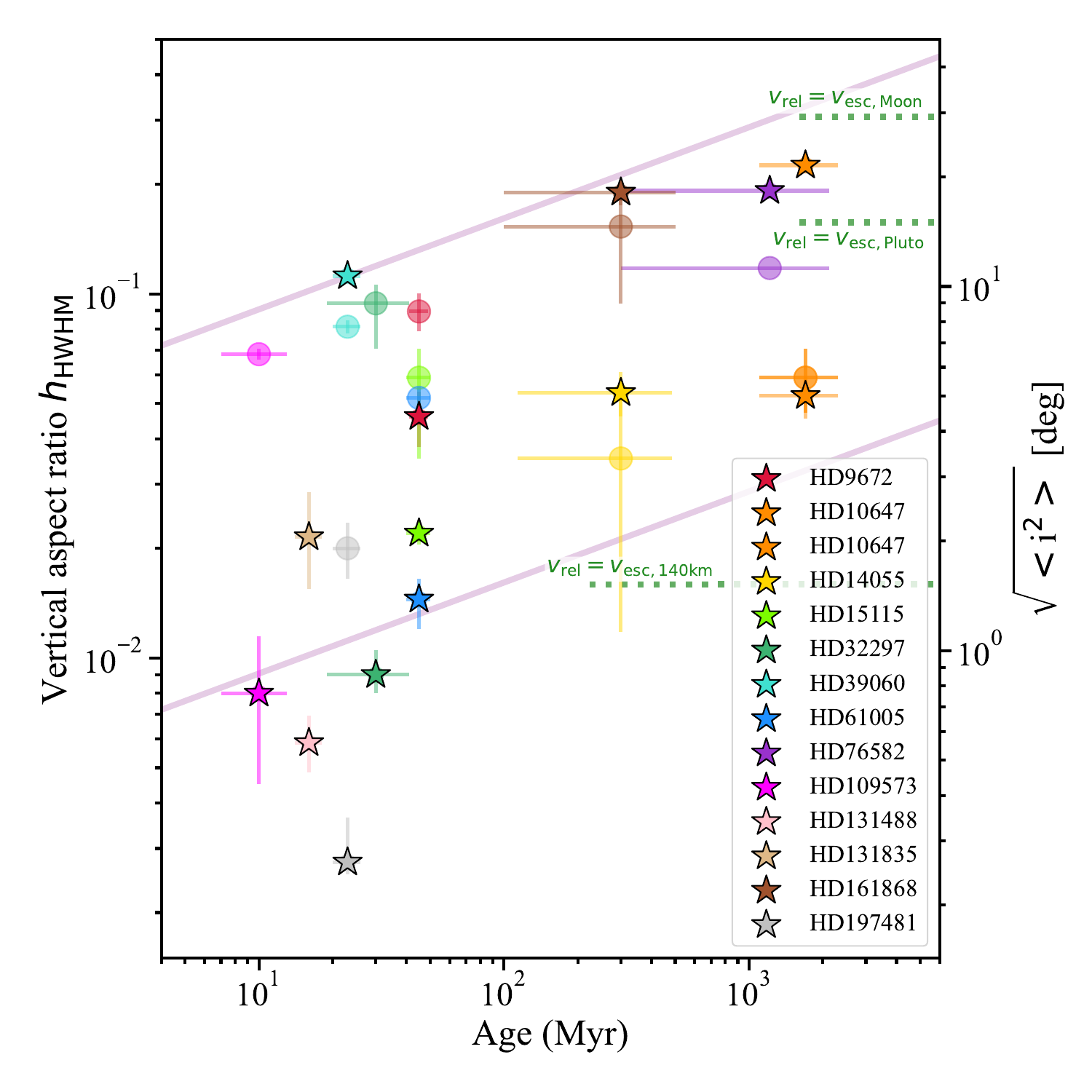}
\caption{Aspect ratio $h_{\rm HWHM}$ of the vertical profiles measured for the ARKS data (stars) compared to the values derived from the REASONS survey \citep[circles, ][]{Matra_2025} with respect to host star age. ARKS values are the best-fit values presented in Table \ref{tab:formsshort}; we note that for HD~10647, two aspect ratios are plotted because the best-fit model is composed of a vertical double Gaussian with two measured aspect ratios. The corresponding rms inclination of dust grains assuming a Rayleigh distribution of inclinations is shown on the right y axis. Purple lines show the $\sqrt{\left\langle i^2\right\rangle} \propto t^{\frac{1}{4}}$ increase with age expected for planetesimal belts stirred by large bodies \citep[assuming that stirring bodies form early;][]{Ida_1993}, with mass $\times$ surface density $\Sigma M = 10^{-2} \: \rm{M_{\oplus}^2}$ au${^2}$ (top) and $\Sigma M = 10^{-6} \: \rm{M_{\oplus}^2}$ au${^2}$ (bottom). Green dotted lines represent the maximum rms inclinations expected from Moon-like, Pluto-like, and 140 km-sized stirring bodies \citep[assuming the median stellar host mass and belt radius of the REASONS sample, ][]{Matra_2025}.}
\label{fig:age_vs_h}
\end{figure}

The REASONS survey uniformly analyzed observations of 74 resolved debris disks \citep{Matra_2025}, many of which are also present in the ARKS sample. The modeling and analysis resulted in scale height measurements or constraints for many of the disks, including 11 out of the 13 disks in the sample presented here. In Figure \ref{fig:age_vs_h}, we compare the REASONS aspect ratios (circles) with the aspect ratios derived from the ARKS parametric modeling (stars).

We find that some sources with well-resolved aspect ratios from REASONS have significantly smaller (but still well resolved) values derived from the ARKS parametric modeling. The disks with the largest discrepancies tend to be some of the thinnest disks in the sample, according to the parametric constraints presented in this work (e.g., HD~197481, HD~109573, HD~32297). Some theoretical predictions suggest an increase in inclination dispersion (or scale height) with age \citep[e.g., without considering damping and following $\sqrt{\left\langle i^2\right\rangle} \propto t^{\frac{1}{4}}$ as in ][assuming that stirrers form early in the disk lifetime]{Ida_1993}. While this trend was not resolved with the REASONS sample, we find some evidence of an increase in $\langle i^2\rangle$ with disk age in the ARKS scale height measurements.

The large differences between the REASONS and ARKS aspect ratios likely stems from differences in the model parametrizations, data resolution, and observing wavelength. \citet{Matra_2025} modeled each of the REASONS disks with a simple model that assumes the disk is both radially and vertically Gaussian. This had the advantage of enabling direct comparison between disks across a large sample of debris disks, including characterizing basic characteristics of the dust belts. However, the ARKS radial analysis reveals a diverse set of belt morphologies, very few of which are consistent with a Gaussian. The work presented here further suggests that vertically, a Gaussian dust distribution may not be representative of many debris disks.

High resolution was not the focus of the REASONS survey, while the ARKS observing strategy focused specifically on resolving the belts well enough to conduct a more detailed analysis of each source  \citep[see][]{Marino_2026}. On average, the resolution of the REASONS observations is several times poorer than the ARKS observations of a given target. In some cases, the angular resolution of the ARKS observations is nearly an order of magnitude higher (e.g., HD~32297 is resolved 0\farcs06 with ARKS and 0\farcs5 with REASONS). This resolution may have resulted in biased scale height measurements, even if they appear to be well-constrained using the simple Gaussian parametrization. Some of the aspect ratios recovered with the ARKS data still appear to be limited by the data resolution to some extent (Figure \ref{fig:vertical_profiles}); follow-up observations of the thinnest disks at higher angular resolution may be useful to fully understand the dependence of measured scale height on resolution. 

Lastly, variations in scale height with observing wavelength have already been observed \citep[e.g., for AU Mic in ALMA Bands 6 and 9,][]{Vizgan_2022}. The REASONS sample includes observations at $0.855 \leq \lambda$ [mm] $\leq 1.36$, so differences in the observing wavelengths could contribute to differences in measured aspect ratios. However, this is likely only a minor contribution, as it is not clear that such a small difference in wavelength would result in a measurable difference in $h$. A list of ARKS targets with measured aspect ratios (both from this work and previous studies) is available in Table \ref{tab:all_h}, along with the corresponding data resolution and observing wavelength.

\begin{figure}
\centering
\includegraphics[width=\linewidth]{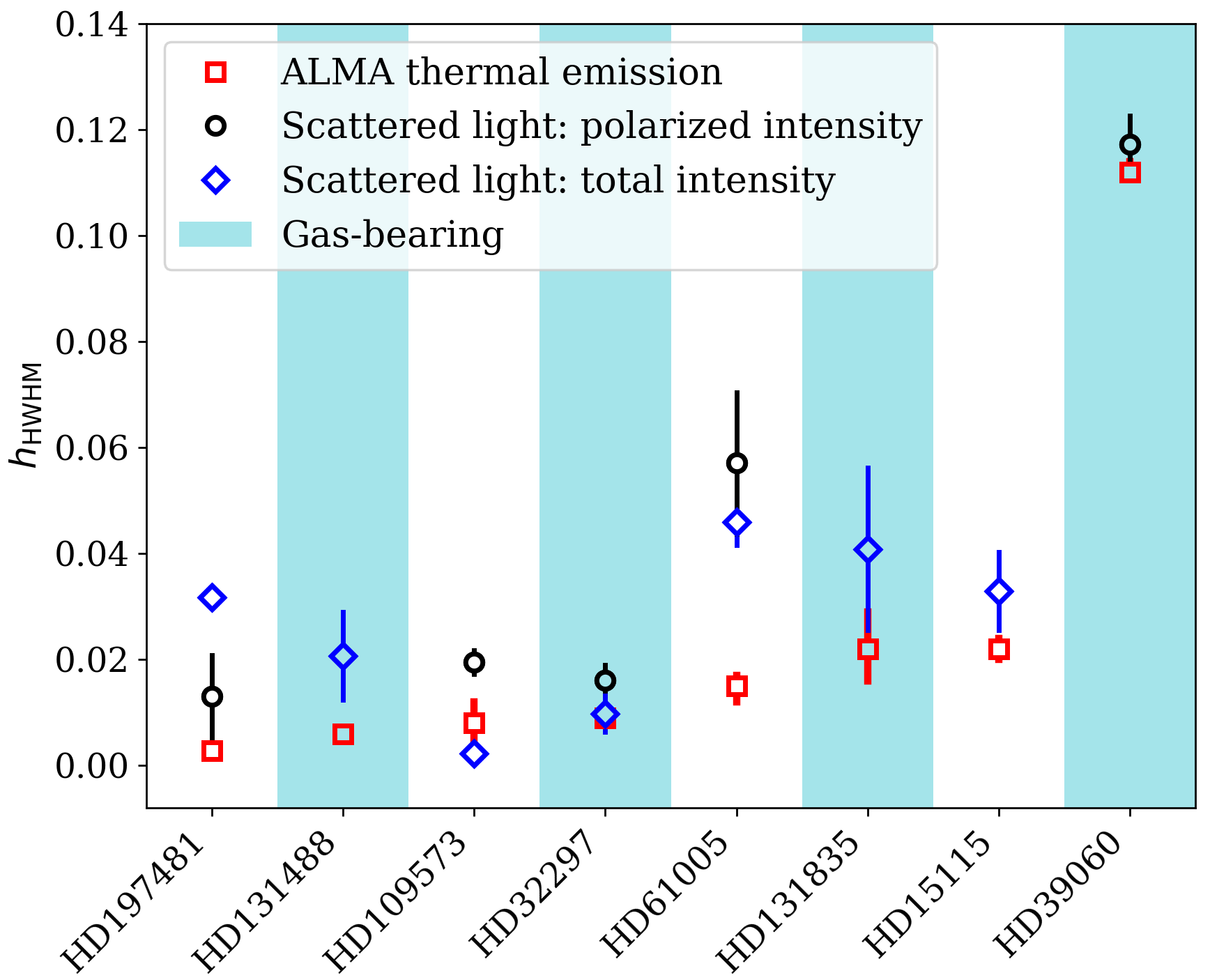}
\caption{Comparison of ALMA $h_{\rm HWHM}$ values (fiducial ARKS values, red squares) with $h_{\rm HWHM}$ constraints from scattered light observations (polarized intensity in black circles and total intensity in blue diamonds). Gas-bearing disks are indicated with a shaded blue background.}
\label{fig:scattered}
\end{figure}

Many ARKS targets have also been detected in scattered light; a detailed comparison of the scattered light and thermal emission of these disks is presented in \citet{Milli_2026}. Here, we briefly discuss the vertical profiles and scale heights from the scattered light observations. The vertical profiles of the disks modeled in scattered light are parametrized with a Gaussian profile of standard deviation $r \tan \psi$ linearly increasing with the radius $r$, where $\psi$ is the vertical opening angle of the disk. The HWHM of such a vertical profile is therefore $\sqrt{2\ln{2}}\times r \tan \psi$ and the corresponding aspect ratio is $h_{\rm HWHM}=\sqrt{2\ln{2}} \tan \psi$. 

This quantity is plotted in Figure \ref{fig:scattered} together with the $h_{\rm HWHM}$ derived from the fiducial ARKS parametric models (Table \ref{tab:formsshort}). Systems are plotted with increasing millimeter scale height, and gas-bearing disks are highlighted with a light blue background. We find that most disks have a similar scale height or are thicker in scattered light compared to thermal imaging. HD~39060 has a substantially larger scale height in both thermal emission and scattered light compared to the rest of the targets.

Theoretical works predict the gas-bearing disks to have a smaller scale heights caused by dust vertical settling \citep{Olofsson_2022}. For second-generation gas present at a low level ($\lesssim0.1M_\oplus$, depending on the dust mass), this settling is expected to be most visible in scattered light while the larger grains seen in thermal images are less affected. On the other hand, we expect larger quantities of gas ($\gtrsim0.1M_\oplus$) to cause significant damping in both millimeter and scattered light observations. In Figure \ref{fig:scattered}, no clear trend emerges from the comparison of the gas-bearing disks shown shaded in blue with the disks containing no detectable amount of CO.

These comparisons come with several caveats. First, not all disks are well resolved vertically in scattered light. For instance, this is the case for HD~32297, as mentioned in \citet{Olofsson_2022_HD32297}. Second, depending on the dataset used to model the disk vertical distribution in scattered light \cite[the spectral filter for HD~15115 or the choice between total or polarized light for most other disks, see details in][]{Milli_2026}, the scale height may be different, mostly because the signal-to-noise of the scattered light observations is not the same.

\subsection{Theoretical implications} \label{sec:discuss:theory}

\begin{figure}[t]
\centering
\includegraphics[width=\linewidth]{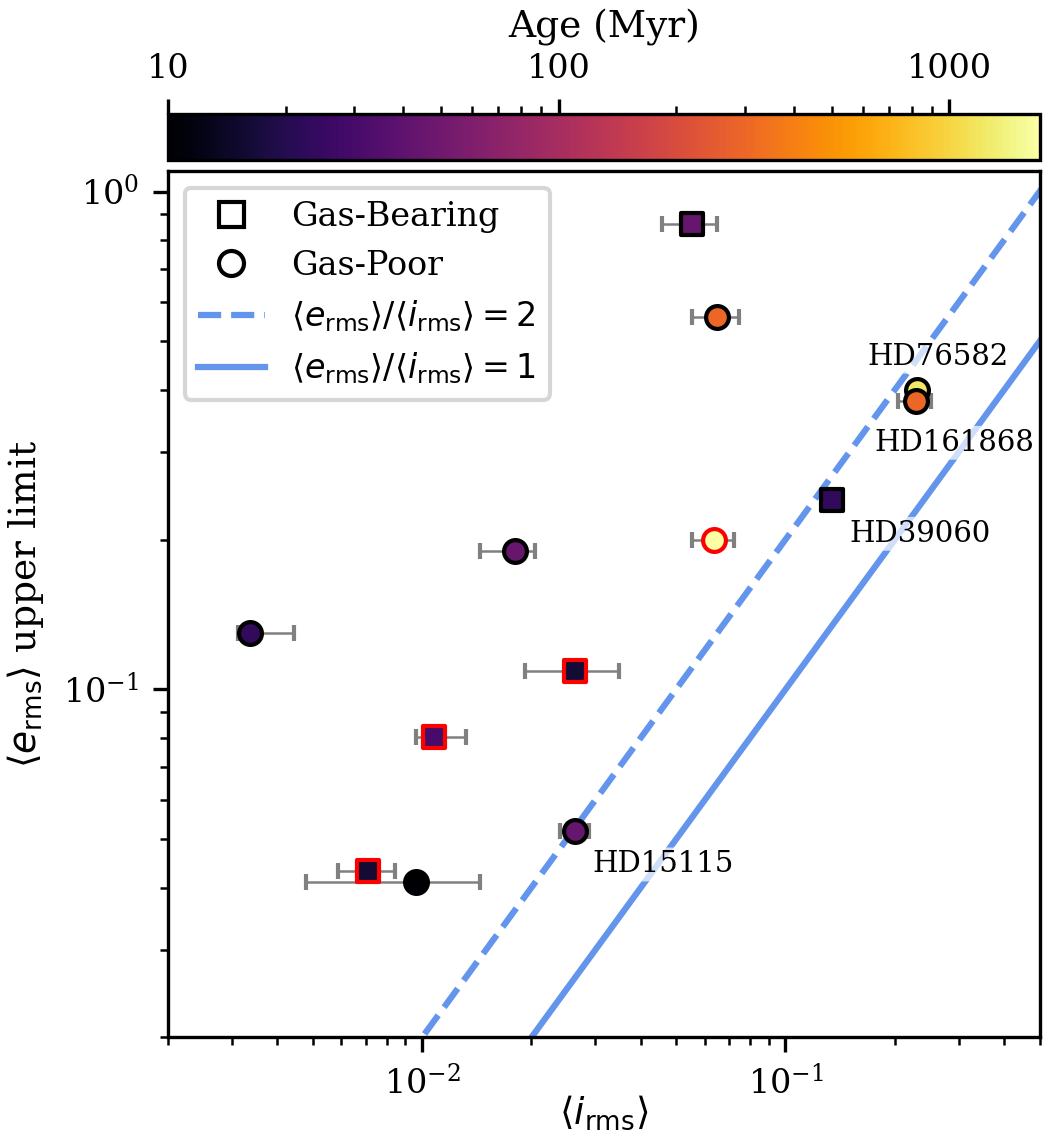}
\caption{Eccentricity and inclination dispersions estimated from the radial and vertical density distributions, respectively. Points with black outlines use the upper limits on $\langle e_{\rm rms} \rangle$ presented in \citet{Han_2026}, while points with red outlines have $\langle e_{\rm rms} \rangle$ estimated using Eqn. \ref{eqn:drr}. Disks may be dynamically dominated by catastrophic collisions when $\langle e_{\rm rms} \rangle / \langle i_{\rm rms} \rangle \lesssim 1$, rather than self-stirred or planet-stirred scenarios that result in a larger ratio of $\langle e_{\rm rms} \rangle / \langle i_{\rm rms}\rangle$. We note that the y-axis shows upper limits of $\langle e_{\rm rms} \rangle$, so we only find upper limits on $\langle e_{\rm rms} \rangle / \langle i_{\rm rms} \rangle$.}
\label{fig:evsi}
\end{figure}

One way to probe which dynamical effects are at work in these disks is to consider the ratio $\langle e_{\rm rms} \rangle / \langle i_{\rm rms}\rangle$. In self-stirred disks an energy equipartition between vertical and radial motion is expected, such that $\langle e_{\rm rms} \rangle / \langle i_{\rm rms}\rangle \sim 2$ \citep{Ida_1992}. If a disk is stirred by a (nearly) co-planar planet, eccentricities will be excited to a greater degree than the inclinations, and we may expect that $\langle e_{\rm rms} \rangle / \langle i_{\rm rms}\rangle>1$ \citep{Pearce_2024}. Alternatively, if a disk is pre-stirred (e.g., if the dynamical excitation is set during the protoplanetary phase), and if the eccentricity and the inclination distributions subsequently evolve due to fragmentation-dominated collisional evolution, we may expect that the ratio decreases to $\langle e_{\rm rms} \rangle / \langle i_{\rm rms}\rangle<1$ with inefficient collisional damping \citep{Jankovic_2024}. The vertical and radial constraints from this work and \citet{Han_2026} enable observational constraints to be placed on this ratio. 

We used the upper limit estimates of $\langle e_{\rm rms} \rangle$ presented in \citet{Han_2026}, which assume a self-stirred disk \citep{Marino_2021, Rafikov_2023}. For disks with no reported $\langle e_{\rm rms} \rangle$, we estimated the upper limit of $\langle e_{\rm rms} \rangle$ using the belt fractional width $\Delta R/R$ as in \citet{Marino_2021},
\begin{equation} \label{eqn:drr}
    \langle e_{\rm rms} \rangle \leq 0.6 \Delta R/R.
\end{equation}
For these sources, we adopted the median fractional widths reported in \citet{Han_2026}. We estimated $\langle i_{\rm rms} \rangle$ as described in Sect. \ref{sec:discuss:solids}. We find that the majority of disks have an upper limit of $\langle e_{\rm rms} \rangle / \langle i_{\rm rms} \rangle > 2$, while four (HD~15115, HD~39060, HD~76582, and HD~161868) have $\langle e_{\rm rms} \rangle / \langle i_{\rm rms} \rangle \lesssim 2$ (Figure \ref{fig:evsi}). While none of the disks have an upper limit of $\langle e_{\rm rms} \rangle / \langle i_{\rm rms} \rangle < 1$, the true ratio could be lower, so we cannot rule out that some disks in the sample may be dynamically dominated by catastrophic collisions.

We emphasize that the values of $\langle e_{\rm rms} \rangle$ and $\langle i_{\rm rms} \rangle$ shown in Figure \ref{fig:evsi} should be taken as estimates. Because $\langle e_{\rm rms} \rangle$ is derived from the radial distribution of material in the disk, there is a degeneracy between the semi-major axis distribution and the eccentricity distribution in the disk. Thus, the true values of $\langle e_{\rm rms} \rangle$ could be smaller. Similarly, while the errors on $\langle i_{\rm rms} \rangle$ shown in Figure \ref{fig:evsi} stem only from uncertainties in the measured scale height, there is additional uncertainty stemming from the approximations and assumptions used to compute $\langle i_{\rm rms} \rangle$. We calculate $\langle i_{\rm rms} \rangle$ using the approximation from \citet{Quillen_2007}, such that  $\langle i_{\rm rms} \rangle \sim \sqrt{2} h_{\rm HWHM}$. \citet{Matra_2019}, on the other hand, assume a Rayleigh distribution of inclinations in a self-stirred disk to derive $\langle i_{\rm rms} \rangle = \sqrt{2} h_{\sigma}$. We note that $h_{\sigma}$ is approximately 85\% of $h_{\rm HWHM}$. However, the non-Gaussianity of most vertical profiles means that neither of these equations are strictly true for this application. 

Another way to compare different theoretical models to the observations is to consider how $h$ varies with wavelength. If a disk is fragmentation-dominated with no significant collisional damping, $h$ is expected not to vary with wavelength, in viscously stirred disks \citep{Kenyon_2002, Kenyon_2004, Kenyon_2008, Kennedy_2010} as well as in pre-stirred disks \citep{Jankovic_2024}. On the other hand, if $h$ increases with wavelength, it could be a sign of efficient collisional damping in the system \citep{Pan_2012}.

We compiled ALMA measurements of $h_{\rm HWHM}$, shown in Table \ref{tab:all_h} and summarized in Figure \ref{fig:all_h}. A few systems show a consistent increase in measured $h_{\rm HWHM}$: HD~15115, HD~32297, HD~61005, and HD~131488. The remaining disks with multiwavelength constraints (HD~9672, HD~10647, HD~14055, HD~107146, HD~161868, and HD~197481\footnote{While not obvious here, there is evidence that HD~197481 is vertically thicker in Band 6 observations compared to Band 9 observations at a similar resolution \citep{Vizgan_2022}.}) appear more consistent with an aspect ratio that remains approximately constant with wavelength. The remaining disks do not have measurements in multiple ALMA bands.

The predictions come with the caveat that results are difficult to directly compare if the $h$ measurements were obtained with different methods or with different data resolutions. Unfortunately, this makes it impossible to draw firm conclusions with the current data and analyses. Follow-up measurements at similar resolutions to the ARKS observations could help to confirm or rule out these tentative trends.

\subsection{Constraints on planetary companions} \label{sec:discuss:planets}

In this work, we have assumed that disks have aspect ratios which are driven by internal processes. However, there is a rich literature exploring alternative scenarios in which planetary (or stellar) companions drive the required dynamical excitation. In these cases, the properties of the hypothetical planet(s) and the disk-to-planet mass ratio ($M_{\rm disk}/m_{\rm pl}$) set the disk particles' eccentricities and inclinations. For example, low mass disks may have observable differences in their vertical structure depending on whether the planet's orbit is inclined with respect to the disk: a coplanar planet would excite the disk radially rather than vertically \citep{Pearce_2014}. On the other hand, initially misaligned, low-inclination planets can generate warps that propagate through the disk over time, causing the disk to settle into a vertically inflated configuration. This can occur regardless of whether the planet lies interior \citep{Mouillet_1997, Dawson_2011, Pearce_2014, Smallwood_2023} or exterior to the disk \citep{Nesvold_2017, Farhat_2023}, assuming the dynamics are not significantly affected by the disk gravity (typically requiring $M_{\rm disk}/m_{\rm pl} \ll 1$; \citealt{Sefilian_2025}). Assuming a given configuration, one can then use the measured vertical scale heights to infer the planet properties. 

For relatively massive disks, the collective gravity of the disk may be significant enough to resist planetary perturbations, maintaining a low dispersion in both radial \citep{Sefilian_2024} and vertical directions \citep{Poblete_2023, Sefilian_2025}. Alternatively, a massive disk could give rise to secular-inclination resonances and localized warps \citep{Sefilian_2025}. This significantly complicates the interpretation of planetary parameters inferred from massless disk models, which typically result in artificially low planetary masses, particularly when the planet lies close to the disk \citep{Sefilian_2025}. A full exploration of planet-driven dynamical excitation scenarios is outside the scope of this study. However, a future study will provide a detailed investigation on planetary inference in the ARKS sample, building on the vertical aspect ratio measurements presented in this work (Jankovic et al. in prep).

\subsection{Limitations} \label{sec:discuss:limits}
We have presented the modeling and analysis of the vertical aspect ratios and dust distributions of the highly inclined ARKS targets. However, the analyses presented here have several limitations which could impact how the results are interpreted. First, we consider only axisymmetric disks with vertical distributions that peak at the disk midplane. The axisymmetric modeling could obscure more complex structures both radially and vertically. For example, the presence of a warp could bias the axisymmetric models to larger scale heights in order to account for the warped material sitting further from the midplane. An analysis of the asymmetric features in the ARKS sample is presented in \citet{Lovell_2026}. 

Second, in this work we assume that aspect ratio is constant with radius. Using a constant aspect ratio is beneficial for enabling direct comparison with the majority of previous vertical structure measurements (see Table \ref{tab:all_h}), as well as for minimizing the number of free parameters in our models. However, recent work has called the validity of this assumption into question, as more disks with evidence of a variable $h$ are identified. For example, \citet{Matra_2019} parametrized the aspect ratio of HD~39060 as $h \propto r^{\beta-1}$. In this parametrization, $\beta=1$ corresponds to a constant aspect ratio (as in our Eqn. \ref{eqn:const_h}), $\beta>1$ to a flared disk, and $\beta=0$ to a constant $H$. The model fit yielded $\beta=0.7 \pm 0.2$, which is consistent with a constant $h(r)$, but also provides marginal evidence (at $<2 \sigma$) of a slowly increasing scale height resulting in a radially varying $h(r)$. This, however, did not result in a significantly improved fit compared to models assuming a constant $h$. Similarly, \cite{Terrill_2023} allowed for a varying aspect ratio with radius when fitting HD~9672 (49~Ceti), and found a loosely constrained value of $\beta=0.79^{+0.29}_{-0.35}$.

More recently, \citet{Han_2025a} fit $H(r)$ nonparametrically, finding several disks that are not consistent with a constant aspect ratio. Some ARKS targets were included in this study: HD~14055 was found to be consistent with a constant aspect ratio, while HD~32297, HD~61005, and HD~197481 were found to have a steeper $H(r)$ (scale height increasing more sharply with radius). Interestingly, HD~15115 was found to have a decreasing $H(r)$, suggesting that the inner ring may be thicker than the outer ring.

From a theoretical perspective, \citet{Sefilian_2025} find that in debris disks perturbed by inner planets, the aspect ratio can follow $h(r) \propto r^{-7/2}$, with the extent of this behavior depending on the disk-to-planet mass ratio, $M_{\rm disk}/M_{\rm planet}$. This effect arises from the disk's ability to mitigate planetary perturbations \citep[see also][]{Sefilian_2024}. When $M_{\rm disk}/M_{\rm planet} \gtrsim 1$, this trend holds across the entire disk; for $M_{\rm disk}/M_{\rm planet} \lesssim 1$, it applies only in the disk's outer regions, while the inner parts maintain $h(r) \approx {\rm const}$. This suggests that planet-disk interactions can produce aspect ratios that vary and decrease with radius. The work presented here only considers the case of a radially constant $h$, but follow-up studies with nonconstant $h$ would be beneficial to fully understand how common scale height variability is in debris disks.

Lastly, the specific radial and vertical density parametrizations can impact the measured scale heights. A poor choice of radial profile can result in biased vertical results, (and vice versa). This may explain some of the differences between $h_{\rm HWHM}$ measurements derived from REASONS and ARKS (Figure \ref{fig:age_vs_h}). Similar discrepancies can be seen in Figure \ref{fig:all_h}, which shows all of the measured $h_{\rm HWHM}$ listed in Table \ref{tab:all_h}. We have attempted to address this issue by leveraging the extensive radial structure analysis presented in \citet{Han_2026} to ensure that we are using the most appropriate parametrizations for the radial dust distributions before extending the exploration to the vertical dust structures. However, it is possible that the radial and/or vertical density distributions adopted in this work do not fully capture the morphology of certain sources.

\section{Conclusions} \label{sec:conclude}

In this paper we have presented the detailed vertical structure analysis of the ARKS debris disks, including constraints from parametric modeling, \frank, and \rave. We successfully measured the scale heights for the 13 most highly inclined targets in the ARKS sample using all three methods, and we recovered scale height constraints for a handful of moderately inclined ARKS targets with \frank and \rave (Table \ref{table:frankraveheight}). Our key results are summarized here:

\begin{itemize}
    \item Many disks are quite thin ($h_{\rm HWHM} \lesssim 0.01$), while others host millimeter-sized dust grains much farther from the midplane ($h_{\rm HWHM} \gtrsim 0.1$).
    \item Lorentzian (or otherwise thick-tailed) vertical dust distributions are common, suggesting that systems may commonly host multiple dynamical populations within the disk. This could hint at Neptune-like migration histories in these systems or could be a remnant of long-term planet-disk interactions in general.
    \item We used our aspect ratio measurements to estimate the total mass of solids in the disk, and we placed limits on the mass of the bodies that could be gravitationally stirring the disks (Table \ref{tab:masses}, Figure \ref{fig:discmasses}). We find that fewer than half of the disks contain less mass than Neptune and that the most massive disks also have the broadest belts radially.
    \item We measure much smaller aspect ratios (thinner disks) than comparable measurements made with lower-resolution data at the same observing wavelength. As a result, we may for the first time be observing the theoretically predicted increase in scale height with disk age. However, because the disks are so thin, follow-up observations at higher resolution could help us better understand this trend.
    \item Aspect ratios obtained from scattered light observations tend to be similar to or larger than the corresponding ALMA millimeter aspect ratios (Figure \ref{fig:scattered}). We do not observe a clear trend when comparing the gas-bearing debris disks against those with no detected gas, though several of the disks are not well resolved vertically in scattered light.
\end{itemize}
These measurements provide new empirical constraints on the vertical structures of debris disks and the range of dynamical processes that are likely to sculpt them. Future high resolution and multiwavelength observations will help to further refine these measurements, strengthening our population level understanding of the debris disk morphologies and dynamics.

\section*{Data Availability}
The ARKS data used in this paper and others can be found in the \href{https://dataverse.harvard.edu/dataverse/arkslp}{ARKS dataverse}. The data products produced by this work can be found at \href{https://doi.org/10.7910/DVN/6NNKLY}{https://doi.org/10.7910/DVN/6NNKLY}. For more information, visit \href{https://arkslp.org}{arkslp.org}.

\begin{acknowledgements}
This paper makes use of the following ALMA data: ADS/JAO.ALMA\# 2022.1.00338.L, 2012.1.00142.S, 2012.1.00198.S, 2015.1.01260.S, 2016.1.00104.S, 2016.1.00195.S, 2016.1.00907.S, 2017.1.00167.S, 2017.1.00825.S, 2018.1.01222.S and 2019.1.00189.S. ALMA is a partnership of ESO (representing its member states), NSF (USA) and NINS (Japan), together with NRC (Canada), MOST and ASIAA (Taiwan), and KASI (Republic of Korea), in cooperation with the Republic of Chile. The Joint ALMA Observatory is operated by ESO, AUI/NRAO and NAOJ. The National Radio Astronomy Observatory is a facility of the National Science Foundation operated under cooperative agreement by Associated Universities, Inc. The project leading to this publication has received support from ORP, that is funded by the European Union’s Horizon 2020 research and innovation programme under grant agreement No 101004719 [ORP]. We are grateful for the help of the UK node of the European ARC in answering our questions and producing calibrated measurement sets. This research used the Canadian Advanced Network For Astronomy Research (CANFAR) operated in partnership by the Canadian Astronomy Data Centre and The Digital Research Alliance of Canada with support from the National Research Council of Canada the Canadian Space Agency, CANARIE and the Canadian Foundation for Innovation. Support for BZ was provided by The Brinson Foundation. This material is based upon work supported by the National Science Foundation Graduate Research Fellowship under Grant No. DGE 2140743. AMH, AJF, and EM acknowledges support from the National Science Foundation under Grant No. AST-2307920. EM acknowledges support from the NASA CT Space Grant. MP acknowledges support from NASA grant 80NSSC21K1334. JM acknowledges funding from the Agence Nationale de la Recherche through the DDISK project (grant No. ANR-21-CE31-0015) and from the PNP (French National Planetology Program) through the EPOPEE project. TDP is supported by a UKRI Stephen Hawking Fellowship and a Warwick Prize Fellowship, the latter made possible by a generous philanthropic donation. A.A.S. is supported by the Heising-Simons Foundation through a 51 Pegasi b Fellowship. MB acknowledges funding from the Agence Nationale de la Recherche through the DDISK project (grant No. ANR-21-CE31-0015). AB acknowledges research support by the Irish Research Council under grant GOIPG/2022/1895. CdB acknowledges support from the Spanish Ministerio de Ciencia, Innovaci\'on y Universidades (MICIU) and the European Regional Development Fund (ERDF) under reference PID2023-153342NB-I00/10.13039/501100011033, from the Beatriz Galindo Senior Fellowship BG22/00166 funded by the MICIU, and the support from the Universidad de La Laguna (ULL) and the Consejer\'ia de Econom\'ia, Conocimiento y Empleo of the Gobierno de Canarias. EC acknowledges support from NASA STScI grant HST-AR-16608.001-A and the Simons Foundation. S.E. is supported by the National Aeronautics and Space Administration through the Exoplanet Research Program (Grant No. 80NSSC23K0288, PI: Faramaz). MRJ acknowledges support from the European Union's Horizon Europe Programme under the Marie Sklodowska-Curie grant agreement no. 101064124 and funding provided by the Institute of Physics Belgrade, through the grant by the Ministry of Science, Technological Development, and Innovations of the Republic of Serbia. This work was also supported by the NKFIH NKKP grant ADVANCED 149943 and the NKFIH excellence grant TKP2021-NKTA-64. Project no.149943 has been implemented with the support provided by the Ministry of Culture and Innovation of Hungary from the National Research, Development and Innovation Fund, financed under the NKKP ADVANCED funding scheme. JBL acknowledges the Smithsonian Institute for funding via a Submillimeter Array (SMA) Fellowship, and the North American ALMA Science Center (NAASC) for funding via an ALMA Ambassadorship. SMM acknowledges funding by the European Union through the E-BEANS ERC project (grant number 100117693), and by the Irish research Council (IRC) under grant number IRCLA- 2022-3788. Views and opinions expressed are however those of the author(s) only and do not necessarily reflect those of the European Union or the European Research Council Executive Agency. Neither the European Union nor the granting authority can be held responsible for them. SM acknowledges funding by the Royal Society through a Royal Society University Research Fellowship (URF-R1-221669) and the European Union through the FEED ERC project (grant number 101162711). JPM acknowledges research support by the National Science and Technology Council of Taiwan under grant NSTC 112-2112-M-001-032-MY3. LM acknowledges funding by the European Union through the E-BEANS ERC project (grant number 100117693), and by the Irish research Council (IRC) under grant number IRCLA- 2022-3788. Views and opinions expressed are however those of the author(s) only and do not necessarily reflect those of the European Union or the European Research Council Executive Agency. Neither the European Union nor the granting authority can be held responsible for them. SP acknowledges support from FONDECYT Regular 1231663 and ANID -- Millennium Science Initiative Program -- Center Code NCN2024\_001. PW acknowledges support from FONDECYT grant 3220399 and ANID -- Millennium Science Initiative Program -- Center Code NCN2024\_001.
\end{acknowledgements}

\bibliographystyle{aa}
\bibliography{refs} 

\appendix
\onecolumn

\section{Summary of parametric modeling} \label{appendix:mongotable}

\begin{table*} [!htbp]
\caption{Summary of the parametric modeling (combinations of radial and vertical functional forms and the number of free parameters $N_p$) for each source.}
\label{tab:formsall}
\centering
\renewcommand{\arraystretch}{1.35}
\begin{tabular}{
    |>{\centering\arraybackslash}m{1.5cm}
    |>{\centering\arraybackslash}m{1cm}
    |>{\centering\arraybackslash}m{1.1cm}
    |>{\centering\arraybackslash}m{0.5cm}
    |>{\centering\arraybackslash}m{1cm}
    |>{\centering\arraybackslash}m{1cm}
    ||>{\raggedright\arraybackslash}m{3.4cm}
    >{\raggedright\arraybackslash}m{3.4cm}|
}
\hline
& & & & & & \multicolumn{2}{c|}{Fiducial Model Parameters} \\ \cline{7-8}
Source & 
Radial Form & 
Vertical Form & 
$N_{p}$ & 
AIC Conf. & 
$\Delta$BIC & 
Best-Fit  & 
50th Percentile \\
\hline
\multirow{10}{*}{HD~9672}   
    &         &          &       &          &        & $\alpha_{\rm{in}}=2.11$          & $\alpha_{\rm{in}}=2.13^{+0.17}_{-0.17}$         \\ 
    &         &          &       &          &        & $\alpha_{\rm{out}}=-1.48$        & $\alpha_{\rm{out}}=-1.45^{+0.09}_{-0.09}$       \\
    & DPL     & vG       & 10    &  7.52    & 60.62  & $R_{\mathrm{c}}=107$ au          & $R_{\mathrm{c}}=107^{+4}_{-4}$ au               \\
    & DG      & vG       & 12    &  $>10$   & 99.18  & $h_{\rm HWHM}=0.041$             & $h_{\rm HWHM}=0.046^{+0.008}_{-0.008}$          \\
    & DPL     & vE       & 11    &  7.76    & 77.54  & $\Sigma_c=-4.170$                & $\Sigma_c=-4.175^{+0.009}_{-0.009}$             \\
    &\bf{DPL} &\bf{vL}   &\bf{10}&\bf{0.27} & \bf{0} & PA$=107.72\degree$               & PA$=107.74$\degree$^{+0.12}_{-0.09}$            \\
    & DPL     & vDG      & 12    & 0        &25.81   & dRA$=0$\farcs$039$               & dRA$=0$\farcs$048^{+0.015}_{-0.016}$            \\ 
    &         &          &       &          &        & dDec$=-0$\farcs$035$             & dDec$=-0$\farcs$037^{+0.006}_{-0.007}$          \\ 
    &         &          &       &          &        & $i=80.03$\degree                 & $i=80.18$\degree$^{+0.12}_{-0.14}$              \\ 
    &         &          &       &          &        & $F_{\star}=5.2\times 10^{-6}$ Jy & $F_{\star}=1.0^{+1.0}_{-0.6}\times 10^{-5}$ Jy  \\ \hline
\multirow{14}{*}{HD~10647}  
    &         &          &       &          &        & $R_1=87.9$ au                    & $R_1=87.6^{+0.9}_{-0.6}$ au                     \\
    &         &          &       &          &        & $R_2=142.4$ au                   & $R_2=143^{+3}_{-3}$ au                          \\
    &         &          &       &          &        & $\sigma_1=10.4$ au               & $\sigma_1=10.6^{+0.8}_{-0.8}$ au                \\
    & DPL     & vG       & 10    & $>10$    & $>100$ & $\sigma_2=36.3$ au               & $\sigma_2=35^{+2}_{-3}$ au                      \\
    & DG      & vG       & 12    & $>10$    & $>100$ & $C_1=0.650$                      & $C_1=0.654^{+0.011}_{-0.013}$                   \\
    & DPL     & vE       & 11    & $>10$    & $>100$ & $h_{\rm HWHM 1}=0.052$           & $h_{\rm HWHM 1}=0.053^{+0.006}_{-0.007}$        \\
    & DG      & vE       & 13    & $>10$    & $>100$ & $h_{\rm HWHM 2}=0.226$           & $h_{\rm HWHM 2}=0.225^{+0.002}_{-0.002}$        \\
    & DPL     & vL       & 10    &6.15      &  0     & $C_{\rm{vert}}=0.91$             & $C_{\rm{vert}}=0.88^{+0.06}_{-0.06}$            \\
    & DG      & vL       & 12    &7.74      & 47.94  & $\Sigma_c=-4.406$                & $\Sigma_c=-4.395^{+0.019}_{-0.019}$             \\
    & DPL     & vDG      & 12    & $>10$    & $>100$ & PA$=57.1$\degree                 & PA$=56.8$\degree$^{+0.29}_{-0.19}$              \\
    &\bf{DG}  &\bf{vDG}  &\bf{14}& \bf{0}   &\bf{9.04}&dRA$=0$\farcs$093$               & dRA$=0$\farcs$092^{+0.010}_{-0.010}$            \\ 
    &         &          &       &          &        & dDec$=0$\farcs$016$              & dDec$=0$\farcs$011^{+0.008}_{-0.009}$           \\
    &         &          &       &          &        & $i=78.74$\degree                 & $i=78.83$\degree$^{+0.21}_{-0.18}$              \\
    &         &          &       &          &        & $F_{\star}=146\times 10^{-6}$ Jy & $F_{\star}=145^{+0.8}_{-1.0}\times 10^{-6}$ Jy  \\ \hline
\multirow{10}{*}{HD~14055}  
    &         &          &     &          &        & $\alpha_{\rm{in}}=1.31$            & $\alpha_{\rm{in}}=1.24^{+0.15}_{-0.12}$         \\
    &\bf{DPL} & \bf{vG}  &\bf{10}&\bf{0.56}&\bf{0} & $\alpha_{\rm{out}}=-3.43$          & $\alpha_{\rm{out}}=-3.2^{+0.7}_{-0.9}$          \\
    & DG      & vG       & 12  &0.27      &25.92   & $R_{\mathrm{c}}=177$ au            & $R_{\mathrm{c}}=180^{+9}_{-12}$ au              \\
    & DPL     & vE       & 11  &1.12      &14.83   & $h_{\rm HWHM}=0.059$               & $h_{\rm HWHM}=0.053^{+0.007}_{-0.008}$          \\ 
    & DG      & vE       & 13  & 0        &38.72   & $\Sigma_c=-5.051$                  & $\Sigma_c=-5.068^{+0.017}_{-0.018}$             \\
    & DPL     & vL       & 10  &1.62      &3.38    & PA$=161.44$\degree                 & PA$=161.59$\degree$^{+0.29}_{-0.16}$            \\
    & DG      & vL       & 12  &1.46      &29.32   & dRA$=-0$\farcs$066$                & dRA$=-0$\farcs$062^{+0.009}_{-0.009}$           \\
    & DPL     & vDG      & 12  &1.59      &29.83   & dDec$=-0$\farcs$064$               & dDec$=-0$\farcs$067^{+0.013}_{-0.011}$          \\
    & DG      & vDG      & 14  &1.07      &54.50   & $i=80.0$\degree                    & $i=80.0$\degree$^{+0.2}_{-0.3}$                 \\
    &         &          &     &          &        & $F_{\star}=1.00\times 10^{-4}$ Jy  & $F_{\star}=1.01^{+0.1}_{-0.1}\times 10^{-4}$ Jy \\ \hline
\end{tabular}
\end{table*}
\begin{table*}
\captionsetup{labelformat=empty}
\caption{Table \ref{tab:formsall} continued.}
\centering
\renewcommand{\arraystretch}{1.3}
\begin{tabular}{
    |>{\centering\arraybackslash}m{1.5cm}
    |>{\centering\arraybackslash}m{1cm}
    |>{\centering\arraybackslash}m{1.1cm}
    |>{\centering\arraybackslash}m{0.5cm}
    |>{\centering\arraybackslash}m{1cm}
    |>{\centering\arraybackslash}m{1cm}
    ||>{\raggedright\arraybackslash}m{3.4cm}
    >{\raggedright\arraybackslash}m{3.4cm}|
}
\hline
& & & & & & \multicolumn{2}{c|}{Fiducial Model Parameters} \\ \cline{7-8}
Source & 
Radial Form & 
Vertical Form & 
$N_{p}$ & 
AIC Conf. & 
$\Delta$BIC & 
Best-Fit  & 
50th Percentile \\
\hline
\multirow{12}{*}{HD~15115}  
    &         &          &     &          &        & $R_1=97.83$ au                     & $R_1=97.88^{+0.05}_{-0.07}$ au                  \\
    &         &          &     &          &        & $R_2=65.94$ au                     & $R_2=66.0^{+0.60}_{-0.40}$ au                   \\
    & DPL     & vG       & 10  & $>10$    &$>100$  & $\sigma_1=4.68$ au                 & $\sigma_1=4.59^{+0.18}_{-0.19}$ au              \\
    & DG      & vG       & 12  & 6.94     &52.55   & $\sigma_2=1.88$ au                 & $\sigma_2=2.3^{+0.9}_{-0.8}$ au                 \\
    & DPL     & vE       & 11  & $>10$    &$>100$  & $C_1=0.78$                         & $C_1=0.81^{+0.04}_{-0.07}$                      \\
    & DG      & vE       & 13  & 6.05     &54.77   & $h_{\rm HWHM}=0.021$               & $h_{\rm HWHM}=0.022^{+0.002}_{-0.002}  $        \\
    & DPL     & vL       & 10  & $>10$    &$>100$  & $\Sigma_c=-3.427$                  & $\Sigma_{c}=-3.437^{+0.038}_{-0.028}$           \\
    &\bf{DG}  & \bf{vL} &\bf{12}& \bf{0}  &\bf{0}  & PA$=98.458$\degree                 & PA$=98.463$\degree$^{+0.013}_{-0.014}$          \\
    & DPL     & vDG      & 12  & $>10$    &$>100$  & dRA$=0$\farcs$038$                 & dRA$=0$\farcs$036^{+0.002}_{-0.003}$            \\
    & DG      & vDG      & 14  & 2.05     &34.56   & dDec$=-0$\farcs$014$               & dDec$=-0$\farcs$014^{+0.001}_{-0.001}$          \\ 
    &         &          &     &          &        & $i=86.74$\degree                   & $i=86.79$\degree$^{+0.05}_{-0.04}$              \\
    &         &          &     &          &        & $F_{\star}=3.3\times 10^{-5}$ Jy   & $F_{\star}=2.5^{+0.6}_{-0.6}\times 10^{-5}$ Jy  \\ \hline
\multirow{10}{*}{HD~32297}  
    &         &          &     &          &        & $\alpha_{\rm{in}}=71$              & $\alpha_{\rm{in}}=66^{+22}_{-18}$               \\
    & DPL     & vG       & 10  & 3.02     &11.96   & $\alpha_{\rm{out}}=-6.4$           & $\alpha_{\rm{out}}=-6.3^{+0.13}_{-0.11}$        \\
    & DG      & vG       & 12  & 5.27     &57.36   & $R_{\mathrm{c}}=105.2$ au          & $R_{\mathrm{c}}=105.4^{+0.5}_{-0.4}$ au         \\
    & DPL     & vE       & 11  & 2.24     &20.24   & $h_{\rm HWHM}=0.0098$              & $h_{\rm HWHM}=0.0092^{+0.0012}_{-0.0012}$       \\
    & DG      & vE       & 13  & 5.09     &68.34   & $\Sigma_c=-2.709$                  & $\Sigma_{c}=-2.718^{+0.012}_{-0.016}$           \\
    &\bf{DPL} & \bf{vL}  &\bf{10}& \bf{0} & \bf{0} & PA$=47.50$\degree                  & PA$=47.50$\degree$^{+0.02}_{-0.04}$             \\
    & DG      & vL       & 12  & 4.33     &47.98   & dRA$=0$\farcs$0201$                & dRA$=0$\farcs$0202^{+0.0015}_{-0.0014}$         \\ 
    & DPL     & vDG      & 12  & 0.97     &27.95   & dDec$=0$\farcs$0152$               & dDec$=0$\farcs$0151^{+0.0013}_{-0.0013}$        \\ 
    &         &          &     &          &        & $i=88.48$\degree                   & $i=88.48$\degree$^{+0.04}_{-0.03}$              \\ 
    &         &          &     &          &        & $F_{\star}=1.1\times 10^{-5}$ Jy   & $F_{\star}=2.3^{+1.2}_{-1.2}\times 10^{-5}$ Jy  \\ \hline
\multirow{10}{*}{\parbox[t]{1.5cm}{\centering HD~39060 (Beta Pic)}}  
    &         &          &     &          &        & $\alpha_{\rm{in}}=1.82$            & $\alpha_{\rm{in}}=1.78^{+0.14}_{-0.10}$         \\
    &         &          &     &          &        & $\alpha_{\rm{out}}=-5.4$           & $\alpha_{\rm{out}}=-5.2^{+0.6}_{-0.5}$          \\
    &         &          &     &          &        & $R_{\mathrm{c}}=125$ au            & $R_{\mathrm{c}}=125^{+2}_{-3}$ au               \\
    & DPL     & vG       & 10  & $>10$    &$>100$  & $h_{\rm HWHM}=0.112$               & $h_{\rm HWHM}=0.112^{+0.002}_{-0.001}$          \\
    & DPL     & vE       & 11  & $>10$    &91.82   & $\Sigma_c=-4.020$                  & $\Sigma_c=-4.031^{+0.014}_{-0.012}$             \\
    & \bf{DPL}& \bf{vL}  &\bf{10}&\bf{1.55}& \bf{0}& PA$=29.76$\degree                  & PA$=29.74$\degree$^{+0.09}_{-0.09}$             \\
    & DPL     & vDG      & 12  &  0       &22.85   & dRA$=-0$\farcs$021$                & dRA$=-0$\farcs$017^{+0.010}_{-0.009}$           \\ 
    &         &          &     &          &        & dDec$=-0$\farcs$007$               & dDec$=-0$\farcs$001^{+0.015}_{-0.015}$          \\ 
    &         &          &     &          &        & $i=87.65$\degree                   & $i=87.66$\degree$^{+0.17}_{-0.13}$              \\
    &         &          &     &          &        & $F_{\star}=7.5\times 10^{-5}$ Jy   & $F_{\star}=7.5^{+1.3}_{-1.2}\times 10^{-5}$ Jy  \\ \hline
\multirow{12}{*}{HD~61005}  
    &         &          &     &          &        & $R_1=69.0$ au                      & $R_1=69.4^{+0.5}_{-0.4}$ au                     \\
    &         &          &     &          &        & $R_2=97$ au                        & $R_2=95^{+2}_{-2}$ au                           \\
    &         &          &     &          &        & $\sigma_1=7.8$ au                  & $\sigma_1=7.5^{+0.5}_{-0.4}$ au                 \\
    & DPL     & vG       & 10  &6.81      &25.66   & $\sigma_2=29.4$ au                 & $\sigma_2=30.8^{+1.7}_{-1.5}$ au                \\
    & DG      & vG       & 12  &2.64      & 9.61   & $C_1=0.839$                        & $C_1=0.837^{+0.011}_{-0.012}$                   \\
    & TG      & vG       & 15  &3.13      &50.28   & $h_{\rm HWHM}=0.0129$              & $h_{\rm HWHM}=0.0143^{+0.0020}_{-0.0021}$       \\
    & DG      & vE       & 13  &3.03      &24.52   & $\Sigma_c=-3.351$                  & $\Sigma_{c}=-3.349^{+0.016}_{-0.015}$           \\
    & \bf{DG} & \bf{vL} &\bf{12}& \bf{0}  &\bf{0}  & PA$=70.286$\degree                 & PA$=70.279$\degree$^{+0.015}_{-0.020}$          \\
    & DG      & vDG      & 14  &1.31      &28.36   & dRA$=0$\farcs$002$                 & dRA$=-0$\farcs$001^{+0.006}_{-0.005}$           \\ 
    &         &          &     &          &        & dDec$=-0$\farcs$017$               & dDec$=-0$\farcs$017^{+0.002}_{-0.003}$          \\
    &         &          &     &          &        & $i=86.23$\degree                   & $i=86.24$\degree$^{+0.03}_{-0.04}$              \\
    &         &          &     &          &        & $F_{\star}=2.3\times 10^{-6}$ Jy   & $F_{\star}=2.0^{+8.0}_{-4.0}\times 10^{-6}$ Jy  \\ \hline
\end{tabular}
\end{table*}
\addtocounter{table}{-1}
\begin{table*}
\captionsetup{labelformat=empty}
\caption{Table \ref{tab:formsall} continued.}
\centering
\renewcommand{\arraystretch}{1.5}
\begin{tabular}{
    |>{\centering\arraybackslash}m{1.5cm}
    |>{\centering\arraybackslash}m{1cm}
    |>{\centering\arraybackslash}m{1.1cm}
    |>{\centering\arraybackslash}m{0.5cm}
    |>{\centering\arraybackslash}m{1cm}
    |>{\centering\arraybackslash}m{1cm}
    ||>{\raggedright\arraybackslash}m{3.4cm}
    >{\raggedright\arraybackslash}m{3.4cm}|
}
\hline
& & & & & & \multicolumn{2}{c|}{Fiducial Model Parameters} \\ \cline{7-8}
Source & 
Radial Form & 
Vertical Form & 
$N_{p}$ & 
AIC Conf. & 
$\Delta$BIC & 
Best-Fit  & 
50th Percentile \\
\hline
\multirow{13}{*}{HD~76582} 
    & DPL     & vG      & 10   &7.99      &68.74   &                                  &                                                \\
    & DG      & vG      & 12   &8.08      &93.10   & $\sigma_{\rm{in}}=47$ au         & $\sigma_{\rm{in}}=45^{+8}_{-8}$ au             \\
    & Erf     & vG      & 10   &7.92      &67.28   & $\sigma_{\rm{out}}=113$ au       & $\sigma_{\rm{out}}=114^{+23}_{-15}$ au         \\
    & AG      & vG      & 10   &7.99      &68.58   & $R_{\mathrm{c}}=182$ au          & $R_{\mathrm{c}} =181^{+11}_{-11}$ au           \\
    & TPL     & vG      & 15   & $>10$    &$>100$  & $h_{\rm HWHM}=0.193$             & $h_{\rm HWHM}=0.192^{+0.009}_{-0.009}$         \\ 
    & Erf     & vE      & 11   &3.81      &29.44   & $\Sigma_c=-4.690$                & $\Sigma_c=-4.692^{+0.019}_{-0.024}$            \\
    & AG      & vE      & 11   &3.50      &27.01   & PA$=103.2$\degree                & PA$=103.9$\degree$^{+1.0}_{-0.7}$              \\
    & DPL     & vL      & 10   &2.09      &6.61    & dRA$=-0$\farcs$029$              & dRA$=-0$\farcs$04^{+0.05}_{-0.05}$             \\
    & Erf     & vL      & 10   &1.63      &4.52    & dDec$=0$\farcs$036$              & dDec$=0$\farcs$04^{+0.03}_{-0.02}$             \\
    & \bf{AG} & \bf{vL} &\bf{10}& \bf{0}  & \bf{0} & $i=72.1$\degree                  & $i=72.6$\degree$^{+0.7}_{-0.6}$                \\
    & DPL     & vDG     & 12   &2.98      &35.04   & $F_{\star}=3.2\times 10^{-5}$ Jy & $F_{\star}=3.6^{+1.8}_{-1.6}\times 10^{-5}$ Jy \\
    & Erf     & vDG     & 12   &2.94      &34.75   &                                  &                                                \\
    & AG      & vDG     & 12   &2.81      &33.92   &                                  &                                                \\ \hline
\multirow{12}{*}{HD~109573} 
    &         &         &      &          &        & $R_1=77.63$ au                   & $R_1=77.61^{+0.03}_{-0.10}$ au                  \\
    &         &         &      &          &        & $R_2=84.46$ au                   & $R_2=83.0^{+1.9}_{-1.7}$ au                     \\
    & DPL     & vG      & 10   &5.29      & 7.99   & $\sigma_1=3.2$ au                & $\sigma_1=3.0^{+0.2}_{-0.3}$ au                 \\
    & \bf{DG} & \bf{vG} &\bf{12}& \bf{0}  & \bf{0} & $\sigma_2=24$ au                 & $\sigma_2=21^{+3}_{-4}$ au                      \\
    & G       & vG      & 9    & $>10$    & 42.41  & $C_1=0.977$                      & $C_1=0.972^{+0.007}_{-0.010}$                   \\
    & AG      & vG      & 10   &7.68      & 39.63  & $h_{\rm HWHM}=0.0065$            & $h_{\rm HWHM}=0.0082^{+0.0033}_{-0.0038}$       \\ 
    & DG      & vE      & 13   &0.95      & 14.06  & $\Sigma_c=-2.810$                & $\Sigma_c=-2.784^{+0.023}_{-0.023}$             \\
    & DG      & vL      & 12   &0.09      & 0.15   & PA$=26.52$\degree                & PA$=26.52$\degree$^{+0.03}_{-0.04}$             \\
    & DG      & vDG     & 14   &1.54      & 28.03  & dRA$=0$\farcs$0130$              & dRA$=0$\farcs$0137^{+0.0008}_{-0.0008}$         \\ 
    &         &         &      &          &        & dDec$=-0$\farcs$0400$            & dDec$=-0$\farcs$0397^{+0.0012}_{-0.0011}$       \\ 
    &         &         &      &          &        & $i=76.57$\degree                 & $i=76.62$\degree$^{+0.09}_{-0.10}$              \\ 
    &         &         &      &          &        & $F_{\star}=2.3\times 10^{-5}$ Jy & $F_{\star}=3.6^{+1.6}_{-1.5}\times 10^{-5}$ Jy  \\ \hline
\multirow{12}{*}{HD~131488} 
    &         &         &      &          &        & $R_1=90.77$ au                   & $R_1=90.86^{+0.14}_{-0.10}$ au                  \\
    &         &         &      &          &        & $R_2=96.8$ au                    & $R_2=95.9^{+0.70}_{-0.50}$ au                   \\
    & DPL     & vG      & 10   & $>10$    &$>100$  & $\sigma_1=1.03$ au               & $\sigma_1=1.17^{+0.18}_{-0.12}$ au              \\
    & \bf{DG} & \bf{vG} &\bf{12}&\bf{0.90}& \bf{0} & $\sigma_2=20.4$ au               & $\sigma_2=19.8^{+1.6}_{-1.3}$ au                \\
    & G       & vG      & 9    & $>10$    &$>100$  & $C_1=0.968$                      & $C_1=0.961^{+0.005}_{-0.006}$                   \\
    & DPL     & vE      & 11   & $>10$    &$>100$  & $h_{\rm HWHM}=0.0048$            & $h_{\rm HWHM}=0.0059^{+0.0011}_{-0.0010}$       \\
    & DG      & vE      & 13   &  0       &11.63   & $\Sigma_c=-2.136$                & $\Sigma_{c}=-2.131^{+0.035}_{-0.063}$    .      \\
    & DPL     & vL      & 10   & $>10$    &$>100$  & PA$=97.304$\degree               & PA$=97.298$\degree$^{+0.013}_{-0.023}$          \\
    & DG      & vL      & 12   &1.26      &1.14    & dRA$=0$\farcs$0017$              & dRA$=0$\farcs$0011^{+0.0007}_{-0.0006}$         \\ 
    &         &         &      &          &        & dDec$=-0$\farcs$007$             & dDec$=-0$\farcs$007^{+0.0002}_{-0.0003}$        \\
    &         &         &      &          &        & $i=84.91$\degree                 & $i=84.87$\degree$^{+0.05}_{-0.05}$              \\
    &         &         &      &          &        & $F_{\star}=4.3\times 10^{-5}$ Jy & $F_{\star}=3.8^{+0.9}_{-0.8}\times 10^{-5}$ Jy  \\ \hline
\end{tabular}
\end{table*}
\addtocounter{table}{-1}
\begin{table*}
\captionsetup{labelformat=empty}
\caption{Table \ref{tab:formsall} continued.} 
\centering
\renewcommand{\arraystretch}{1.5}
\begin{tabular}{
    |>{\centering\arraybackslash}m{1.5cm}
    |>{\centering\arraybackslash}m{1cm}
    |>{\centering\arraybackslash}m{1.1cm}
    |>{\centering\arraybackslash}m{0.5cm}
    |>{\centering\arraybackslash}m{1cm}
    |>{\centering\arraybackslash}m{1cm}
    ||>{\raggedright\arraybackslash}m{3.4cm}
    >{\raggedright\arraybackslash}m{3.4cm}|
}
\hline
& & & & & & \multicolumn{2}{c|}{Fiducial Model Parameters} \\ \cline{7-8}
Source & 
Radial Form & 
Vertical Form & 
$N_{p}$ & 
AIC Conf. & 
$\Delta$BIC & 
Best-Fit  & 
50th Percentile \\
\hline
\multirow{12}{*}{HD~131835} 
    &         &         &      &          &        & $R_1=69.0$ au                    & $R_1=69.0^{+0.4}_{-0.3}$ au                     \\
    &         &         &      &          &        & $R_2=107.7$ au                   & $R_2=109.1^{+2.2}_{-1.9}$ au                    \\
    &         &         &      &          &        & $\sigma_1=4.7$ au                & $\sigma_1=4.8^{+0.6}_{-0.6}$ au                 \\
    & DPL     & vG      & 10   & $>10$    &$>100$  & $\sigma_2=42.7$ au               & $\sigma_2=42.0^{+1.4}_{-1.8}$ au                \\
    & DG      & vG      & 12   &0.55      & 1.08   & $C_1=0.877$                      & $C_1=0.877^{+0.012}_{-0.012}$                   \\
    & TG      & vG      & 15   &1.67      &43.94   & $h_{\rm HWHM}=0.017$             & $h_{\rm HWHM}=0.022^{+0.007}_{-0.006}$          \\ 
    & DG      & vE      & 13   &1.26      &16.22   & $\Sigma_c=-2.915$                & $\Sigma_{c}=-2.912^{+0.047}_{-0.043}$           \\
    & \bf{DG} & \bf{vL} &\bf{12}&\bf{0}   &\bf{0}  & PA$=58.90$\degree                & PA$=58.96$\degree$^{+0.18}_{-0.18}$             \\
    & DG      & vDG     & 14   &1.46      &30.01   & dRA$=0$\farcs$0007$              & dRA$=0$\farcs$0018^{+0.0018}_{-0.0018}$         \\ 
    &         &         &      &          &        & dDec$=-0$\farcs$0063$            & dDec$=-0$\farcs$0056^{+0.0014}_{-0.0014}$       \\
    &         &         &      &          &        & $i=74.5$\degree                  & $i=74.6$\degree$^{+0.2}_{-0.2}$                 \\
    &         &         &      &          &        & $F_{\star}=4.3\times 10^{-6}$ Jy & $F_{\star}=1.0^{+1.2}_{-0.7}\times 10^{-5}$ Jy  \\ \hline
\multirow{10}{*}{HD~161868} 
    &         &         &      &          &        & $\alpha_{\rm{in}}=1.24$           & $\alpha_{\rm{in}}=1.21^{+0.19}_{-0.15}$           \\
    &         &         &      &          &        & $\alpha_{\rm{out}}=-5.3$          & $\alpha_{\rm{out}}=-5.3^{+1.8}_{-2.5}$            \\
    & DPL     & vG      & 10   &3.48      &12.04   & $R_{\mathrm{c}}=160$ au           & $R_{\mathrm{c}}=160^{+11}_{-13}$ au               \\
    & DG      & vG      & 12   &3.29      &34.57   & $h_{\rm HWHM}=0.192$              & $h_{\rm HWHM}=0.192^{+0.012}_{-0.016}$            \\ 
    & DPL     & vE      & 11   &1.94      &14.71   & $\Sigma_c=-4.896$                 & $\Sigma_c=-4.902^{+0.025}_{-0.029}$               \\
    & \bf{DPL}& \bf{vL} &\bf{10}&\bf{1.26}& \bf{0} & PA$=57.8$\degree                  & PA$=57.3$\degree$^{+1.0}_{-1.2}$                  \\
    & DPL     & vDG     & 12   &   0      &20.79   & dRA$=-0$\farcs$004$               & dRA$=-0$\farcs$01^{+0.03}_{-0.02}$                \\ 
    &         &         &      &          &        & dDec$=-0$\farcs$08$               & dDec$=-0$\farcs$08^{+0.02}_{-0.02}$               \\
    &         &         &      &          &        & $i=66.7$\degree                   & $i=66.7$\degree$^{+0.9}_{-0.9}$                   \\
    &         &         &      &          &        & $F_{\star}=1.7\times 10^{-4}$ Jy  & $F_{\star}=1.6^{+0.2}_{-0.2}\times 10^{-4}$ Jy    \\ \hline
\multirow{12}{*}{HD~197481} 
    &         &         &      &          &        & $\alpha_{\rm{in}}=4.5$            & $\alpha_{\rm{in}}=4.5^{+0.3}_{-0.2}$               \\
    &         &         &      &          &        & $\alpha_{\rm{out}}=-12.2$         & $\alpha_{\rm{out}}=-12.2^{+1.6}_{-1.9}$            \\
    & DPL    & vG       & 12   &6.21      &34.01   & $R_{\mathrm{c}}=37.4$ au          & $R_{\mathrm{c}}=37.2^{+0.6}_{-0.7}$ au             \\
    & DG     & vG       & 14   & 3.35     &26.87   & $h_{\rm HWHM}=0.0026$             & $h_{\rm HWHM}=0.0028^{+0.0009}_{-0.0002}$          \\
    & DPL    & vE       & 13   & 3.13     &14.65   & $\Sigma_c=-3.756$                 & $\Sigma_{c}=-3.711^{+0.045}_{-0.052}$              \\
    & DG     & vE       & 15   &   0      &23.33   & PA$=128.73$\degree                & PA$=128.73$\degree$^{+0.02}_{-0.02}$               \\
    &\bf{DPL}& \bf{vL} &\bf{12}&\bf{2.49} & \bf{0} & dRA$=-0$\farcs$019$               & dRA$=-0$\farcs$021^{+0.007}_{-0.007}$              \\
    & DG     & vL       & 14   & 2.95     &24.17   & dDec$=-0$\farcs$012$              & dDec$=-0$\farcs$009^{+0.006}_{-0.005}$             \\
    & DPL    & vDG      & 14   & 1.20     &15.57   & $i=88.16$\degree                  & $i=88.13$\degree$^{+0.08}_{-0.09}$                 \\
    & DG     & vDG      & 16   & 1.09     &36.58   & $F_{\star1}=3.5\times 10^{-4}$ Jy & $F_{\star1}=3.5^{+0.2}_{-0.2}\times 10^{-4}$ Jy    \\ 
    &         &         &      &          &        & $F_{\star2}=1.5\times 10^{-4}$ Jy & $F_{\star2}=1.5^{+0.3}_{-0.2}\times 10^{-4}$ Jy    \\
    &         &         &      &          &        & $F_{\star3}=2.0\times 10^{-4}$ Jy & $F_{\star3}=2.0^{+0.2}_{-0.2}\times 10^{-4}$ Jy   \\ \hline
\end{tabular}
\tablefoot{We use the AIC and BIC to identify a fiducial model for each source, shown in bold. We show the $\Delta$AIC value in terms of a confidence level from the best-fit AIC model (the model with $\Delta$AIC$=0$), and we show the $\Delta$BIC as a difference from the best-fit BIC model (the model with $\Delta$BIC$=0$). The rightmost columns show the best-fit and 50th percentile values for the fiducial models, with errors showing the 16th and 84th percentiles. The surface density normalization factor $\Sigma_c$ is parametrized as an exponent; the units of $10^{\Sigma_c}$ are g\;cm$^{-2}$.}
\end{table*}
\addtocounter{table}{-1}
\clearpage

\section{\frank and \rave vertical inference} \label{appendix:frankraveappend}

\begin{figure}[h!]
\centering
\includegraphics[width=0.6\linewidth]{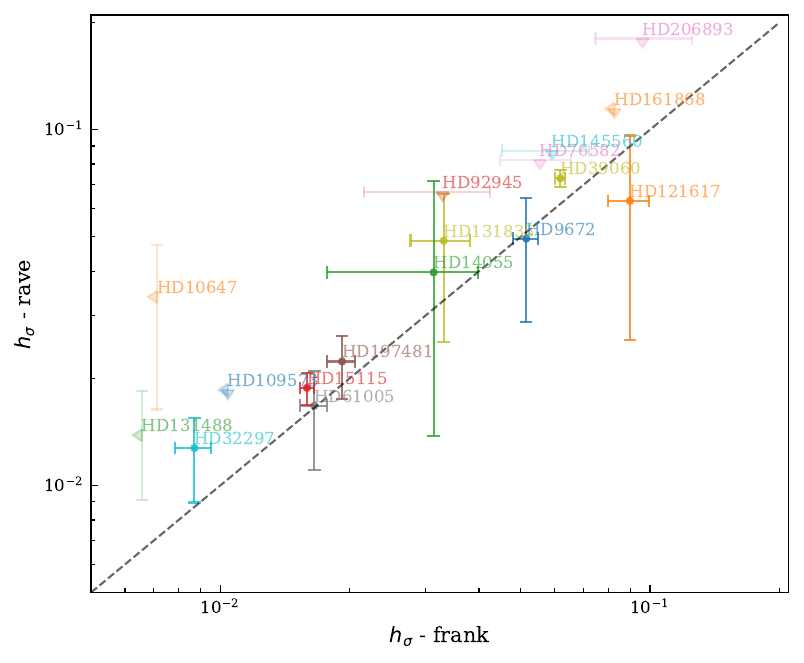}
\caption{Comparison between the scale height aspect ratios $h_{\sigma}$ fitted with \frank and \rave. Circular points indicate measurements with both methods, while triangles indicate constraints that are only upper limits (leftward and downward triangles for upper limits with \frank and \rave, respectively). The dashed line indicates where the fitted values from the two methods are equal. }
    \label{fig:h_comparison}
\end{figure}

\begin{figure*}[h!]
\centering
\includegraphics[width=\linewidth]{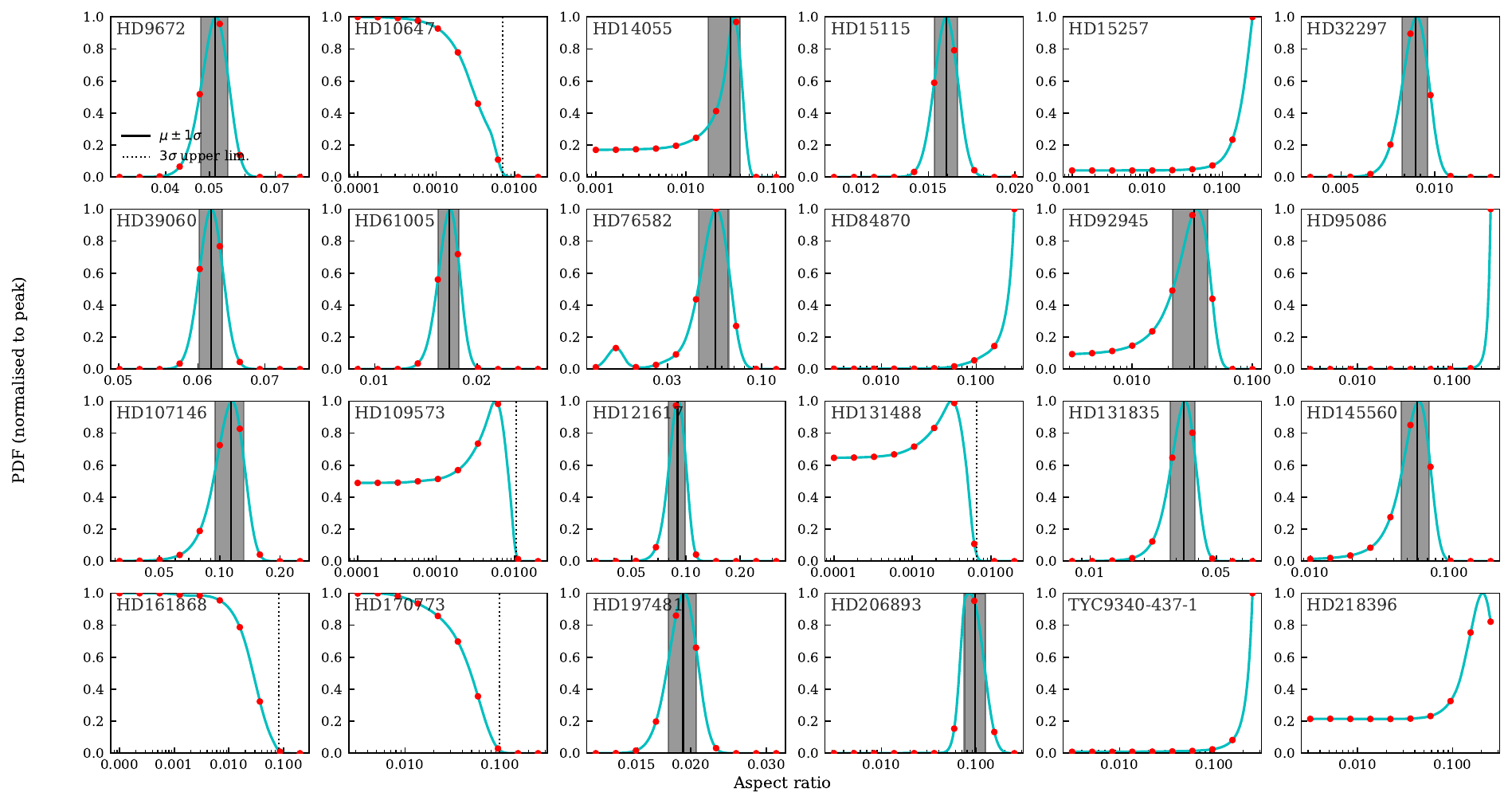}
\caption{Probability density distributions of $h_{\sigma}$ fitted with \frank. The red dots indicate sample points where the \frank model was fitted and the squared residuals computed, based on which the distribution was interpolated to obtain the cyan lines. The vertical solid black lines and shaded region indicate the median and 1$\sigma$ uncertainty region for distributions with a clear peak, whereas black dotted lines indicate the 3$\sigma$ upper limit for disks where only an upper limit can be placed. }
\label{fig:frank_h}
\end{figure*}

\begin{figure*}[h!]
\centering
\includegraphics[width=\linewidth]{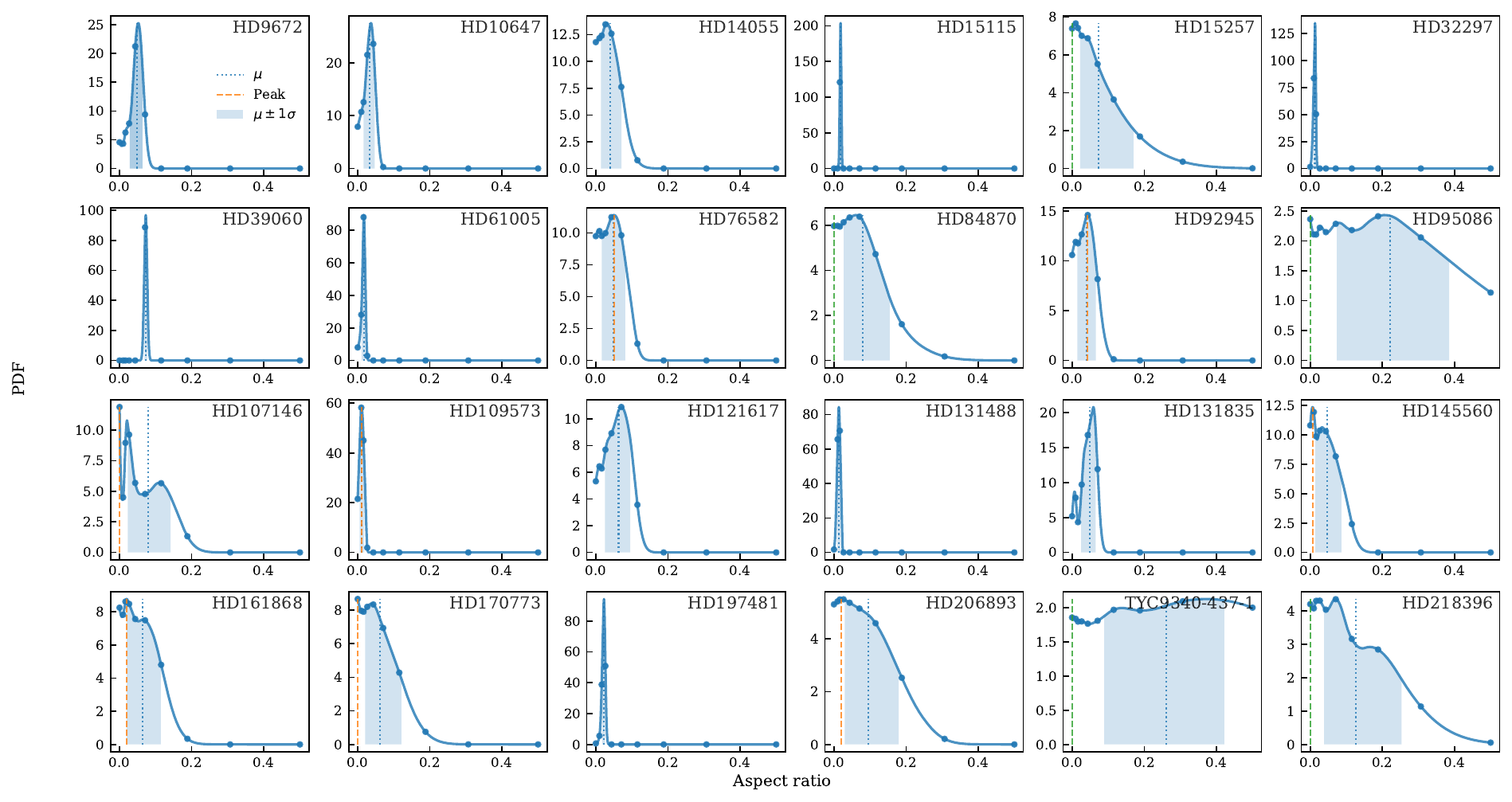}
    \caption{Probability density distributions of $h_{\sigma}$ fitted with \rave. The points in each panel indicate sample points where the \rave model was fitted and the squared residuals computed, based on which the distribution was interpolated. The vertical blue dotted black lines and shaded region indicate the median and 1$\sigma$ uncertainty region for distribution of each disk. An orange dashed line, when present in a panel, indicates the mode of the distribution, and is used as the height assumption when fitting the final model where only an upper bound can be placed. A green dashed line at $h_{\sigma}$=0, when present in a panel, indicates the height is unconstrained and a flat disk was assumed when fitting the final \rave model. For all other disks, the median was used to estimate the scale height and fit the final \rave model. }
    \label{fig:rave_h}
\end{figure*}

\section{Summary of new and archival aspect ratio measurements} \label{appendix:h}

\begin{table*}[h!]
\caption{Compilation of $h$ measurements for each source, along with the resolution and wavelength of the observations used.}
\label{tab:all_h}
\centering
\renewcommand{\arraystretch}{1.3}
\begin{tabular}{
    |>{\centering\arraybackslash}m{1.7cm}
    |>{\centering\arraybackslash}m{2.0cm}
    |>{\centering\arraybackslash}m{2.0cm}
    |>{\centering\arraybackslash}m{2.0cm}
    |>{\centering\arraybackslash}m{1.6cm}
    |>{\centering\arraybackslash}m{1cm}
    |>{\arraybackslash}m{3.6cm}|
}
\hline
Source & 
$h_{\sigma}$ & 
$h_{\rm HWHM}$ & 
$h_{\rm FWHM}$ & 
Resolution (arcsec) & 
$\lambda$ (mm) &
Reference \\
\hline
\multirow{9}{*}{\centering HD~9672}
    & $0.039 \pm 0.007$              & $\bf{0.046 \pm 0.008}$    & $0.092 \pm 0.016$         & 0.16 & 0.89  & This work (parametric) \\ 
    & $\bf{0.051^{+0.004}_{-0.003}}$ & $0.060^{+0.005}_{-0.004}$ & $0.120^{+0.009}_{-0.007}$ & 0.16 & 0.89  & This work (\frank)     \\ 
    & $\bf{0.051^{+0.033}_{-0.034}}$ & $0.060^{+0.039}_{-0.040}$ & $0.120^{+0.078}_{-0.080}$ & 0.16 & 0.89  & This work (\rave)      \\ 
    & $\bf{0.075 \pm 0.004}$         & $0.088 \pm 0.005$         & $0.177 \pm 0.009$         & 0.47 & 0.614 & \citet{Delamer_2021}   \\ 
    & $\bf{0.050 \pm 0.007}$         & $0.059 \pm 0.008$         & $0.118 \pm 0.016$         & 0.47 & 0.614 & \citet{Terrill_2023}   \\ 
    & $\bf{0.033^{+0.014}_{-0.018}}$ & $0.039^{+0.016}_{-0.021}$ & $0.078^{+0.033}_{-0.042}$ & 0.47 & 0.614 & \citet{Han_2025a}      \\ 
    & $\bf{0.069^{+0.005}_{-0.007}}$ & $0.081^{+0.006}_{-0.008}$ & $0.162^{+0.012}_{-0.016}$ & 0.5  & 0.868 & \citet{Delamer_2021}   \\ 
    & $\bf{0.076 \pm 0.009}$         & $0.089 \pm 0.011$         & $0.179 \pm 0.021$         & 0.5  & 0.868 & \citet{Matra_2025}     \\ 
    & $\bf{0.15 \pm 0.02}$           & $0.18 \pm 0.02$           & $0.35 \pm 0.05$           & 0.85 & 1.32  & \citet{Delamer_2021}   \\ 
\hline
\end{tabular}
\end{table*}
\addtocounter{table}{-1}
\begin{table*}[h!]
\captionsetup{labelformat=empty}
\caption{Table \ref{tab:all_h} continued.}
\centering
\renewcommand{\arraystretch}{1.3}
\begin{tabular}{
    |>{\centering\arraybackslash}m{1.7cm}
    |>{\centering\arraybackslash}m{2.0cm}
    |>{\centering\arraybackslash}m{2.0cm}
    |>{\centering\arraybackslash}m{2.0cm}
    |>{\centering\arraybackslash}m{1.6cm}
    |>{\centering\arraybackslash}m{1cm}
    |>{\arraybackslash}m{3.6cm}|
}
\hline
Source & 
$h_{\sigma}$ & 
$h_{\rm HWHM}$ & 
$h_{\rm FWHM}$ & 
Resolution (arcsec) & 
$\lambda$ (mm) &
Reference \\
\hline
\multirow{8}{*}{\centering HD~10647}
    & $0.045^{+0.006}_{-0.005}$      & $\bf{0.053^{+0.006}_{-0.007}}$     & $0.106^{+0.014}_{-0.012}$ & 0.43 & 0.89 & This work (parametric)\tablefootmark{a} \\ 
    & $0.191 \pm 0.002$              & $\bf{0.225 \pm 0.002}$             & $0.450 \pm 0.004$         & 0.43 & 0.89 & This work (parametric)\tablefootmark{b}\\ 
    & $\bf{<0.007}$                  & $<0.008$                           & $<0.016$                  & 0.43 & 0.89 & This work (\frank)           \\ 
    & $\bf{0.036^{+0.023}_{-0.024}}$ & $0.042^{+0.027}_{-0.028}$          & $0.085^{+0.054}_{-0.057}$ & 0.43 & 0.89 & This work (\rave)            \\ 
    & $\bf{0.048 \pm 0.004}$         & $0.057^{+0.005}_{-0.005}$          & $0.113^{+0.009}_{-0.009}$ & 0.5  & 0.86 & \citet{Lovell_2021}          \\ 
    & $\bf{0.037^{+0.007}_{-0.008}}$ & $0.044^{+0.008}_{-0.009}$          & $0.087^{+0.016}_{-0.019}$ & 0.5  & 0.86 & \citet{Terrill_2023}         \\ 
    & $\bf{0.05 \pm 0.01}$           & $0.06 \pm 0.01$                    & $0.12 \pm 0.02$           & 1.0  & 1.26 & \citet{Matra_2025}           \\ 
    & $\bf{0.046^{+0.009}_{-0.010}}$ & $0.054^{+0.011}_{-0.012}$          & $0.108^{+0.021}_{-0.024}$ & 0.5  & 0.86 & \citet{Han_2025a}            \\ 
\hline
\multirow{6}{*}{\centering HD~14055}
    & $0.045^{+0.006}_{-0.007}$      & $\bf{0.053^{+0.007}_{-0.008}}$ & $0.106^{+0.014}_{-0.016}$ & 0.23 & 0.89 & This work (parametric) \\ 
    & $\bf{0.031^{+0.014}_{-0.008}}$ & $0.037^{+0.016}_{-0.009}$      & $0.073^{+0.033}_{-0.019}$ & 0.23 & 0.89 & This work (\frank)     \\ 
    & $\bf{0.040^{+0.026}_{-0.032}}$ & $0.047^{+0.031}_{-0.038}$      & $0.094^{+0.061}_{-0.075}$ & 0.23 & 0.89 & This work (\rave)      \\ 
    & $\bf{<0.10}$                   & $<0.12$                        & $<0.024$                  & 1.85 & 1.27 & \citet{Terrill_2023}   \\ 
    & $\bf{0.03 \pm 0.02}$           & $0.04 \pm 0.02$                & $0.07 \pm 0.05$           & 1.85 & 1.27 & \citet{Matra_2025}     \\ 
    & $\bf{0.111^{+0.062}_{-0.064}}$ & $0.131^{+0.073}_{-0.075}$      & $0.261^{+0.146}_{-0.151}$ & 1.85 & 1.27 & \citet{Han_2025a}      \\ 
\hline
\multirow{6}{*}{\centering HD~15115}
    & $0.019^{+0.002}_{-0.002}$      & $\bf{0.022 \pm 0.002}$         & $0.044^{+0.004}_{-0.004}$ & 0.12 & 0.89 & This work (parametric) \\ 
    & $\bf{0.016 \pm 0.001}$         & $0.019 \pm 0.001$              & $0.038 \pm 0.002$         & 0.12 & 0.89 & This work (\frank)     \\ 
    & $\bf{0.018^{+0.007}_{-0.005}}$ & $0.021^{+0.008}_{-0.006}$      & $0.042^{+0.016}_{-0.012}$ & 0.12 & 0.89 & This work (\rave)      \\ 
    & $\bf{0.048 \pm 0.007}$         & $0.057 \pm 0.008$              & $0.113^{+0.016}_{-0.016}$ & 0.6  & 1.3  & \citet{Terrill_2023}   \\ 
    & $\bf{0.05^{+0.01}_{-0.02}}$    & $0.06^{+0.01}_{-0.02}$         & $0.12^{+0.02}_{-0.05}$    & 0.6  & 1.3  & \citet{Matra_2025}     \\ 
    & $\bf{0.067^{+0.030}_{-0.034}}$ & $0.079^{+0.035}_{-0.040}$      & $0.158^{+0.071}_{-0.080}$ & 0.6  & 1.3  & \citet{Han_2025a}      \\ 
\hline
\multirow{7}{*}{\centering HD~32297} 
    & $0.008^{+0.001}_{-0.002}$    & $\bf{0.009^{+0.002}_{-0.001}}$ & $0.018^{+0.002}_{-0.004}$ & 0.06 & 0.89 & This work (parametric) \\ 
    & $\bf{0.009 \pm 0.001}$       & $0.011 \pm 0.001$              & $0.021 \pm 0.002$         & 0.06 & 0.89 & This work (\frank)     \\ 
    & $\bf{0.012 \pm 0.007}$       & $0.014 \pm 0.008$              & $0.028 \pm 0.016$         & 0.06 & 0.89 & This work (\rave)      \\ 
    & $\bf{0.08 \pm 0.01}$         & $0.09 \pm 0.01$                & $0.19 \pm 0.02$           & 0.5  & 1.34 & \citet{Terrill_2023}   \\ 
    & $\bf{0.015 \pm 0.002}$       & $0.018 \pm 0.002$              & $0.035 \pm 0.005$         & 0.5  & 1.34 & \citet{Worthen_2024}   \\ 
    & $\bf{0.08^{+0.01}_{-0.02}}$  & $0.09^{+0.01}_{-0.02}$         & $0.19^{+0.02}_{-0.05}$    & 0.5  & 1.34 & \citet{Matra_2025}     \\ 
    & $\bf{0.118^{+035}_{-0.044}}$ & $0.139^{+0.041}_{-0.052}$      & $0.278^{+0.082}_{-0.104}$ & 0.5  & 1.34 & \citet{Han_2025a}      \\ 
\hline
\multirow{5}{*}{\centering HD~39060}  
    & $0.095^{+0.001}_{-0.002}$      & $\bf{0.112^{+0.002}_{-0.001}}$ & $0.224^{+0.002}_{-0.004}$ & 0.22 & 1.33 & This work (parametric) \\ 
    & $\bf{0.062 \pm 0.002}$         & $0.073 \pm 0.002$              & $0.146 \pm 0.005$         & 0.22 & 1.33 & This work (\frank)     \\ 
    & $\bf{0.073 \pm 0.011}$         & $0.086 \pm 0.013$              & $0.172 \pm 0.026$         & 0.22 & 1.33 & This work (\rave)      \\ 
    & $\bf{0.014 \pm 0.006}$         & $0.016 \pm 0.007$              & $0.033 \pm 0.014$         & 0.3  & 1.33 & \citet{Matra_2019}\tablefootmark{c} \\ 
    & $\bf{0.110^{+0.008}_{-0.006}}$ & $0.130^{+0.009}_{-0.007}$      & $0.259^{+0.019}_{-0.014}$ & 0.3  & 1.33 & \citet{Matra_2019}\tablefootmark{d} \\ 
\hline
\multirow{6}{*}{\centering HD~61005}  
    & $0.013^{+0.002}_{-0.003}$      & $\bf{0.015^{+0.002}_{-0.003}}$ & $0.030^{+0.004}_{-0.006}$ & 0.15 & 0.89 & This work (parametric) \\ 
    & $\bf{0.017 \pm 0.001}$         & $0.020 \pm 0.001$              & $0.040 \pm 0.002$         & 0.15 & 0.89 & This work (\frank)     \\ 
    & $\bf{0.016 \pm 0.010}$         & $0.019 \pm 0.012$              & $0.038 \pm 0.024$         & 0.15 & 0.89 & This work (\rave)      \\ 
    & $\bf{0.039^{+0.003}_{-0.004}}$ & $0.046^{+0.004}_{-0.005}$      & $0.092^{+0.007}_{-0.009}$ & 0.46 & 1.29  & \citet{Terrill_2023}  \\ 
    & $\bf{0.044^{+0.003}_{-0.004}}$ & $0.052^{+0.004}_{-0.005}$      & $0.104^{+0.007}_{-0.009}$ & 0.46 & 1.29 & \citet{Matra_2025}     \\ 
    & $\bf{0.029^{+0.007}_{-0.009}}$ & $0.034^{+0.008}_{-0.011}$      & $0.068^{+0.016}_{-0.021}$ & 0.46 & 1.29 & \citet{Han_2025a}      \\ 
\hline
\multirow{3}{*}{\centering HD~76582}  
    & $0.163^{+0.008}_{-0.008}$      & $\bf{0.192^{+0.010}_{-0.009}}$ & $0.384^{+0.020}_{-0.018}$ & 0.56 & 0.89 & This work (parametric) \\ 
    & $\bf{0.055^{+0.011}_{-0.010}}$ & $0.065^{+0.013}_{-0.012}$      & $0.130^{+0.026}_{-0.024}$ & 0.56 & 0.89 & This work (\frank)     \\ 
    & $\bf{<0.116}$                  & $<0.137$                       & $<0.273$                  & 0.56 & 0.89 & This work (\rave)      \\ 
\hline
\end{tabular}
\end{table*}
\addtocounter{table}{-1}
\begin{table*}[h!]
\captionsetup{labelformat=empty}
\caption{Table \ref{tab:all_h} continued.}
\centering
\renewcommand{\arraystretch}{1.3}
\begin{tabular}{
    |>{\centering\arraybackslash}m{1.7cm}
    |>{\centering\arraybackslash}m{2.0cm}
    |>{\centering\arraybackslash}m{2.0cm}
    |>{\centering\arraybackslash}m{2.0cm}
    |>{\centering\arraybackslash}m{1.6cm}
    |>{\centering\arraybackslash}m{1cm}
    |>{\arraybackslash}m{3.6cm}|
}
\hline
Source & 
$h_{\sigma}$ & 
$h_{\rm HWHM}$ & 
$h_{\rm FWHM}$ & 
Resolution (arcsec) & 
$\lambda$ (mm) &
Reference \\
\hline
\multirow{2}{*}{\centering HD~92945}  
    & $\bf{0.033^{+0.011}_{-0.010}}$ & $0.039^{+0.013}_{-0.012}$ & $0.078^{+0.026}_{-0.024}$ & 0.40 & 0.86 & This work (\frank)     \\ 
    & $\bf{<0.096}$                  & $<0.113$                  & $<0.226$                  & 0.40 & 0.86 & This work (\rave)      \\ 
\hline
\multirow{2}{*}{\centering HD~107146}  
    & $\bf{0.113^{+0.019}_{-0.018}}$ & $0.133^{+0.022}_{-0.021}$ & $0.266^{+0.045}_{-0.042}$ & 0.59 & 0.86             & This work (\frank)     \\ 
    & $\bf{<0.184}$                  & $<0.217$                  & $<0.433$                  & 0.59 & 1.00\tablefootmark{e} & This work (\rave)      \\ 
\hline
\multirow{7}{*}{\centering HD~109573}
    & $<0.0099$              & $\bf{<0.0116}$    & $<0.0232$         & 0.08 & 0.89 & This work (parametric) \\ 
    & $\bf{<0.010}$          & $<0.012$          & $<0.024$          & 0.08 & 0.89 & This work (\frank)     \\ 
    & $\bf{<0.025}$          & $<0.029$          & $<0.059$          & 0.08 & 0.89 & This work (\rave)      \\ 
    & $\bf{0.038 \pm 0.005}$ & $0.045 \pm 0.006$ & $0.089 \pm 0.012$ & 0.17 & 0.88 & \citet{Kennedy_2018}   \\ 
    & $\bf{0.052 \pm 0.003}$ & $0.061 \pm 0.004$ & $0.122 \pm 0.007$ & 0.17 & 0.88 & \citet{Terrill_2023}   \\ 
    & $\bf{0.058 \pm 0.002}$ & $0.068 \pm 0.002$ & $0.137 \pm 0.005$ & 0.17 & 0.88 & \citet{Matra_2025}     \\ 
    & $\bf{0.048^{+0.014}_{-0.019}}$ & $0.057^{+0.016}_{-0.022}$ & $0.113^{+0.033}_{-0.045}$ & 0.17 & 0.88 & \citet{Han_2025a} \\ 
\hline
\multirow{2}{*}{\centering HD~121617}  
    & $\bf{0.090^{+0.010}_{-0.009}}$ & $0.106^{+0.012}_{-0.011}$ & $0.212^{+0.024}_{-0.021}$ & 0.12 & 0.89 & This work (\frank)     \\
    & $\bf{0.071^{+0.046}_{-0.049}}$ & $0.084^{+0.054}_{-0.058}$ & $0.167^{+0.108}_{-0.115}$ & 0.12 & 0.89 & This work (\rave)      \\
\hline
\multirow{4}{*}{\centering HD~131488}
    & $\bf{<0.0059}$                 & $<0.0070$                          & $<0.014$                     & 0.04 & 0.89 & This work (parametric)     \\ 
    & $\bf{<0.007}$                  & $<0.008$                           & $<0.016$                     & 0.04 & 0.89 & This work (\frank)     \\ 
    & $\bf{0.014 \pm 0.008}$         & $0.016 \pm 0.009$                  & $0.033 \pm 0.019$            & 0.04 & 0.89 & This work (\rave)      \\ 
    & $\bf{0.08 \pm 0.04}$           & $0.09 \pm 0.05$                    & $0.19 \pm 0.09$              & 0.54 & 1.33 & \citet{Worthen_2024}   \\ 
\hline
\multirow{3}{*}{\centering HD~131835}  
    & $0.019^{+0.006}_{-0.005}$      & $\bf{0.022^{+0.007}_{-0.006}}$ & $0.044^{+0.014}_{-0.012}$ & 0.13 & 0.89 & This work (parametric) \\ 
    & $\bf{0.033 \pm 0.005}$         & $0.039 \pm 0.006$              & $0.078 \pm 0.012$         & 0.13 & 0.89 & This work (\frank)     \\ 
    & $\bf{0.049^{+0.032}_{-0.031}}$ & $0.058^{+0.038}_{-0.036}$      & $0.115^{+0.075}_{-0.073}$ & 0.13 & 0.89 & This work (\rave)      \\ 
\hline
\multirow{2}{*}{\centering HD~145560}  
    & $\bf{0.059^{+0.014}_{-0.013}}$ & $0.069^{+0.016}_{-0.015}$ & $0.139^{+0.033}_{-0.031}$ & 0.09 & 0.89 & This work (\frank)     \\
    & $\bf{<0.123}$                  & $<0.145$                  & $<0.290$                  & 0.09 & 0.89 & This work (\rave)      \\
\hline
\multirow{6}{*}{\centering HD~161868}  
    & $0.163^{+0.010}_{-0.014}$      & $\bf{0.192^{+0.012}_{-0.016}}$ & $0.384^{+0.024}_{-0.032}$ & 0.52 & 0.89 & This work (parametric) \\
    & $\bf{<0.083}$                  & $<0.098$                       & $<0.195$                  & 0.52 & 0.89 & This work (\frank)     \\
    & $\bf{<0.187}$                  & $<0.220$                       & $<0.440$                  & 0.52 & 0.89 & This work (\rave)      \\
    & $\bf{0.13^{+0.04}_{-0.05}}$    & $0.15^{+0.05}_{-0.06}$         & $0.31^{+0.09}_{-0.12}$    & 1.4  & 1.27 & \citet{Matra_2025}     \\
    & $\bf{0.15^{+0.03}_{-0.04}}$    & $0.18^{+0.04}_{-0.05}$         & $0.35^{+0.07}_{-0.09}$    & 1.4  & 1.27 & \citet{Terrill_2023}   \\
    & $\bf{0.137^{+0.037}_{-0.042}}$ & $0.161^{+0.044}_{-0.049}$      & $0.323^{+0.087}_{-0.099}$ & 1.4  & 1.27 & \citet{Han_2025a}      \\
\hline
\multirow{2}{*}{\centering HD~170773}  
    & $\bf{<0.099}$ & $<0.117$ & $<0.233$ & 0.33 & 0.89 & This work (\frank)     \\
    & $\bf{<0.202}$ & $<0.238$ & $<0.476$ & 0.33 & 0.89 & This work (\rave)      \\
\hline
\multirow{9}{*}{\centering HD~197481}  
    & $0.0024^{+0.0008}_{-0.0002}$   & $\bf{0.0028^{+0.0009}_{-0.0002}}$ & $0.0056^{+0.0018}_{-0.0004}$ & 0.35 & 1.35 & This work (parametric) \\
    & $\bf{0.019 \pm 0.001}$         & $0.022 \pm 0.001$                 & $0.045 \pm 0.002$            & 0.35 & 1.35 & This work (\frank)     \\
    & $\bf{0.021 \pm 0.013}$         & $0.025 \pm 0.015$                 & $0.049 \pm 0.031$            & 0.35 & 1.35 & This work (\rave)      \\
    & $\bf{0.031^{+0.005}_{-0.004}}$ & $0.036^{+0.006}_{-0.005}$         & $0.073^{+0.012}_{-0.009}$    & 0.35 & 1.35 & \citet{Daley_2019}     \\
    & $\bf{0.025^{+0.008}_{-0.002}}$ & $0.029^{+0.009}_{-0.002}$         & $0.059^{+0.019}_{-0.005}$    & 0.35 & 1.35 & \citet{Vizgan_2022}    \\
    & $\bf{0.017 \pm 0.003}$         & $0.020 \pm 0.004$                 & $0.040 \pm 0.007$            & 0.35 & 1.35 & \citet{Matra_2025}     \\
    & $\bf{0.020 \pm 0.002}$         & $0.024 \pm 0.002$                 & $0.047 \pm 0.005$            & 0.35 & 1.35 & \citet{Terrill_2023}   \\
    & $\bf{0.022 \pm 0.003}$         & $0.026 \pm 0.004$                 & $0.052 \pm 0.007$            & 0.35 & 1.35 & \citet{Han_2025a}      \\
    & $\bf{0.019^{+0.006}_{-0.001}}$ & $0.022^{+0.007}_{-0.001}$         & $0.045^{+0.014}_{-0.002}$    & 0.23 & 0.45 & \citet{Vizgan_2022}    \\
\hline
\multirow{2}{*}{\centering HD~206893}  
    & $\bf{0.096^{+0.021}_{-0.029}}$ & $0.113^{+0.025}_{-0.034}$ & $0.226^{+0.049}_{-0.068}$ & 0.45 & 1.35 & This work (\frank)     \\
    & $\bf{<0.271}$                  & $<0.319$                  & $<0.638$                  & 0.45 & 1.35 & This work (\rave)      \\
\hline
\end{tabular}
\tablefoot{The originally reported value is shown in bold, while conversions from this value are provided such that $h_{\sigma}$, $h_{\rm HWHM}$, and $h_{\rm FWHM}$ are shown for each source. We note that for some entries, $h_{\sigma}$ may not strictly be a standard deviation if a standard deviation is undefined for the method used (e.g., as in the parametric vertical Lorentzian fits shown in this work).
For HD~10647, the contributions of the multicomponent fit contain \tablefoottext{a}{$88 \pm 6$\% of the mass} and \tablefoottext{b}{$12 \pm 6$\% of the mass.} For HD~39060, the contributions of the multicomponent fit from \citet{Matra_2019} contain \tablefoottext{c}{$20^{+5}_{-4}$\% of the mass} and \tablefoottext{d}{$80^{+4}_{-5}$\% of the mass.} \tablefoottext{e}{The \rave analysis of HD~10746 used combined Band 6 and Band 7 observations.}
}
\end{table*}
\clearpage

\begin{figure*}[h!]
\centering
\includegraphics[width=\linewidth]{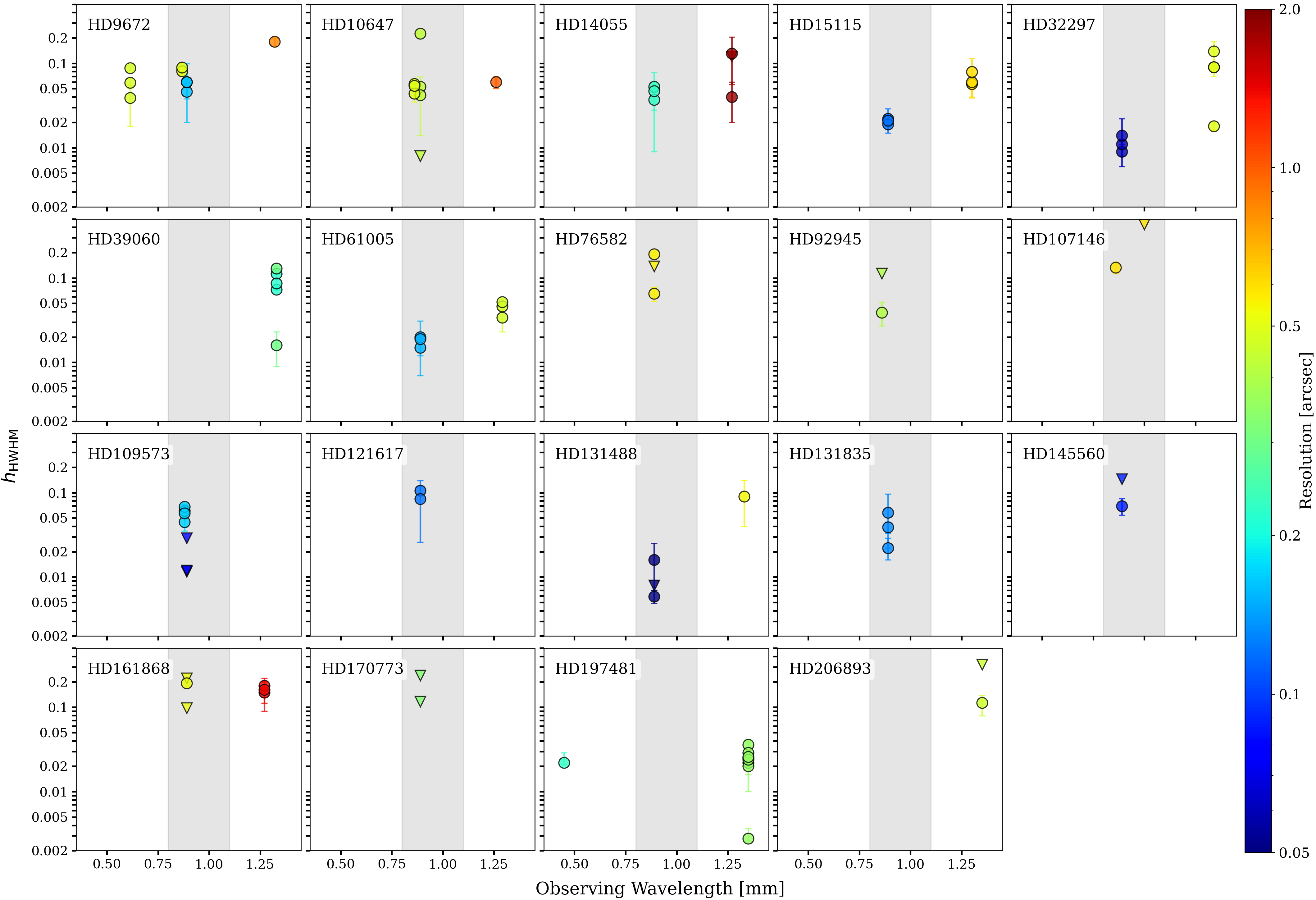}
\caption{All values of $h_{\rm HWHM}$ measured in this work and previous studies (Table \ref{tab:all_h}) with respect to observing wavelength. The gray shaded region shows the wavelength range of ALMA Band 7. The color bar shows the angular resolution of the observations used for each measurement. Triangular points indicate measurements which are upper limits, while circular points show the upper and lower $1\sigma$ uncertainties associated with the measurement.}
\label{fig:all_h}
\end{figure*}

\section{Comparison of disk mass estimates} \label{appendix:krivovcompare}

\begin{table*}[htbp]
    \centering
    \renewcommand{\arraystretch}{1.3}
\caption{Comparison of disk mass estimates from \citet{Pan_2012} (1) and \citetalias{Krivov_2021} (2), as well as the assumed millimeter dust masses $M_{\rm mm}$ used in each of these models.}
    \label{table:mdisc}
\rowcolors{2}{gray!15}{white}
\begin{tabular}{l|cc|cc|c|c}
        \hline
        & $M_{\rm disk}$ [$M_{\oplus}$] & $M_{\rm disk}^{\rm min}$ [$M_{\oplus}$] 
        & $M_{\rm mm}$ [$M_{\oplus}$] & $M_{\rm mm}$ [$M_{\oplus}$] 
        & Rescaled $M_{\rm disk}^{\rm min}$
        & Compatible? \\
        Target & (1) & (2) & (1) & (2) & [$M_{\oplus}$] & \\
        \hline
HD~9672   & 86.6  & 14.3  & $4.95\times 10^{-1}$  & $2.8\times 10^{-1}$ & 25.3 & yes           \\
HD~15115  & 3.72  & 7.0   & $6.17\times 10^{-3}$  & $8.5\times 10^{-2}$ & 0.51 & yes, rescaled \\
HD~39060  & 104   & 5.5   & $1.83\times 10^{-1}$  & $7.9\times 10^{-2}$ & 12.7 & yes           \\
HD~61005  & 10.2  & 14.2  & $1.26\times 10^{-1}$  & $1.3\times 10^{-1}$ & 13.8 & no            \\
HD~131488 & 9.01  & 85.8  & $2.75\times 10^{-1}$  & $7.2\times 10^{-1}$ & 32.8 & no            \\
HD~131835 & 49.7  & 39.1  & $4.08\times 10^{-1}$  & $3.8\times 10^{-1}$ & 42.0 & yes           \\
HD~197481 & 0.830 & 1.8   & $1.49\times 10^{-2}$  & $1.4\times 10^{-2}$ & 1.92 & no            \\
\hline
\end{tabular}
\tablefoot{We rescale the originally reported $M_{\rm disk}^{\rm min}$ to use the $M_{\rm mm}$ obtained from our fiducial models. The rescaled $M_{\rm disk}^{\rm min}$ are compatible with the $M_{\rm disk}$ obtained from the \citet{Pan_2012} model for HD~9672, HD~15115, HD~39060, and HD~131835, and incompatible for HD~61005, HD~131488, and HD~197481. These discrepancies may stem from differences in model assumptions (see Sect. \ref{sec:discuss:solids}). For the \citet{Pan_2012} model, difficulties constraining the scale height aspect ratio (which sets model parameter $\left\langle v_{\rm rel}\right\rangle$) may also contribute, particularly for HD~131488 and HD~197481 which are both noted to have $h$ measurements limited by the resolution of the observations.}
\end{table*}

\end{document}